\documentclass[times,twocolumn]{aastex62}

\usepackage{CJK}
\usepackage{amsmath}
\usepackage{multirow}
\usepackage{cases}
\usepackage{graphicx}
\usepackage{subfigure}
\usepackage{natbib}
\usepackage{color}
\usepackage{tabularx}
\usepackage{bm}
\usepackage{threeparttable}
\usepackage{gensymb}
\usepackage{booktabs}
\usepackage{array}
\usepackage{placeins}

\newcommand{\thetae}{\theta_{\rm E}}
\newcommand{\thetas}{\theta_{*}}

\newcommand{\pie}{\pi_{\rm E}}

\newcommand{\murel}{\mu_{\rm rel}}
\newcommand{\te}{t_{\rm E}}

\newcommand{\eventa}{KMT-2023-BLG-1810}
\newcommand{\eventb}{KMT-2023-BLG-0084}
\newcommand{\eventc}{KMT-2023-BLG-0584}
\newcommand{\eventd}{KMT-2023-BLG-1118}
\newcommand{\evente}{KMT-2023-BLG-1697}
\newcommand{\eventf}{KMT-2023-BLG-2218}

\DeclareGraphicsExtensions{.pdf,.png,.jpg}

\shorttitle{}
\shortauthors{Li et al.}

\begin{document}
\begin{CJK*}{UTF8}{gbsn}
\title{{\large Mass Production of 2023 KMTNet Microlensing Planets. III: Three Planets from the Subprime Field}}

\correspondingauthor{Hongyu Li, Weicheng Zang}
\email{lihongyu25@mails.tsinghua.edu.cn, zangweicheng@westlake.edu.cn}

\author{Hongyu Li}
\affiliation{Department of Astronomy, Westlake University, Hangzhou 310030, Zhejiang Province, China}
\affiliation{Department of Astronomy, Tsinghua University, Beijing 100084, China}

\author{Zhixing Li}
\affiliation{Department of Astronomy, Westlake University, Hangzhou 310030, Zhejiang Province, China}

\author[0000-0001-6000-3463]{Weicheng Zang}
\affiliation{Department of Astronomy, Westlake University, Hangzhou 310030, Zhejiang Province, China}

\author[0000-0001-9823-2907]{Yoon-Hyun Ryu} 
\affiliation{Korea Astronomy and Space Science Institute, Daejeon 34055, Republic of Korea}

\author[0000-0001-5207-5619]{Andrzej Udalski}
\affiliation{Astronomical Observatory, University of Warsaw, Al. Ujazdowskie 4, 00-478 Warszawa, Poland}

\author{Takahiro Sumi}
\affiliation{Department of Earth and Space Science, Graduate School of Science, Osaka University, Toyonaka, Osaka 560-0043, Japan}

\author[0000-0003-0626-8465]{Hongjing Yang}
\affiliation{Westlake Institute for Advanced Study, Hangzhou 310030, Zhejiang Province, China}
\affiliation{Department of Astronomy, Westlake University, Hangzhou 310030, Zhejiang Province, China}

\author[0000-0002-1279-0666]{Jiyuan Zhang}
\affiliation{Department of Astronomy, Tsinghua University, Beijing 100084, China}

\author[0000-0001-8317-2788]{Shude Mao}
\affiliation{Department of Astronomy, Westlake University, Hangzhou 310030, Zhejiang Province, China}

\collaboration{(Leading Authors)}

\author[0000-0003-3316-4012]{Michael D. Albrow}
\affiliation{University of Canterbury, School of Physical and Chemical Sciences, Private Bag 4800, Christchurch 8020, New Zealand}

\author[0000-0001-6285-4528]{Sun-Ju Chung}
\affiliation{Korea Astronomy and Space Science Institute, Daejeon 34055, Republic of Korea}

\author{Andrew Gould} 
\affiliation{Max-Planck-Institute for Astronomy, K\"onigstuhl 17, 69117 Heidelberg, Germany}
\affiliation{Department of Astronomy, Ohio State University, 140 W. 18th Ave., Columbus, OH 43210, USA}

\author{Cheongho Han}
\affiliation{Department of Physics, Chungbuk National University, Cheongju 28644, Republic of Korea}

\author[0000-0002-9241-4117]{Kyu-Ha Hwang}
\affiliation{Korea Astronomy and Space Science Institute, Daejeon 34055, Republic of Korea}

\author[0000-0002-0314-6000]{Youn Kil Jung}
\affiliation{Korea Astronomy and Space Science Institute, Daejeon 34055, Republic of Korea}
\affiliation{National University of Science and Technology (UST), Daejeon 34113, Republic of Korea}

\author[0000-0002-4355-9838]{In-Gu Shin}
\affiliation{Department of Astronomy, Westlake University, Hangzhou 310030, Zhejiang Province, China}

\author[0000-0003-1525-5041]{Yossi Shvartzvald}
\affiliation{Department of Particle Physics and Astrophysics, Weizmann Institute of Science, Rehovot 7610001, Israel}

\author[0000-0001-9481-7123]{Jennifer C. Yee}
\affiliation{Center for Astrophysics $|$ Harvard \& Smithsonian, 60 Garden St.,Cambridge, MA 02138, USA}

\author[0000-0002-7511-2950]{Sang-Mok Cha} 
\affiliation{Korea Astronomy and Space Science Institute, Daejeon 34055, Republic of Korea}
\affiliation{School of Space Research, Kyung Hee University, Yongin, Kyeonggi 17104, Republic of Korea} 

\author{Dong-Jin Kim}
\affiliation{Korea Astronomy and Space Science Institute, Daejeon 34055, Republic of Korea}

\author[0000-0003-0562-5643]{Seung-Lee Kim} 
\affiliation{Korea Astronomy and Space Science Institute, Daejeon 34055, Republic of Korea}

\author[0000-0003-0043-3925]{Chung-Uk Lee}
\affiliation{Korea Astronomy and Space Science Institute, Daejeon 34055, Republic of Korea}

\author[0009-0000-5737-0908]{Dong-Joo Lee} 
\affiliation{Korea Astronomy and Space Science Institute, Daejeon 34055, Republic of Korea}

\author[0000-0001-7594-8072]{Yongseok Lee} 
\affiliation{Korea Astronomy and Space Science Institute, Daejeon 34055, Republic of Korea}
\affiliation{School of Space Research, Kyung Hee University, Yongin, Kyeonggi 17104, Republic of Korea}

\author[0000-0002-6982-7722]{Byeong-Gon Park}
\affiliation{Korea Astronomy and Space Science Institute, Daejeon 34055, Republic of Korea}

\author[0000-0003-1435-3053]{Richard W. Pogge} 
\affiliation{Department of Astronomy, Ohio State University, 140 West 18th Ave., Columbus, OH  43210, USA}
\affiliation{Center for Cosmology and AstroParticle Physics, Ohio State University, 191 West Woodruff Ave., Columbus, OH 43210, USA}

\collaboration{(The KMTNet Collaboration)}

\author[0000-0001-7016-1692]{Przemek Mr\'{o}z}
\affiliation{Astronomical Observatory, University of Warsaw, Al. Ujazdowskie 4, 00-478 Warszawa, Poland}

\author[0000-0002-0548-8995]{Micha{\l}~K. Szyma\'{n}ski}
\affiliation{Astronomical Observatory, University of Warsaw, Al. Ujazdowskie 4, 00-478 Warszawa, Poland}

\author[0000-0002-2335-1730]{Jan Skowron}
\affiliation{Astronomical Observatory, University of Warsaw, Al. Ujazdowskie 4, 00-478 Warszawa, Poland}

\author[0000-0002-9245-6368]{Radoslaw Poleski}
\affiliation{Astronomical Observatory, University of Warsaw, Al. Ujazdowskie 4, 00-478 Warszawa, Poland}

\author[0000-0002-7777-0842]{Igor Soszy\'{n}ski}
\affiliation{Astronomical Observatory, University of Warsaw, Al. Ujazdowskie 4, 00-478 Warszawa, Poland}

\author[0000-0002-2339-5899]{Pawe{\l} Pietrukowicz}
\affiliation{Astronomical Observatory, University of Warsaw, Al. Ujazdowskie 4, 00-478 Warszawa, Poland}

\author[0000-0003-4084-880X]{Szymon Koz{\l}owski}
\affiliation{Astronomical Observatory, University of Warsaw, Al. Ujazdowskie 4, 00-478 Warszawa, Poland}

\author[0000-0002-9326-9329]{Krzysztof A. Rybicki}
\affiliation{Astronomical Observatory, University of Warsaw, Al. Ujazdowskie 4, 00-478 Warszawa, Poland}
\affiliation{Department of Particle Physics and Astrophysics, Weizmann Institute of Science, Rehovot 76100, Israel}

\author[0000-0002-6212-7221]{Patryk Iwanek}
\affiliation{Astronomical Observatory, University of Warsaw, Al. Ujazdowskie 4, 00-478 Warszawa, Poland}

\author[0000-0001-6364-408X]{Krzysztof Ulaczyk}
\affiliation{Department of Physics, University of Warwick, Gibbet Hill Road, Coventry, CV4~7AL,~UK}

\author[0000-0002-3051-274X]{Marcin Wrona}
\affiliation{Astronomical Observatory, University of Warsaw, Al. Ujazdowskie 4, 00-478 Warszawa, Poland}
\affiliation{Villanova University, Department of Astrophysics and Planetary Sciences, 800 Lancaster Ave., Villanova, PA 19085, USA}

\author[0000-0002-1650-1518]{Mariusz Gromadzki}
\affiliation{Astronomical Observatory, University of Warsaw, Al. Ujazdowskie 4, 00-478 Warszawa, Poland}

\author{Mateusz J. Mr\'{o}z}
\affiliation{Astronomical Observatory, University of Warsaw, Al. Ujazdowskie 4, 00-478 Warszawa, Poland}

\collaboration{(The OGLE Collaboration)}

\author{Fumio Abe}
\affiliation{Institute for Space-Earth Environmental Research, Nagoya University, Nagoya 464-8601, Japan}

\author{David P. Bennett}
\affiliation{Code 667, NASA Goddard Space Flight Center, Greenbelt, MD 20771, USA}
\affiliation{Department of Astronomy, University of Maryland, College Park, MD 20742, USA}

\author{Aparna Bhattacharya}
\affiliation{Code 667, NASA Goddard Space Flight Center, Greenbelt, MD 20771, USA}
\affiliation{Department of Astronomy, University of Maryland, College Park, MD 20742, USA}

\author{Ian A. Bond}
\affiliation{Institute of Natural and Mathematical Sciences, Massey University, Auckland 0745, New Zealand}

\author{Ryusei Hamada}
\affiliation{Department of Earth and Space Science, Graduate School of Science, Osaka University, Toyonaka, Osaka 560-0043, Japan}

\author{Yuki Hirao}
\affiliation{Department of Earth and Space Science, Graduate School of Science, Osaka University, Toyonaka, Osaka 560-0043, Japan}

\author{Stela Ishitani Silva}
\affiliation{Department of Physics, The Catholic University of America, Washington, DC 20064, USA}
\affiliation{Code 667, NASA Goddard Space Flight Center, Greenbelt, MD 20771, USA}

\author{Shota Miyazaki}
\affiliation{Department of Earth and Space Science, Graduate School of Science, Osaka University, Toyonaka, Osaka 560-0043, Japan}

\author{Yasushi Muraki}
\affiliation{Institute for Space-Earth Environmental Research, Nagoya University, Nagoya 464-8601, Japan}

\author{Kansuke Nunota}
\affiliation{Department of Earth and Space Science, Graduate School of Science, Osaka University, Toyonaka, Osaka 560-0043, Japan}

\author{Greg Olmschenk}
\affiliation{Code 667, NASA Goddard Space Flight Center, Greenbelt, MD 20771, USA}

\author{Cl\'ement Ranc}
\affiliation{Sorbonne Universit\'e, CNRS, Institut d'Astrophysique de Paris, IAP, F-75014, Paris, France}

\author{Nicholas J. Rattenbury}
\affiliation{Department of Physics, University of Auckland, Private Bag 92019, Auckland, New Zealand}

\author[0000-0002-1228-4122]{Yuki K. Satoh}
\affiliation{College of Science and Engineering, Kanto Gakuin University, Yokohama, Kanagawa 236-8501, Japan}

\author{Daisuke Suzuki}
\affiliation{Department of Earth and Space Science, Graduate School of Science, Osaka University, Toyonaka, Osaka 560-0043, Japan}

\author{Takuto Tamaoki}
\affiliation{Department of Earth and Space Science, Graduate School of Science, Osaka University, Toyonaka, Osaka 560-0043, Japan}

\author{Sean Terry}
\affiliation{Code 667, NASA Goddard Space Flight Center, Greenbelt, MD 20771, USA}
\affiliation{Department of Astronomy, University of Maryland, College Park, MD 20742, USA}
\author{Paul J. Tristram}
\affiliation{University of Canterbury Mt.\ John Observatory, P.O. Box 56, Lake Tekapo 8770, New Zealand}

\author{Aikaterini Vandorou}
\affiliation{Code 667, NASA Goddard Space Flight Center, Greenbelt, MD 20771, USA}
\affiliation{Department of Astronomy, University of Maryland, College Park, MD 20742, USA}
\author{Hibiki Yama}
\affiliation{Department of Earth and Space Science, Graduate School of Science, Osaka University, Toyonaka, Osaka 560-0043, Japan}

\collaboration{(The MOA Collaboration)}

\begin{abstract}
To complete the analysis of the 2023 KMTNet subprime-field microlensing planetary events identified by its AlertFinder system, we present the analysis of six events, KMT-2023-BLG-(1810, 0084, 1118, 0584, 1697, 2218). We find that the first three events are securely confirmed as planetary, with inferred mass ratios of $\log q \sim -1.9$, $-2.0$, and $-2.6$, respectively. The remaining three events exhibit the well-known degeneracy between binary-lens/single-source (2L1S) and single-lens/binary-source (1L2S) models, and two of these also admit viable stellar binary solutions. A Bayesian analysis indicates that the companions in the confirmed planetary events are likely either super-Jupiters orbiting beyond the snow line of M- or K-dwarf hosts or, for two degenerate solutions of KMT-2023-BLG-1118, Saturn-mass planets orbiting late-type M dwarfs. To date, the 2023 KMTNet sample contains 25 unambiguous planetary events, and its mass-ratio distribution is consistent with that of the KMTNet planetary sample from 2016--2019.

\end{abstract}

\section{Introduction}\label{sec:intro}

Since 2016, the Korea Microlensing Telescope Network (KMTNet; \citealt{KMT2016}) has conducted a wide-field photometric survey covering approximately $97~\mathrm{deg}^2$ toward the Galactic bulge to search for exoplanets via the gravitational microlensing technique \citep{Shude1991, Andy1992}. KMTNet consists of three identical 1.6-m telescopes equipped with $4~\mathrm{deg}^2$ cameras, located in Chile (KMTC), South Africa (KMTS), and Australia (KMTA). This global configuration enables continuous monitoring of microlensing events throughout the bulge observing season. Of the total survey area, about $13~\mathrm{deg}^2$ is monitored at a cadence of $\Gamma \geq 2~\mathrm{hr}^{-1}$, while the remaining $\sim 84~\mathrm{deg}^2$ is observed at a lower cadence of $\Gamma \leq 1~\mathrm{hr}^{-1}$. These regions are referred to as the prime and subprime fields, respectively (see Figure 12 of \citealt{KMTeventfinder} for the field layout and observing cadences). To date, KMTNet has played a major or decisive role in the discovery of more than 200 microlensing planets, out of a total of about 280 known detections \citep{NASAExo}.

The publication of KMTNet planets proceeds mainly through two approaches. First, during the observing season and within a few years afterward, planets identified through by-eye inspection are analyzed and published by interested researchers. These studies often result in single-event papers, particularly for planets with notable characteristics, such as two-planet events (e.g., \citealt{OB181011}) or planets with very low planet-to-host mass ratios ($q$) (e.g., \citealt{KB180029}). Second, several years after each observing season, the semi-automated KMTNet AnomalyFinder system \citep{OB191053,2019_prime} systematically searches for planetary signals. Unpublished by-eye planets, together with newly identified planets sharing similar properties, such as event location (prime or subprime), mass ratios, and observing season, are then grouped into a single paper. This ``mass production'' approach has produced 14 papers, which collectively complete the publication of all KMTNet planets observed between 2016 and 2019 (see \citealt{2017_subprime} and references therein).

\cite{OB160007} constructed a statistical sample of 63 planets, including all planets with $\log q < -4$ from 2016--2019 and planets with $\log q > -4$ from the 2018 and 2019 seasons, representing the largest microlensing planetary sample to date. This sample reveals a possible ``mass-ratio desert'' at $-3.6 < \log q < -3.0$, which separates two populations of microlensing planets: one corresponding to super-Earths or mini-Neptunes and the other to gas giants. The observed distribution was reproduced, within the uncertainties, by \cite{Guo2026} using a Monte Carlo-based planet population synthesis model combined with N-body simulations. Furthermore, the potential desert was confirmed in the 2016--2017 sample for planets with $\log q > -4$. However, current uncertainties in both the strength of the desert and the frequency of planets across different mass-ratio ranges remain large, emphasizing the need for a larger sample. Moreover, a larger sample could also reveal additional features in the mass-ratio distribution.

Therefore, we have been conducting the AnomalyFinder search, together with the systematic analysis and publication of KMTNet planets observed after the 2020 season\footnote{The 2020 season is not included in this study due to the shutdown of KMTC and KMTS caused by COVID-19.}. An introduction to the current analysis plan can be found in \cite{2023_prime}. In this work, we present the discovery and analysis of six candidate planetary events from the 2023 season. These events are located in the KMTNet subprime fields and were first identified by the KMTNet AlertFinder system \citep{KMTAF}.

The structure of the paper is as follows. In Section~\ref{sec:obser}, we describe the observations and data reduction for these events. Section~\ref{sec:model} details the light-curve analysis, and Section~\ref{sec:lens} presents the source and lens properties. Finally, we summarize the planetary sample from the 2023 season events discovered by the KMTNet AlertFinder system.

\section{Observations}\label{sec:obser}

\begin{table*}
    \renewcommand\arraystretch{1.35}
    \centering
    \caption{Event Names, Alert, Locations, and Cadences for the five events analyzed in this paper}
    \begin{tabular}{c c c c c c c}
    \hline
    \hline
    Event Name & First Alert Date & ${\rm RA}_{\rm J2000}$ & ${\rm Decl.}_{\rm J2000}$ & $\ell$ & $b$ & Cadence \\
    \hline
    \eventa & 25 Jul 2023 & 18:02:27.53 & $-$30:16:39.40 & +0.6941 & $-$3.8356	 & $1.0~{\rm hr}^{-1}$ \\
    \hline
    \eventb & 17 Mar 2023 & 18:15:46.06 & $-$25:13:46.20 & +6.5488 & $-$3.9888 & $0.4~{\rm hr}^{-1}$ \\
    MOA-2023-BLG-105 & & & & & & $\sim 0.4~{\rm hr}^{-1}$ \\
    \hline
    \eventc & 27 Apr 2023 & 17:32:12.29 & $-$27:16:01.74 & $-$0.1643 & +3.3925 & $1.2~{\rm hr}^{-1}$  \\
    \hline 
    \eventd & 02 Jun 2023 & 18:10:27.44 & $-$26:16:21.61 & +5.0591 & $-$3.4339 & $0.4~{\rm hr}^{-1}$ \\
    OGLE-2023-BLG-0752 & & & & & & $1~{\rm night}^{-1}$ \\
    \hline
    \evente & 17 Jul 2023 & 17:46:35.95 & $-$24:09:53.10 & +4.1911 & +2.2930 & $1.0~{\rm hr}^{-1}$ \\
    OGLE-2023-BLG-1051 & & & & & & $1~{\rm night}^{-1}$ \\
    \hline
    \eventf & 14 Sep 2023 & 17:37:57.05 & $-$29:06:52.63 & $-$1.0368 & +1.3295 & $1.0~{\rm hr}^{-1}$ \\
    OGLE-2023-BLG-1291 & & & & & & $1~{\rm night}^{-1}$ \\
    \hline
    \hline
    \end{tabular}
    \label{event_info}
\end{table*}

All six events were initially identified by the KMTNet AlertFinder system. Subsequently, three of them were independently reported by the Early Warning System of  \citep{Udalski1994, Udalski2003} and one event, KMT-2023-BLG-0084/MOA-2023-BLG-105, was later recognized by the MOA survey. Table \ref{event_info} summarizes the basic observational characteristics of these events, including their designations, first alert dates, equatorial and Galactic coordinates, and the cadences of the contributing surveys. We designate them by their first discovery names. 

As introduced above, KMTNet observations were obtained with three 1.6 m telescopes. OGLE conducted observations with its 1.3 m telescope in Chile, equipped with a $1.4,{\rm deg}^2$ camera \citep{OGLEIV}. The MOA survey used a 1.8 m telescope in New Zealand with a $2.2,{\rm deg}^2$ camera. Most images from KMTNet and OGLE were obtained in the $I$ band, while MOA primarily employed its broad MOA-Red filter, which roughly corresponds to the combined Cousins $R$ and $I$ bands. All three surveys also obtained a fraction of $V$-band images to enable color measurements of the source stars.

For the light-curve analysis, we used the data products generated by each survey's difference image analysis pipeline: \cite{pysis, Yang_TLC, Yang_TLC2} for KMTNet, \cite{Wozniak2000} for OGLE, and \cite{Bond2001} for MOA. The photometric uncertainties were then renormalized following the prescription of \cite{MB11293}, so that the reduced $\chi^{2}$/dof of each data set is unity, where dof is the number of degrees of freedom.

\section{Light-curve Analysis}\label{sec:model}

\begin{table}[htb]
  \renewcommand\arraystretch{1.20}
  \centering
  \caption{Lensing Parameters for \eventa}
  \begin{tabular}{c|c c}
    \hline
    \hline
    Parameter & A & B \\
    \hline
    $\chi^2$/dof & 1417.2/1412 & 1423.7/1412 \\ \hline
    $t_0$ (HJD$'$) & $152.07 \pm 0.26$ & $151.95 \pm 0.21$ \\
    $u_0$ & $0.223 \pm 0.024$ & $0.204 \pm 0.027$ \\
    $t_{\rm E}$ (days) & $40.1 \pm 3.1$ & $43.1 \pm 3.5$ \\
    $\rho (10^{-3})$ & $1.08 \pm 0.27$ & $1.16 \pm 0.27$ \\
    $\alpha$ (deg) & $240.1 \pm 1.1$ & $239.2 \pm 1.2$ \\
    $s$ & $1.257 \pm 0.009$ & $1.044 \pm 0.019$ \\
    $\log q$ & $-1.903 \pm 0.046$ & $-1.974 \pm 0.054$ \\
    $I_{\rm S,KMTC}$ & $20.953 \pm 0.023$ & $21.083 \pm 0.022$ \\
    \hline\hline
  \end{tabular}
  \tablecomments{HJD$^\prime = \mathrm{HJD} -2460000$.}
  \label{tab:parm-a}
\end{table}

\begin{figure}
    \centering
    \includegraphics[width=0.47\textwidth]{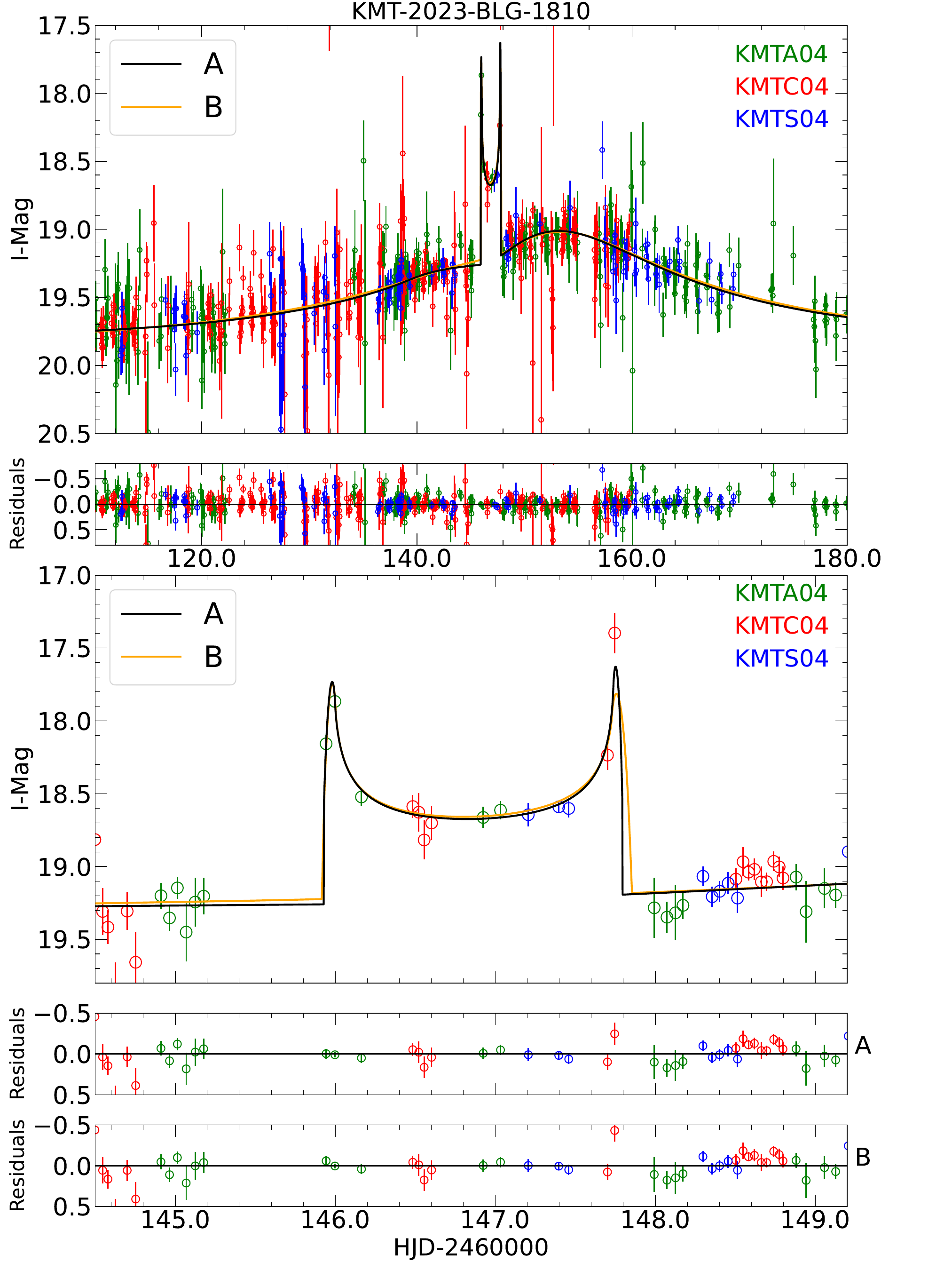}
    \caption{Light curve of \eventa\ with the 2L1S models shown as solid black and yellow curves. The lower panel provides a zoomed-in view of the U-shaped anomaly. Data from different observatories are plotted in different colors. The corresponding 2L1S model parameters are listed in Table~\ref{tab:parm-a}.}
\label{fig:lc1810}
\end{figure}

\begin{figure*}
    \centering
    \includegraphics[width=0.95\textwidth]{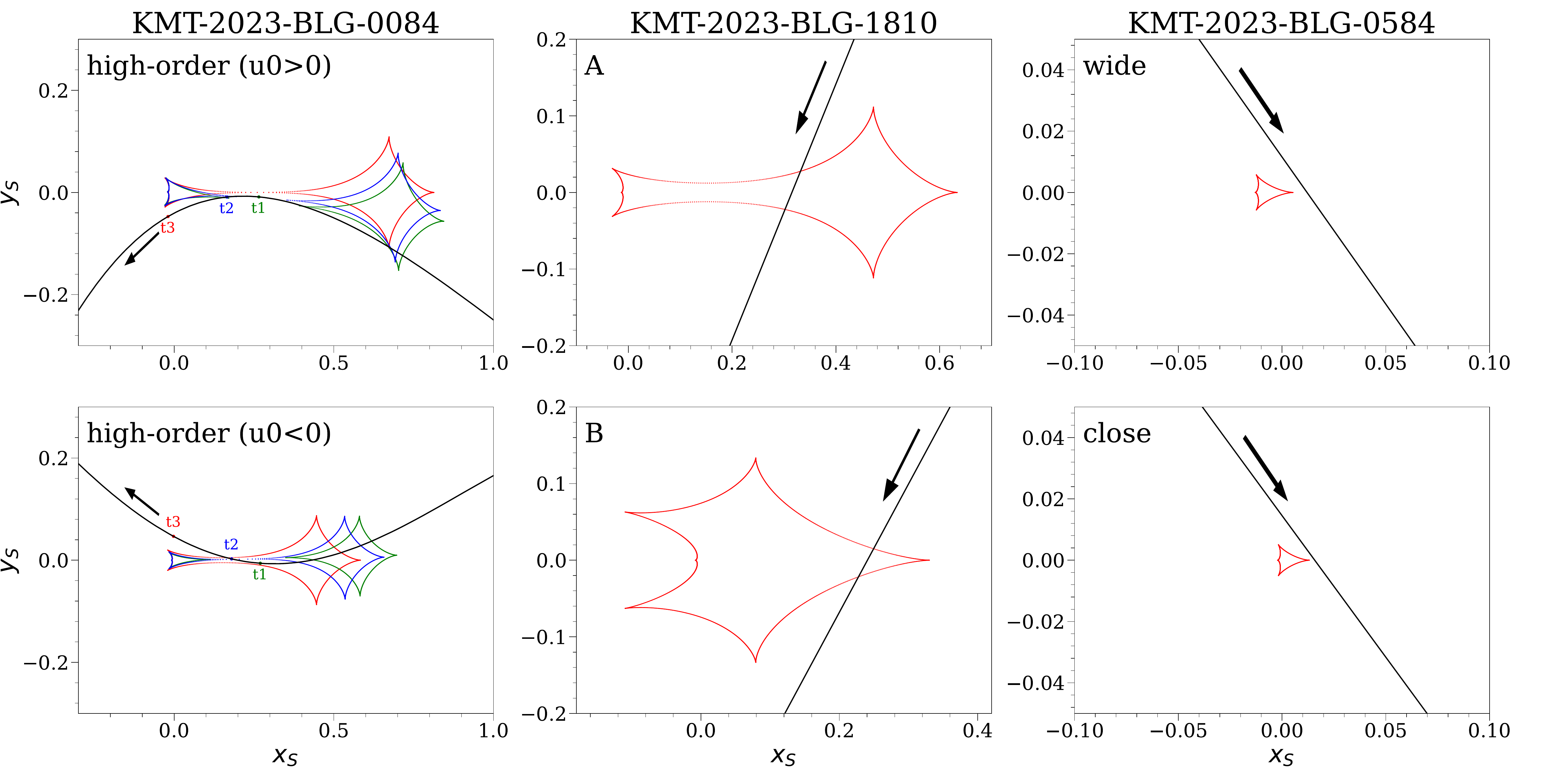}
    \caption{Geometries of the 2L1S models for events \eventb, \eventa, and \eventc. For each event, the geometries of different solutions are arranged vertically. For \eventb, the source positions and the caustic structures at $t_1=10.0$, $t_2=20.0$ and $t_3=40.0$ are presented in green blue and red, respectively. $t_1$ is an epoch before the anomaly, $t_2$ corresponds to the anomaly, and $t_3$ corresponds to the primary peak in the modeling light-curve (Figure~\ref{fig:lc0084}). For \eventa\ and \eventc, the red curves indicate the caustic structures, the black curve is the source trajectory, and the arrow marks the direction of the source motion.}
    \label{fig:cau1}
\end{figure*}

\subsection{Preamble}\label{model_preamble}

In this section, we analyze the light curves following an approach broadly similar to that used in \cite{-4planet}, but we summarize the key model parameters and procedures here to avoid repeating material from earlier work. Readers interested in a more extensive discussion can consult \cite{-4planet}.

Each event is first modeled with a static binary-lens single-source (2L1S) framework. This parameterization consists of seven quantities: the three standard point-source point-lens (PSPL, \citealt{Paczynski1986}) parameters $(t_0, u_0, \te)$, the time of closest approach, the impact parameter expressed in units of the angular Einstein radius $\thetae$, and the Einstein timescale, respectively,
\begin{equation}
\theta_{\rm E} = \sqrt{\kappa M_{\rm L}\pi_{\rm rel}}, \qquad
\te = \frac{\theta_{\rm E}}{\mu_{\rm rel}},
\label{equ:1}
\end{equation}
where $\kappa = 4G/(c^2{\rm au}) \simeq 8.144,{\rm mas}M_{\odot}^{-1}$, $M_{\rm L}$ is the total lens mass, and $(\pi_{\rm rel}, \mu_{\rm rel})$ denote the lens-source relative parallax and proper motion. The binary lens configuration is encoded by $(s, q, \alpha)$, representing the projected separation normalized to $\thetae$, the binary mass ratio, and the trajectory angle. The final parameter, $\rho = \theta_*/\thetae$, is the normalized source radius, where $\theta_*$ is the angular source radius. 

Magnifications for 2L1S models, $A(t)$, are evaluated using the \texttt{VBBinaryLensing} contour-integration algorithm \citep{Bozza2010,Bozza2018,VBMicrolensing2025}. For each data set $i$, we adopt two linear flux parameters, $f_{{\rm S},i}$ and $f_{{\rm B},i}$, corresponding to the source and blend fluxes, so the observed flux is fitted as
\begin{equation}
f_i(t) = f_{{\rm S},i} A(t) + f_{{\rm B},i}.
\label{equ:2}
\end{equation}

Our fitting strategy begins with a coarse grid search over $(\log q, \log s, \rho)$, while $(t_0, u_0, \te, \alpha)$ are allowed to vary. The local minima are then explored with a dense grid search over $(\log q, \log s, \rho)$. Each retained minimum is refined using an Markov chain Monte Carlo (MCMC) exploration of the $\chi^2$ surface with the \texttt{emcee} sampler \citep{emcee2,emcee}, followed by downhill minimization using \texttt{SciPy} \citep{scipy}. Around all local minima, we then allow all parameters to vary freely. Reported quantities correspond to the posterior medians with 68\% credible intervals.

For all events, we also examine whether microlensing parallax \citep{Gould1992,Gould2000,Gouldpies2004} provides additional constraints. The parallax vector,
\begin{equation}
\boldsymbol{\pi}_{\rm E} = \frac{\pi_{\rm rel}}{\theta_{\rm E}} \frac{\boldsymbol{\mu}_{\rm rel}}{\mu_{\rm rel}},
\label{equ:3}
\end{equation}
is expressed through its north and east components $(\pi_{\rm E,N}, \pi_{\rm E,E})$. When parallax is included, we additionally consider orbital motion of the lens \citep{MB09387,OB09020} and evaluate both $u_0 > 0$ and $u_0 < 0$ configurations to account for the ecliptic degeneracy \citep{Jiang2004,Poindexter2005}.

For events lacking sharp caustic crossings and showing a bump-type anomaly, we also check whether a single-lens binary-source model (1L2S) can reproduce the observed anomaly \citep{Gaudi1998}. For a static 1L2S configuration, the effective magnification in band $\lambda$ is given by \citep{MB12486}
\begin{equation}
A_{\lambda}(t) = \frac{A_{1}(t) f_{{\rm S},1,\lambda} + A_{2}(t) f_{{\rm S},2,\lambda}}{f_{{\rm S},1,\lambda}+f_{{\rm S},2,\lambda}}
= \frac{A_{1}(t) + q_{f,\lambda}A_{2}(t)}{1 + q_{f,\lambda}},
\end{equation}
where $q_{f,\lambda} = f_{{\rm S},2,\lambda}/f_{{\rm S},1,\lambda}$ is the flux ratio, $A_j(t)$ is the magnification of each source, and $j=1,2$ label the primary and secondary sources.

\cite{OB160007} classified a model as degenerate when its $\Delta\chi^2$ relative to the best-fit solution is less than 10, and several papers in the AnomalyFinder series have followed this convention (e.g., \citealt{2016_prime}). For this work we adopt a more permissive exclusion threshold and remove models with $\Delta\chi^2>20$. Because we report the $\Delta\chi^2$ for every candidate, readers are free to apply alternate cutoffs when assembling a planetary sample.

\subsection{\eventa}

Figure~\ref{fig:lc1810} presents the observed light curve of \eventa\ together with the best-fit 2L1S models. About five days prior to the primary peak, the light curve exhibits a pronounced U-shaped anomaly lasting about two days, characteristic of a caustic-crossing feature. We find that the 2L1S models including higher-order effects yield unphysical parallax values of $\sim 7$, with the parallax signal arising solely from the KMTC data set. We therefore restrict our analysis to the KMTC data within $100 < {\rm HJD}^{\prime} < 160$, where ${\rm HJD}^{\prime} = {\rm HJD} - 2460000$.

The grid search identifies two local minima in parameter space, whose MCMC-derived parameters are listed in Table~\ref{tab:parm-a} and are labeled as solutions ``A'' and ``B''. As illustrated in Figure~\ref{fig:cau1}, the U-shaped anomaly arises from the source star traversing a resonant caustic. Owing to the coverage during the first caustic crossing, finite-source effects \citep{1994ApJ...421L..75G,Shude1994,Nemiroff1994} are constrained for both solutions. Solution ``A'' yields the lowest $\chi^2$, while solution ``B'' is disfavored by $\Delta\chi^2 = 6.5$. Inspection of Figure~\ref{fig:lc1810} indicates that this difference is driven primarily by a single KMTC data point near the caustic exit. The inferred mass ratio, $\log q \sim -1.9$, places the companion in the super-Jupiter mass-ratio regime. 

Including higher-order effects yields a $\chi^2$ improvement of only about 1 and places a constraint on the parallax component $\pi_{\rm E,\parallel} = -0.01 \pm 0.17$ and $0.00\pm0.13$ for the $u_0 > 0$ and $u_0 < 0$ solutions, respectively, where $\pi_{\rm E,\parallel} \simeq \pi_{\rm E,E}$ corresponds to the minor axis of the parallax error ellipse and is approximately aligned with the direction of Earth's acceleration. In contrast, the component along the major axis of the parallax ellipse, $\pi_{\rm E,\bot} \simeq \pi_{\rm E,N}$, remains essentially unconstrained, with $\sigma(\pi_{\rm E,\bot}) \sim 0.6$, much larger than its typical expected value of $\sim 0.1$. We therefore adopt the above measurements of $\pi_{\rm E,\parallel}$ in the Bayesian analysis presented in Section~\ref{sec:lens} to place constraints on the physical properties of the lens system. 

\subsection{\eventb}

\begin{figure}
    \centering
    \includegraphics[width=0.47\textwidth]{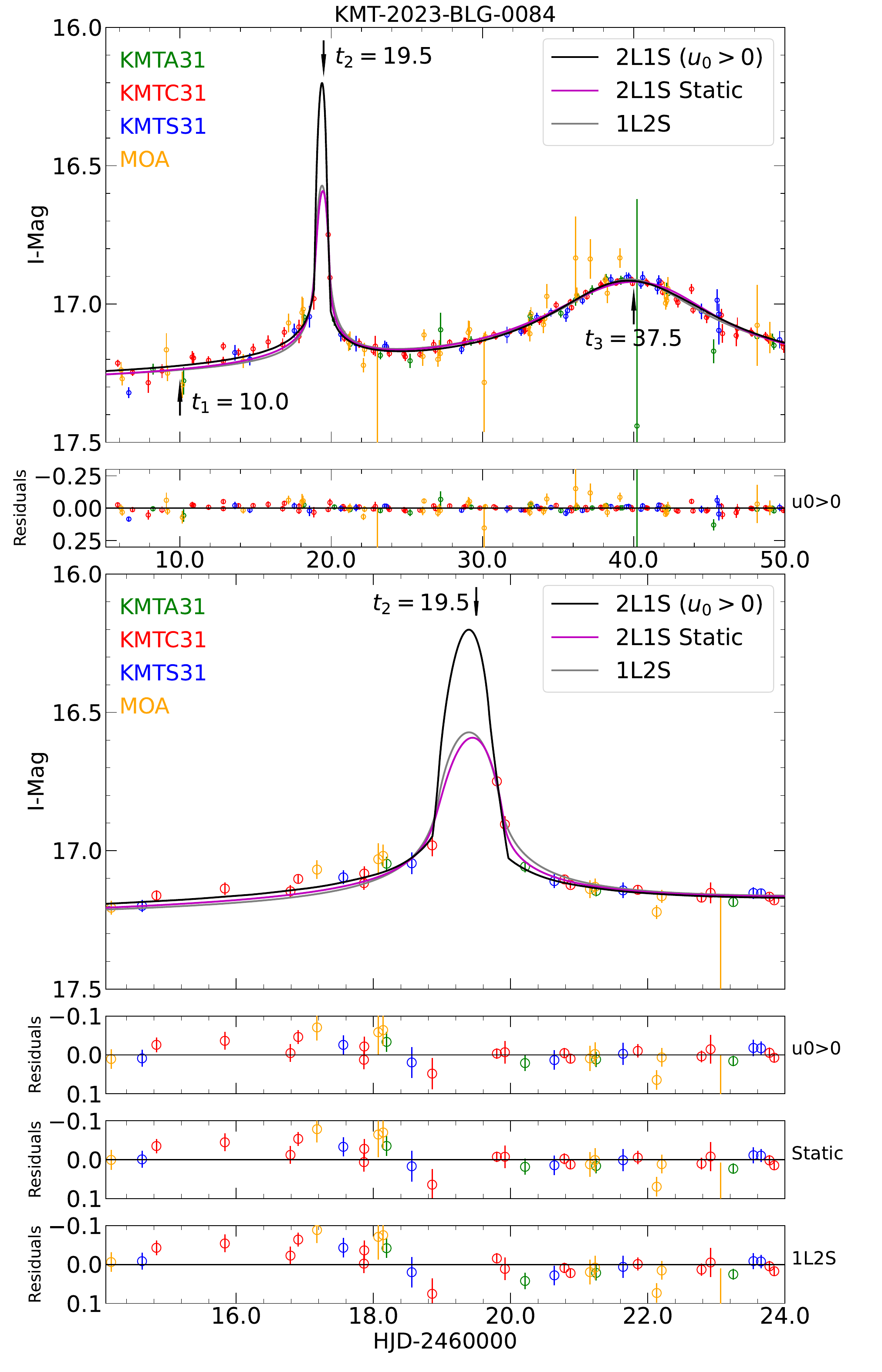}
    \caption{Light curve and models for \eventb. The symbols are the same as those in Figure~\ref{fig:lc1810}. The three arrows indicate the epochs corresponding to the caustic structures shown in Figure~\ref{fig:cau1}.}
\label{fig:lc0084}
\end{figure}

Figure~\ref{fig:lc0084} presents the observed light curve of KMT-2023-BLG-0084 together with the best-fit static 2L1S model and the models including higher-order effects, namely parallax and orbital motion. The light curve exhibits a sharp, bump-like anomaly about 20 days prior to the primary peak. The grid search reveals only one local minimum in parameter space, corresponding to the ``static" solution listed in Table~\ref{tab:parm-b}. Figure~\ref{fig:cau1} illustrates that the anomaly arises from the source star traversing the central caustic. 

We find that including higher-order effects improves the fit by $\Delta \chi^2 = 58.9$, with the majority of the improvement occurring before and during the anomaly, a region primarily covered by KMTC data. In models that include higher-order effects, the source approaches the extended magnification region associated with the cusp, resulting in a slightly higher magnification than in the static model prior to the prominent bump caused by the caustic crossing (see Figure~\ref{fig:lc0084}). This leads to the observed improvement in $\Delta\chi^2$ relative to the static model. Owing to the caustic crossing, finite-source effects are measured. The normalized source radius, $\rho = \theta_*/\theta_{\rm E}$, is measured to be $(6.3 \pm 1.4) \times 10^{-4}$ for the best-fit high-order ($u_0>0$) model (Table~\ref{tab:parm-b}). Compared to the static model, the relative uncertainty in $\rho$ is about twice as large, which arises from correlations between parallax and $\rho$, because most of the $\chi^2$ improvement occurs around the anomaly.

\begin{figure}
    \centering
    \includegraphics[width=0.47\textwidth]{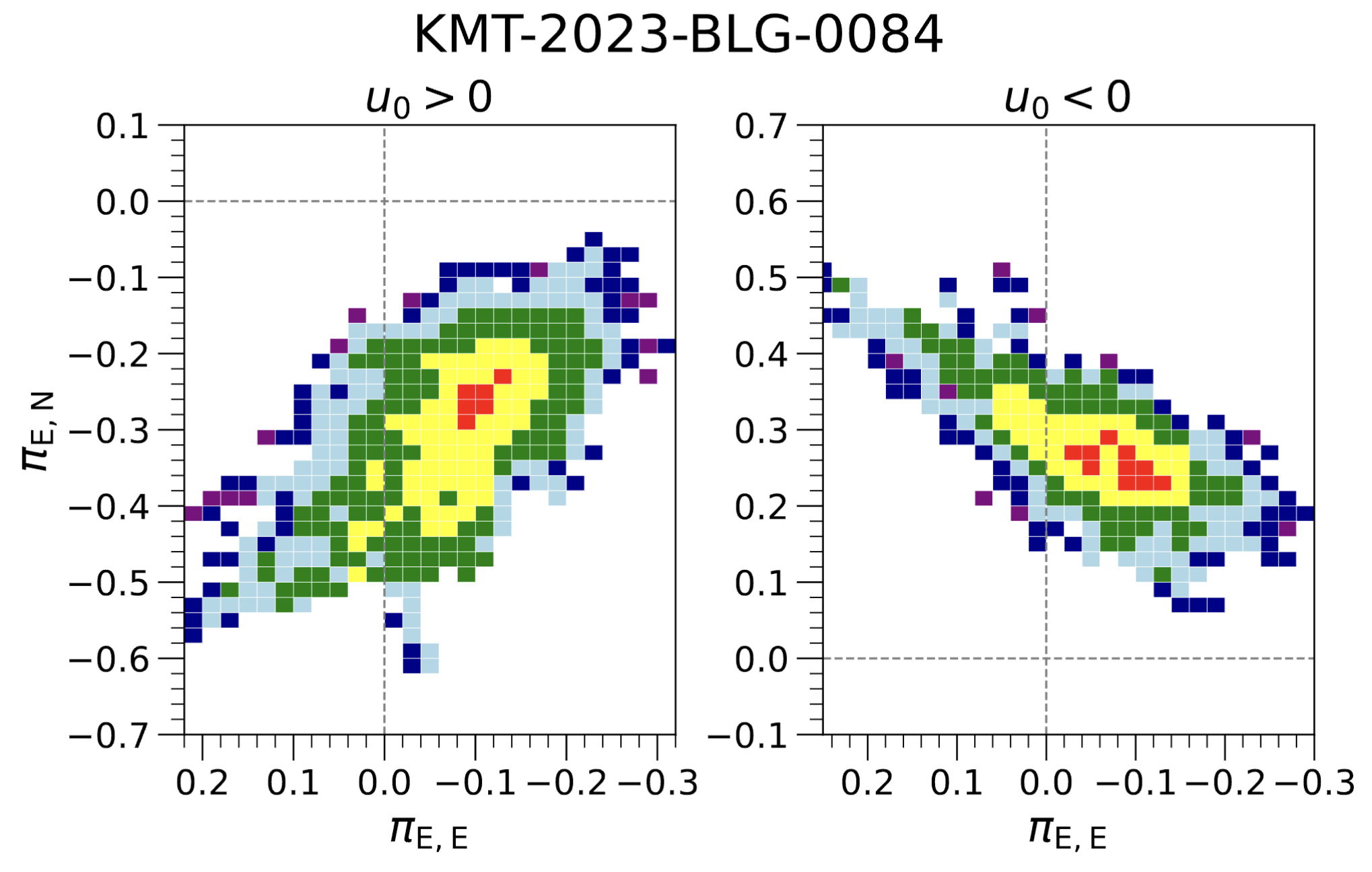}
    \caption{Parallax contour of the 2L1S higher-order models for \eventb. Red, yellow, green, light blue, dark blue, and purple denote regions with differences in likelihood ratios $[-2\Delta\ln{\mathcal{L}/\mathcal{L}_{max}]}< 1, 4, 9, 16, 25, 36$, respectively.}
\label{fig:pie0084}
\end{figure}

The 2L1S model including high-order effects places constraints on the parallax components. Figure \ref{fig:pie0084} shows the parallax contours for the two parallax solutions. The magnitude of $\pi_{\rm E}$ is constrained to $0.31 \pm 0.08$ and $0.31 \pm 0.07$ for the ``$u_0>0$'' and ``$u_0<0$'' solutions, respectively. The constraints on $\pie$ will be adopted in the Bayesian analysis presented in Section~\ref{sec:lens} to constrain the physical properties of the lens system. Among the two degenerate solutions, the $u_0>0$  solution yields the lowest $\chi^2$, while the $u_0<0$ solution is disfavored by $\Delta\chi^2 = 3.8$. The inferred mass ratio, $\log q \sim -2.0$, places the companion in the super-Jupiter mass-ratio regime.

The bump-like anomaly suggests the possibility of a 1L2S interpretation. However, the 1L2S model with parallax for this event is strongly disfavored, with $\Delta\chi^2 = 49.1$ relative to the 2L1S model that includes higher-order effects. Moreover, the secondary source in the 1L2S model exhibits finite-source effects, with $\rho_2 = (4.36\pm0.92) \times 10^{-3}$ from which we infer a lens-source relative proper motion of $\mu_{\rm rel} \sim 0.1~\mathrm{mas~yr^{-1}}$. This value is extremely small. According to \citet{MASADA, 2019_subprime}, only $\sim 0.04\%$ of microlensing planetary events are expected to have proper motions smaller than this value. For these reasons, we reject the 1L2S interpretation for this event.

\begin{table*}[htb]
  \renewcommand\arraystretch{1.20}
  \centering
  \caption{Lensing Parameters for \eventb}
  \begin{tabular}{c | c  c c c}
    \hline
    \hline
    Parameters & Static & \multicolumn{2}{c}{Parallax + Orbital Motion}&1L2S \\
               &       & $u_0>0$ & $u_0<0$& \\
    \hline
    $\chi^2/\mathrm{dof}$ & 1536.4/1503 &1477.5/1503 & 1481.3/1503& 1526.6/1503\\ \hline
    $t_0$ (HJD$^\prime$) & $37.37 \pm 0.21$ & $37.51 \pm 0.36$ & $38.58 \pm 0.43$&$39.77\pm0.07$ \\
    $u_0$ & $-0.047 \pm 0.005$ & $0.042 \pm 0.003$ & $-0.045 \pm 0.002$&$0.035\pm0.008$ \\
    $t_{\rm E}$ (days) & $103.4 \pm 4.7$ & $107.7 \pm 6.0$ & $103.5 \pm 3.4$&$111.0\pm24.1$ \\
    $\rho (10^{-4})$ & $13.3 \pm 1.3$ & $6.3 \pm 1.4$ & $6.3 \pm 1.2$& ...\\
    $\alpha$ (deg) & $164.8 \pm 0.3$ & $199.9 \pm 0.8$ & $161.3 \pm 0.9$& ...\\
    $s$ & $1.479 \pm 0.019$ &$1.306 \pm 0.046$ & $1.323 \pm 0.053$&... \\
    $\log q$ & $-1.64 \pm 0.14$ & $-1.97 \pm 0.12$ & $-1.93 \pm 0.14$&... \\
    $\pi_{\rm E,N}$ & ... & $-0.29 \pm 0.09$ & $0.29 \pm 0.07$&$0.62\pm0.22$ \\
    $\pi_{\rm E,E}$ & ... & $-0.07 \pm 0.08$ & $-0.02 \pm 0.10$&$-0.05\pm0.06$ \\
    $\pi_{\rm E}$ & ... & $0.31 \pm 0.08$ & $0.31\pm 0.07$&... \\
    $ds/dt$ ($\rm yr^{-1}$) & ... & $-0.81 \pm 0.32$ & $-0.69 \pm 0.40$&... \\
    $d\alpha/dt$ ($\rm deg\,yr^{-1}$) & ... & $-16.4 \pm 21.9$ & $1.0 \pm 7.2$ &...\\
    $t_{0,2}$(HJD$^\prime$)&...&...&...&$19.78\pm0.44$\\
    $u_{0,2}$&...&...&...&$0.039\pm0.014$\\
    $\rho_2 (10^{-3})$&...&...&...&$4.36\pm0.92$\\
    $q_f$&...&...&...&$0.103\pm0.015$\\
    $I_{\rm S,KMTC}$ & $21.448 \pm 0.008$ & $21.526 \pm 0.008$ & $21.445 \pm 0.008$&$21.427\pm0.052$ \\
    \hline
    \hline
  \end{tabular}
  \label{tab:parm-b}
\end{table*}

\subsection{\eventc}

\begin{figure}
    \centering
    \includegraphics[width=0.47\textwidth]{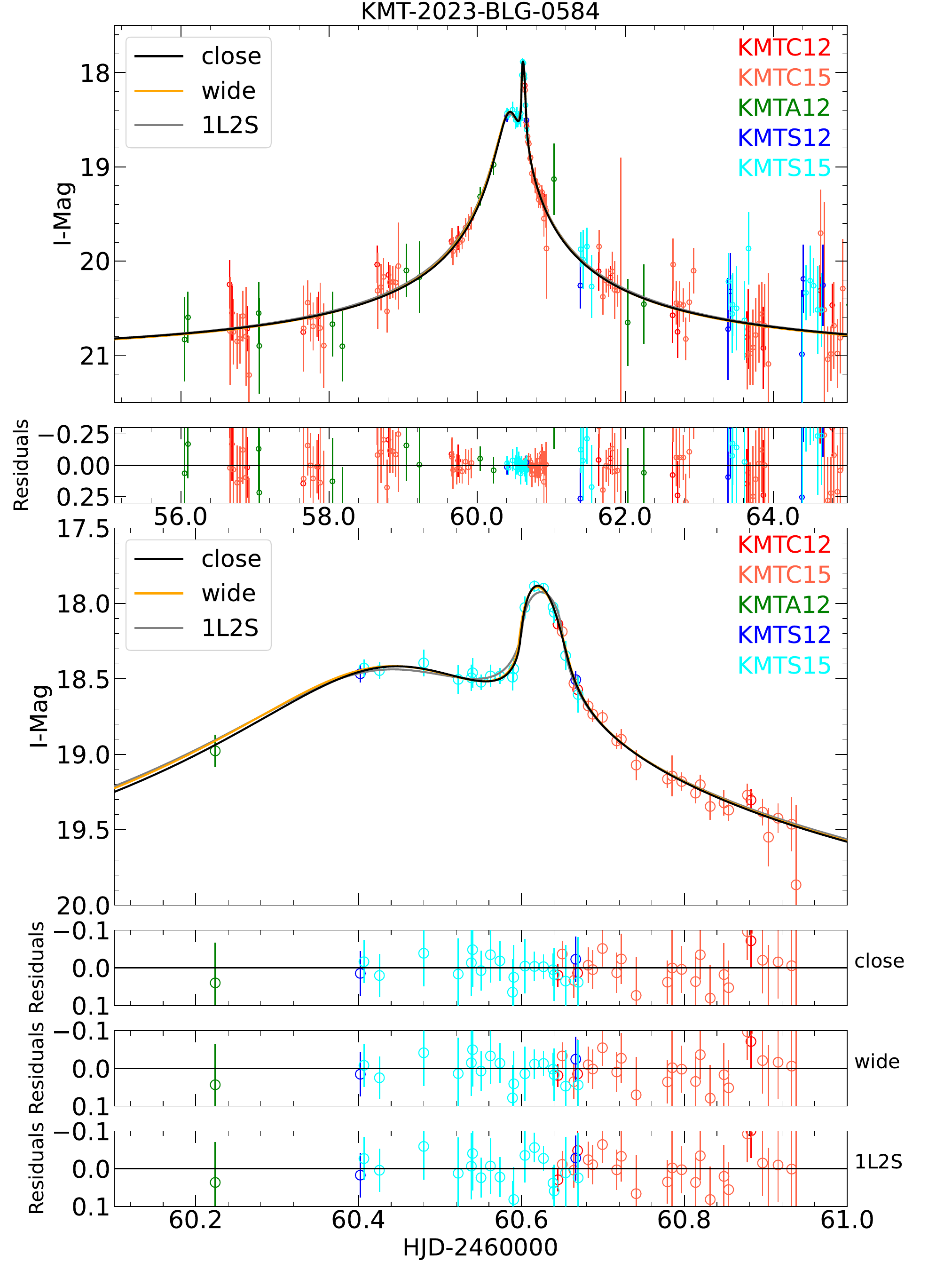}
    \caption{Light curve and models for \eventc. The symbols are the same as those in Figure~\ref{fig:lc1810}. This event lies in the overlap region of two KMTNet fields.}
\label{fig:lc0584}
\end{figure}

Figure~\ref{fig:lc0584} displays the observed light curve of KMT-2023-BLG-0584 together with the best-fit 2L1S models and the 1L2S model. The event exhibits a bump-like anomaly occurring $\sim0.2$~days after the otherwise PSPL peak, with a short duration of $\lesssim0.1$~days. Thanks to the good coverage of the anomaly by the KMTS and KMTC data sets, only two viable 2L1S solutions are identified from a grid search. Table~\ref{tab:2L1S_params_0584} lists the lensing parameters.  The corresponding caustic geometries are shown in Figure~\ref{fig:cau1}. The two solutions exhibit the ``Close/Wide'' degeneracy \citep{Griest1998}, for which the caustic structures of the central caustic and all parameters except the normalized projected separation are nearly identical between the two models. The ``Close'' solution is marginally preferred, with $\Delta\chi^2 = 0.4$ relative to the ``Wide'' solution. Both models yield an inferred mass ratio of $\log q \sim -2.2$, placing the companion in the super-Jupiter mass-ratio regime.

We perform the 1L2S modeling for this event and obtain a $\Delta\chi^2 = 8.6$ (see Table~\ref{tab:2L1S_params_0584}), indicating that the model is slightly disfavored relative to the best-fit 2L1S model. Finite-source effects for the secondary source are only weakly constrained and are consistent with $\rho_2 = 0$ at $\Delta\chi^2 \sim 7$. If we adopt the best-fit value of $\rho_2$, we obtain $\mu_{\rm rel} \sim 6~{\rm mas~yr^{-1}}$, which is consistent with typical values for microlensing events. Because there are no $V$-band observations during the anomaly, the 1L2S interpretation cannot be tested using the color information of the secondary source \citep{Shude1991,Gaudi1998}. Therefore, we cannot rule out the 1L2S interpretation. Because this is only a candidate planetary event, we do not pursue further analysis.

\begin{table*}[htb]
  \renewcommand\arraystretch{1.20}
  \centering
  \caption{Lensing Parameters for \eventc}
  \label{tab:2L1S_params_0584}
  \begin{tabular}{c | c c c}
    \hline\hline
    Parameter & Wide & Close&1L2S \\
    \hline
    $\chi^2$/dof & 2248.8/2249 & 2248.4/2249& 2257.0/2249\\ \hline
    $t_0$ (HJD$'$) & $60.482 \pm 0.010$ & $60.488 \pm 0.038$& $60.429 \pm 0.015$\\
    $u_0$ & $0.013 \pm 0.003$ & $0.013 \pm 0.003$&$0.017\pm0.007$ \\
    $t_{\rm E}$ (days) & $10.5 \pm 1.9$ & $10.5 \pm 2.1$&$9.4\pm4.2$ \\
    $\rho(10^{-3})$ & $1.99 \pm 0.44$ & $2.01 \pm 0.65$&... \\
    $\alpha$ (deg) & $317.3 \pm 2.3$ & $315.9 \pm 2.2$&... \\
    $s$ & $1.859 \pm 0.177$ & $0.580 \pm 0.085$&... \\
    $\log q$ & $-2.10 \pm 0.16$ & $-2.20 \pm 0.13$&... \\
    $t_{02}$ (HJD$'$)&...&...&$60.626\pm0.002$\\
    $u_{02}$ (HJD$'$)&...&...&$0.0012\pm0.0008$\\
    $\rho_2(10^{-3})$&...&...&$2.80\pm1.04$\\
    $q_f$&...&...&$0.104\pm0.025$\\
    $I_{\rm S,KMTC}$ & $22.989 \pm 0.032$ & $22.989 \pm 0.032$&$22.978\pm0.207$ \\
    \hline\hline
  \end{tabular}
\end{table*}

\subsection{\eventd}

\begin{figure}

    \centering
    \includegraphics[width=0.47\textwidth]{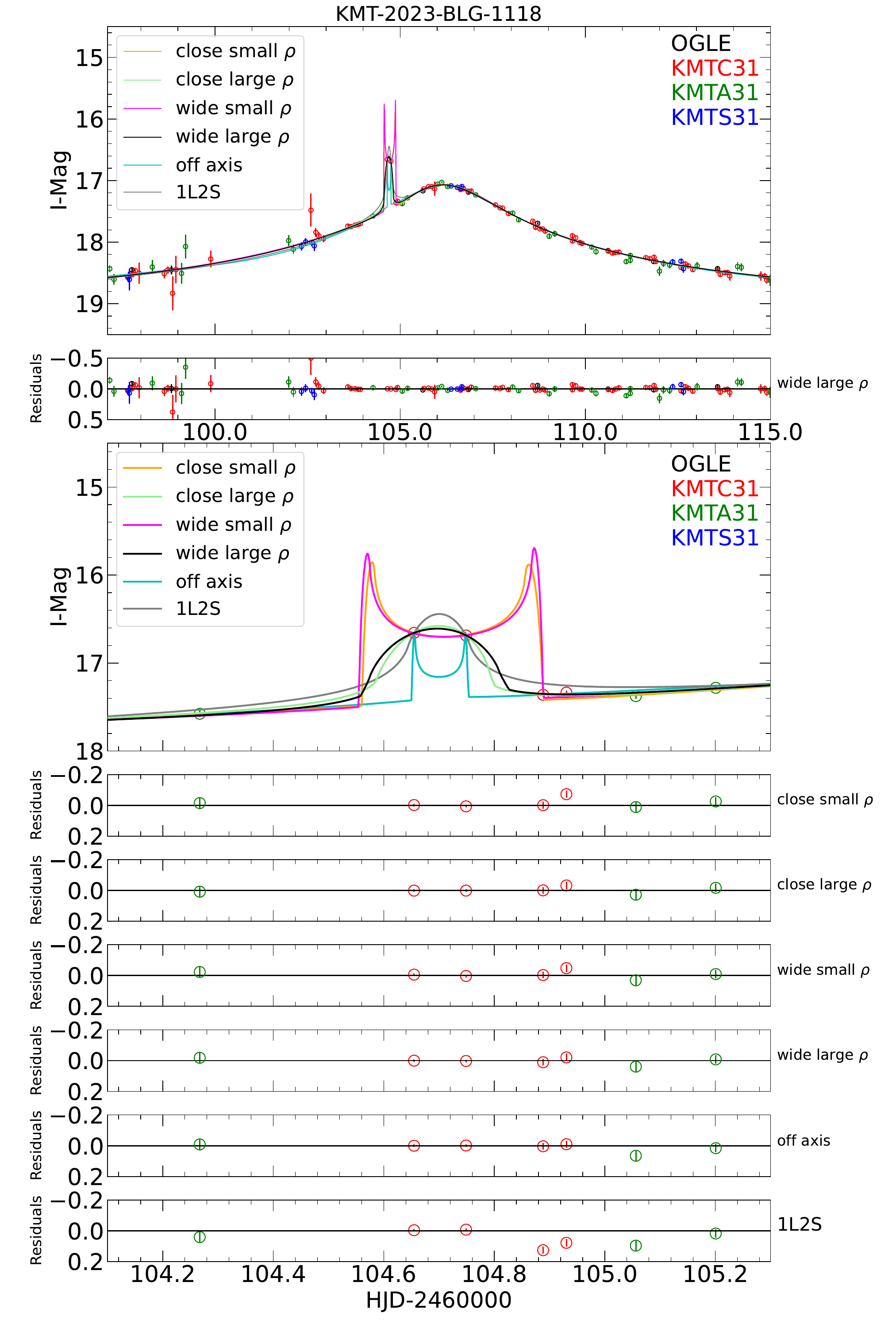}
    \caption{Light curve and models for \eventd. The symbols are the same as those in Figure~\ref{fig:lc1810}.}
\label{fig:lc1118}
\end{figure}

\begin{table*}[htb]
    \renewcommand\arraystretch{1.20}
    \caption{Lensing Parameters for \eventd}
    \centering
    \label{tab:2L1S_params_1118}
        \begin{tabular}{c|cccccc}
        \hline\hline
        Parameter & Close Small $\rho$ & Close Large $\rho$ & Wide Small $\rho$& Wide Large $\rho$ & Off-axis & 1L2S \\
        \hline
        $\chi^2/\mathrm{dof}$ 
        & 1184.4/1175 & 1177.0/1175 & 1179.9/1175 & 1175.8/1175 & 1191.1/1175&1257.3/1175 \\ \hline
        $t_0$ (HJD$^\prime$-100) 
        & $5.987\pm0.019$ & $6.091\pm0.017$ & $5.976\pm0.017$ & $6.082\pm0.016$ & $6.073\pm0.015$&$6.272\pm0.013$ \\
        $u_0$ 
        & $0.079\pm0.006$ & $0.079\pm0.004$ & $0.084\pm0.005$ & $0.081\pm0.005$ & $0.073\pm0.002$&$0.067\pm0.004$ \\
        $t_{\rm E}$ (days) 
        & $16.62\pm0.87$ & $17.20\pm0.63$ & $16.10\pm0.68$ & $16.96\pm0.79$ & $18.31\pm0.37$&$19.63\pm0.93$ \\
        $\rho (10^{-3})$ & $<1.9$ & $5.2\pm1.2$ & $<2.5$ & $6.1\pm1.1$ & $0.158\pm0.016$&... \\
        $\alpha$ (deg) 
        & $225.8\pm2.0$ & $224.3\pm0.6$ & $227.0\pm0.6$ & $224.8\pm0.6$ & $183.8\pm1.1$&... \\
        $s$ 
        & $0.932\pm0.011$ & $0.937\pm0.016$ & $1.212\pm0.011$ & $1.196\pm0.016$ & $1.040\pm0.001$&... \\
        $\log q$ 
        & $-2.25\pm0.10$ & $-2.66\pm0.09$ & $-2.17\pm0.06$ & $-2.62\pm0.09$ & $-2.48\pm0.03$&... \\
        $t_{0,2}$(HJD$^\prime-100$) &...&...&...&...&...&$4.700\pm0.001$\\
        $u_{0,2} (10^{-3})$&...&...&...&...&...&$0.20\pm0.16$\\
        $\rho_2 (10^{-3})$&...&...&...&...&...&$2.92\pm0.24$\\
        $q_f$&...&...&...&...&...&$0.0247\pm0.0007$\\
        $I_{\rm S,OGLE}$ 
        & $19.844\pm0.012$ & $19.895\pm0.012$ & $19.768\pm0.012$ & $19.869\pm0.012$ & $20.036\pm0.012$&$20.130\pm0.064$ \\
        \hline\hline
        \end{tabular}
        \tablecomments{The upper limit on $\rho$ is $3\sigma$.}
\end{table*}

Figure~\ref{fig:lc1118} shows the observed light curve of \eventd\ together with the best-fit 2L1S models and the 1L2S model. The anomaly appears about 1.5~days before the primary peak and lasts for about 0.2~days, during which it is covered by only two KMTC data points. Owing to the limited temporal coverage of this short-timescale anomaly, the grid search yields five viable models that can account for the observed data.

A pair of ``Close/Wide'' solutions with smaller source radii exhibit U-shaped anomalies in the model light curves, with the two anomalous data points lying near the center of the U-shaped feature. Because there are no data points during the caustic crossing for this geometry, $\rho$ is not well constrained, with a $3\sigma$ upper limit of $\sim 2\times10^{-3}$. In contrast, the pair of ``Close/Wide'' solutions with larger source radii display bump-like anomalies in the model light curves, with detectable finite-source effects that constrain the normalized source radius to a relatively large value of $\rho \sim 6 \times 10^{-3}$.

In the fifth configuration, the anomaly is produced by an off-axis caustic crossing involving the resonant caustic; we therefore label this solution as ``Off-axis''. Similar off-axis solutions have been reported for another 2023 planet-candidate event, KMT-2023-BLG-0486 \citep{2023_prime}.

The parameters of all these solutions are listed in Table~\ref{tab:2L1S_params_1118}. Among the five solutions, the ``Wide Large $\rho$'' solution provides the best fit to the data. The ``Close Small $\rho$'',`` Close Large $\rho$'', and ``Wide Small $\rho$'' solutions are disfavored by $\Delta \chi^2 = 8.6$, $1.2$, and $4.1$, respectively, while the ``Off-axis'' solution is highly disfavored by $\Delta \chi^2 = 15.3$. In addition, this solution exhibits a U-shaped anomaly, with the two KMTC data points lying directly on the caustic-crossing peaks. As a result, the normalized source radius is well constrained to be $\rho = (0.158 \pm 0.016)\times10^{-3}$, but this value is extremely small. According to the source-property analysis in Section~\ref{sec:lens}, this implies an extremely high lens-source relative proper motion of $\sim 60~{\rm mas~yr^{-1}}$. Therefore, the ``Off-axis'' solution is rejected. Among the four remaining solutions, the mass ratios lie in the gas-giant regime, with $-2.7 < \log q < -2.1$.

We also test the 1L2S model and find that it is strongly disfavored: the best-fitting 1L2S model is worse than the preferred 2L1S model by $\Delta\chi^2 = 81.5$. As shown in Figure~\ref{fig:lc1118}, although the 1L2S model can fit the two KMTC points during the anomaly, it cannot reproduce the KMTA and KMTC points outside the anomaly. Thus, we reject the 1L2S model.

Due to the short $\te$ and the faintness of the event, including higher-order effects improves the fit by only $\Delta\chi^2 < 1$ and yields a large uncertainty of $\sigma(\pi_{\rm E,\parallel}) > 0.3$, rendering these effects uninformative for the Bayesian analysis.

\begin{figure*}
    \centering
    \includegraphics[width=0.93\textwidth]{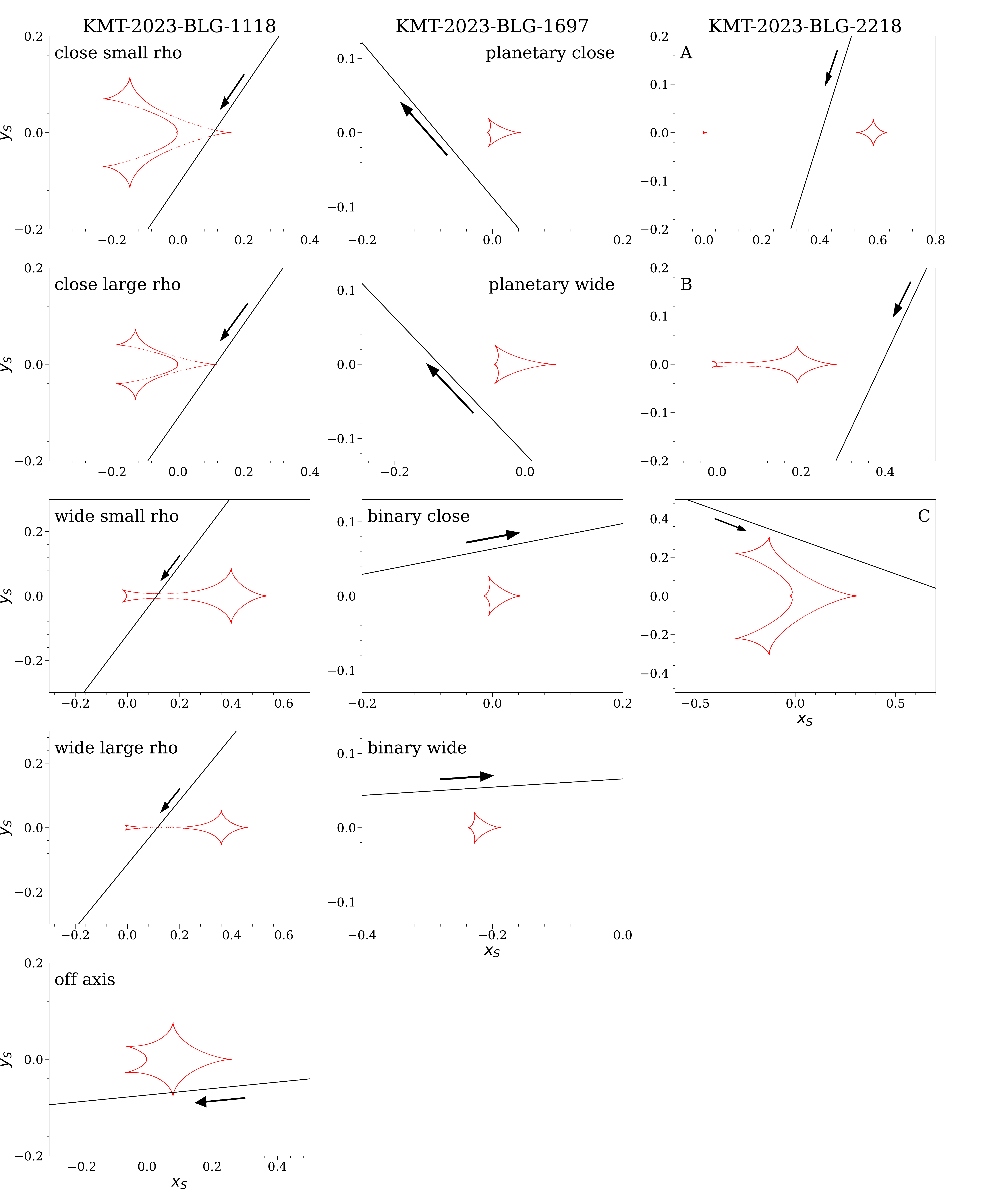}
    \caption{Geometries of the 2L1S models for event \eventd, \evente, and \eventf. For each event, the geometries of different solutions are arranged vertically.  The
symbols are the same as those in Figure \ref{fig:cau1}.}
    \label{fig:cau2}
\end{figure*}

\subsection{\evente}

\begin{figure}
    \centering
    \includegraphics[width=0.47\textwidth]{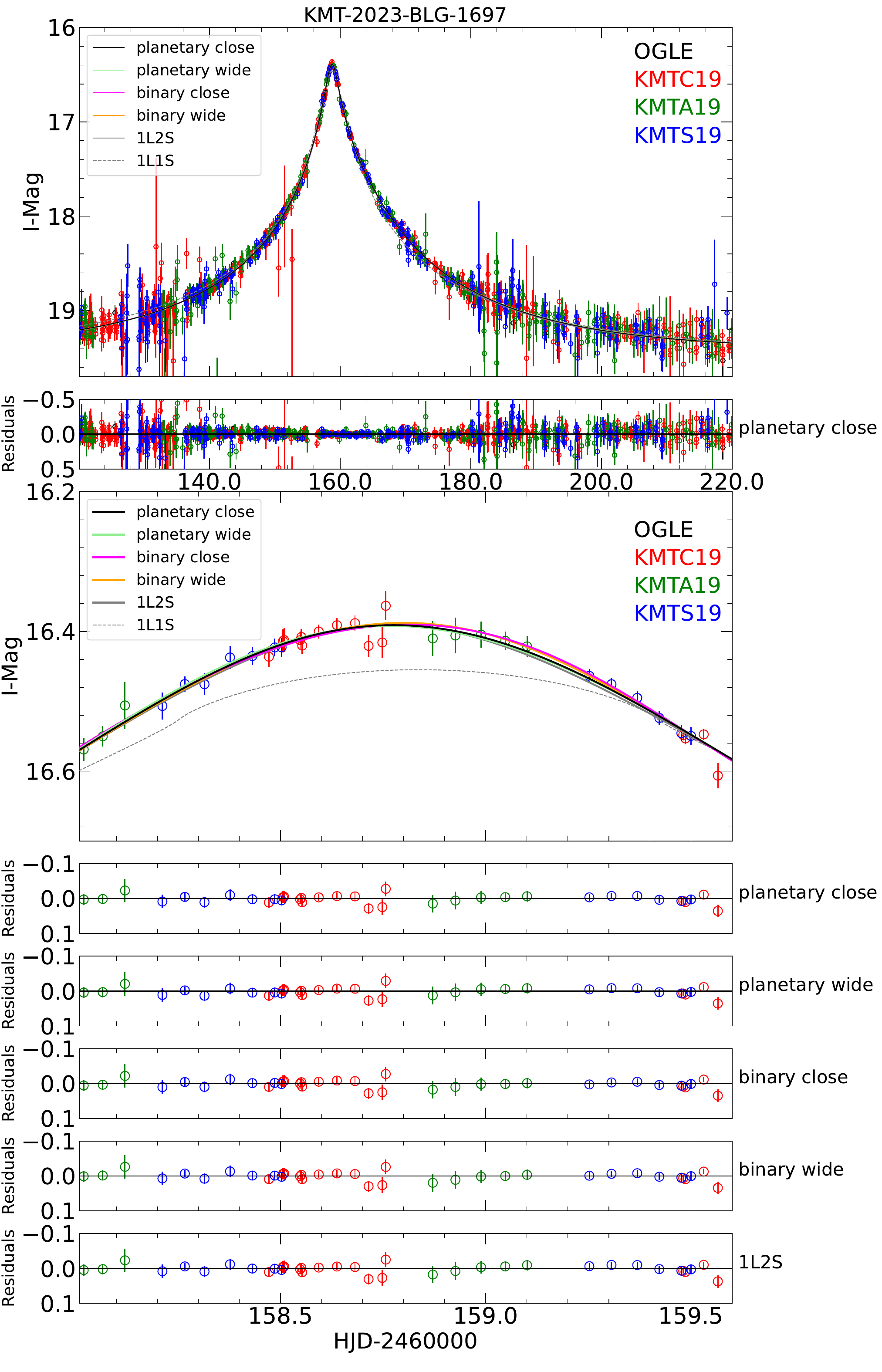}
    \caption{Light curve and models for \evente. The symbols are the same as those in Figure~\ref{fig:lc1810}.}
\label{fig:lc1697}
\end{figure}

Figure~\ref{fig:lc1697} displays the observed light curve of KMT-2023-BLG-1697 along with the best-fit 2L1S models and the corresponding 1L2S model and 1L1S model. The anomaly appears near the peak of the event and produces a slight asymmetry in the peak region. A grid search in parameter space identifies four degenerate solutions.

The four solutions can be divided into two groups according to their inferred mass ratios, with each group exhibiting the Close/Wide'' degeneracy. One pair of solutions has an inferred mass ratio of $\log q \sim -1.5$, which is near the planet/brown dwarf boundary; we label these the ``Planetary Close'' and ``Planetary Wide'' solutions. The other pair, referred to as the ``Binary'' solutions, yields inferred mass ratios of $\log q = -1.2$ and $-1.0$. Figure~\ref{fig:cau2} illustrates the source trajectory and caustic geometry of the event, showing that the anomaly arises from the source passing near the central caustic.

Among these four solutions, the best-fit model is the ``Planetary Close'' solution, while the ``Planetary Wide'', ``Binary Close'', and ``Binary Wide'' solutions are disfavored by $\Delta\chi^2 = 4.4$, $8.8$, and $50.9$, respectively. The $\Delta\chi^2$ for the ``Planetary Wide'' solution arises mainly from the wing after the peak, during which the wide companion produces weak microlensing signatures. Because the ``Binary Close'' solution cannot be rejected, both planetary and binary interpretations can explain the observed data.

\begin{table*}[htb] 
    \renewcommand\arraystretch{1.20}
    \caption{Lensing Parameters for \evente}
    \centering
        \begin{tabular}{c|ccccc}
        \hline\hline
        Parameter & Planetary Close & Planetary Wide & Binary Close & Binary Wide&1L2S  \\
        \hline
        $\chi^2/\mathrm{dof}$ & 2783.1/2785 & 2787.5/2785 & 2791.9/2785 & 2834.0/2785&2776.2/2785  \\ \hline
        $t_0$ (HJD$^\prime$-150) & $9.174\pm0.011$ & $9.155\pm0.012$ & $9.100\pm0.014$ & $9.150\pm0.010$&$8.760\pm0.007$  \\
        $u_0$ & $0.061\pm0.002$ & $0.063\pm0.002$ & $0.063\pm0.002$ & $0.053\pm0.001$&$0.045\pm0.003$  \\
        $t_{\rm E}$ (days) & $35.49\pm0.80$ & $37.83\pm0.81$ & $37.44\pm0.81$ & $38.90\pm0.76$&$32.59\pm0.57$\\
        $\rho (10^{-2})$ & $2.05 \pm 0.20$ & $2.23 \pm 0.18$ & $2.18 \pm 0.12$ & $2.05 \pm 0.15$& $4.21\pm0.63$ \\
        $\alpha$ (deg) & $133.8\pm0.5$ & $137.3\pm0.8$ & $10.1\pm1.1$ & $3.3\pm0.7$&... \\
        $s$ & $0.545 \pm 0.005$ & $1.750 \pm 0.020$ & $0.446 \pm 0.012$ & $2.904 \pm 0.076$&...  \\
        $\log q$ & $-1.577\pm0.023$ & $-1.499\pm0.028$ & $-1.190\pm0.025$ & $-1.022\pm0.032$& ... \\
        $t_{0,2}$ (HJD$^\prime$-150)&...&...&...&...&$10.560\pm0.142$\\
        $u_{0,2}$&...&...&...&...&$0.168\pm0.006$\\
        $\rho_2 (10^{-2})$&...&...&...&...&$2.45\pm1.46$\\
        $q_f$&...&...&...&...&$1.025\pm0.096$\\
        $I_{\rm S,OGLE}$ & $19.628 \pm 0.006$ & $19.617 \pm 0.006$ & $19.647 \pm 0.006$ & $19.708 \pm 0.006$&$19.440\pm0.007$ \\
        \hline\hline
        \end{tabular}
\end{table*}

We also explore this event using a 1L2S model and find that it provides a better fit than the best-fit 2L1S model by $\Delta\chi^2 = 6.9$. Finite-source effects are detected with only a $\Delta\chi^2 = 2$ improvement and therefore do not provide a meaningful constraint on the lens-source relative proper motion. As a result, the 1L2S model cannot be excluded based on the inferred proper motion. The two sources have roughly equal fluxes, and the colors of both sources are consistent with those of bulge stars. Thus, we cannot reject the 1L2S interpretation. Because the lens-source configuration can be described by either a 2L1S or a 1L2S model, this event remains only a planetary candidate, and we therefore do not pursue further analysis.

\subsection{\eventf}

\begin{figure}
    \centering
    \includegraphics[width=0.47\textwidth]{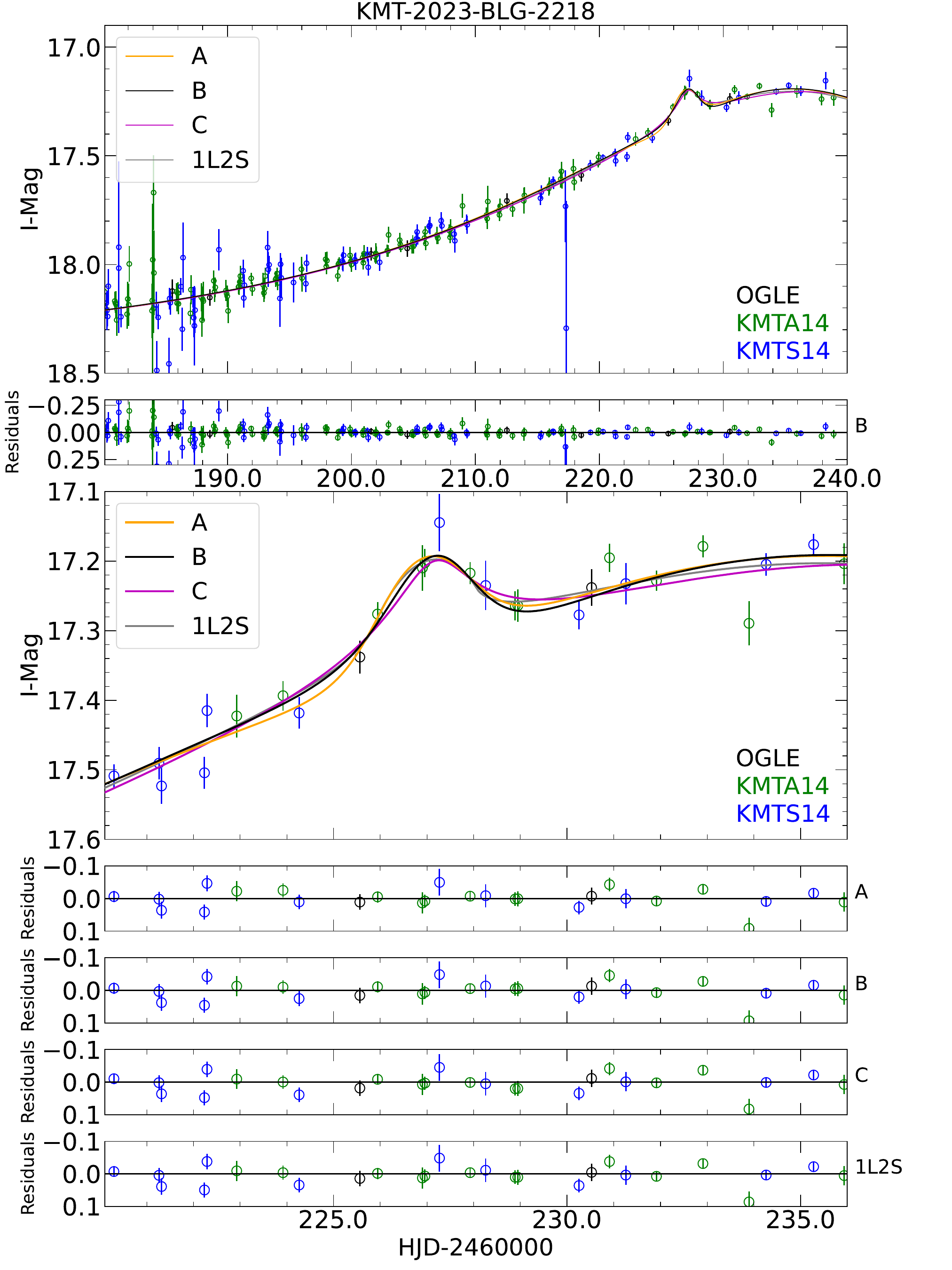}
    \caption{Light curve and models for \eventf. The symbols are the same as those in Figure~\ref{fig:lc1810}.}
\label{fig:lc2218}
\end{figure}

Figure~\ref{fig:lc2218} shows the observed light curve of \eventf\ together with the best-fit 2L1S models, as well as the 1L2S model. Because the 2023 bulge season ended just after $t_0$, the falling side of the light curve was not observed. The light curve exhibits a bump-type anomaly about 7~days prior to the primary peak, mainly covered by KMTA and KMTS data. No KMTC data are available for this event because the source lies near the edge of a CCD chip.

The grid search reveals three local minima, which we label as solutions ``A'', ``B'', and ``C''. Figure~\ref{fig:cau2} shows the caustic structure and source trajectory for these solutions. Among them, the ``A'' and ``B'' solutions are subject to the ``Inner/Outer'' degeneracy \citep{GG1997}, in which the source passes between the planetary and central caustics (the ``Inner'' solution) or outside the entire set of caustic structures (the ``Outer'' solution). For solution ``C'', the topology is ``Off-axis'', as can be seen from Figure~\ref{fig:cau2}. None of the three solutions involves a caustic crossing, and the coverage of the anomaly is sparse. As a result, finite-source effects are not measured, and the data provide only a $3\sigma$ upper limit on the normalized source radius of $\rho \lesssim 4\times10^{-2}$.

The 2L1S parameters are summarized in Table~\ref{tab:parm_2218}. Among the three solutions, solution~``A'' yields the lowest $\chi^2$, while solutions~``B'' and ``C'' are disfavored by $\Delta\chi^2 = 0.7$ and $9.3$, respectively. Solutions~``A'' and ``B'' yield mass ratios of $\log q \sim -2.9$, consistent with a Jupiter-mass-ratio companion, whereas solution~``C'' gives $\log q \sim -1.4$, indicating a brown-dwarf companion.

\begin{table*}[htb]
  \renewcommand\arraystretch{1.20}
  \centering
  \caption{Lensing Parameters for \eventf}
  \label{tab:parm_2218}
  \begin{tabular}{c|c c c c}
    \hline\hline
    Parameter & A & B& C&1L2S \\
    \hline
    $\chi^2$/dof &1348.2/1348  &1348.9/1348  &1357.5/1348 &1350.9/1348\\ \hline
    $t_0$ (HJD$'$) &$235.14 \pm 0.55$  &$235.38\pm0.56$  &$239.15\pm1.21$&$235.69\pm0.45$\\
    $u_0$ &$0.345 \pm 0.036$  &$0.345 \pm 0.033$  &$0.276\pm0.023$&$0.333\pm0.013$\\
    $t_{\rm E}$ (days) & $45.0\pm2.4$ &$45.0\pm2.3$  &$53.0\pm2.3$&$47.1\pm1.2$\\
    $\rho (10^{-2})$ &$\textless4.07$  &$\textless4.07$  &$\textless3.89$&...\\
    $\alpha$ (deg) &$242.5 \pm 0.9$  &$241.8 \pm 1.0$  &$342.1\pm6.4$&...\\
    $s$ &$1.369\pm0.035$  &$1.090\pm0.028$  &$0.921\pm0.040$&...\\
    $\log q$ &$-2.890\pm0.110$  &$-2.967\pm0.121$  &$-1.435\pm0.096$&...\\
    $t_{02}$ (HJD$'$)&...&...&...&$227.08\pm0.16$\\
    $u_{02}$&...&...&...&$0.011\pm0.006$\\
    $\rho_2 (10^{-2})$&...&...&...&$2.28\pm0.94$\\
    $q_f$&...&...&...&$0.0048\pm0.0015$\\
        $I_{\rm S,OGLE}$ &$18.386\pm0.022$  &$18.374\pm0.022$  &$18.665\pm0.022$&$18.427\pm0.052$\\
    \hline\hline
  \end{tabular}
\end{table*}

The resulting parameters from the 1L2S modeling are listed in Table~\ref{tab:parm_2218}. The 1L2S model is slightly disfavored, with $\Delta\chi^2 = 2.7$ relative to the best-fit 2L1S model. Finite-source effects are only marginally constrained, with a $\Delta\chi^2 = 6$ improvement relative to the point-source model, and therefore cannot be used to reject the 1L2S interpretation based on the inferred lens-source relative proper motion. In addition, there are no $V$-band data points during the anomaly, so the 1L2S interpretation cannot be tested using the color of the secondary source. Given the degeneracy between the 2L1S and 1L2S models, the planetary interpretation is not uniquely favored. We therefore refrain from further analysis.

\begin{figure*}
    \centering
    \includegraphics[width=0.99\textwidth]{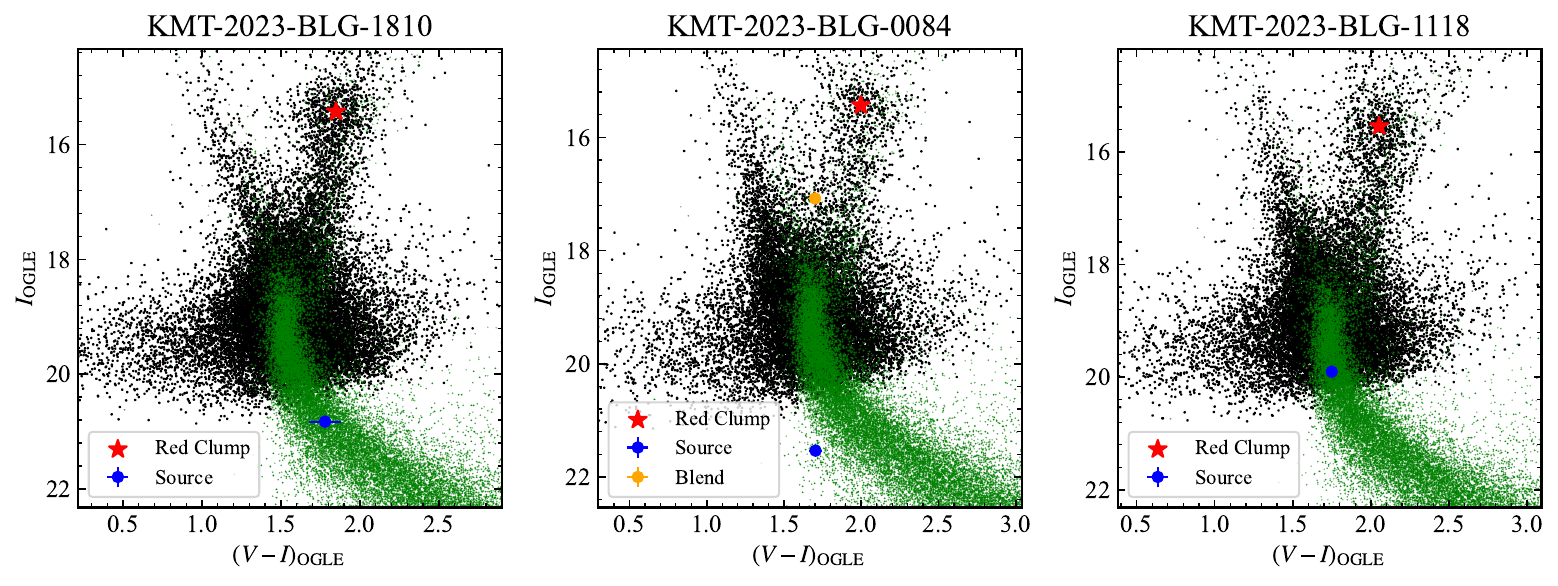}
    \caption{Color-magnitude diagrams for the three unambiguous planetary events. All CMDs are constructed using the OGLE-III star catalog \citep{OGLEIII}. In each panel, the red asterisk and the blue dot represent the centroids of the red clump and the source star, respectively. On the CMD of \eventb, the orange dot shows the blend. The green dots show the HST CMD from \citet{HSTCMD}, for which the centroids of the red clump, $(V-I,I)_{\rm cl,HST} = (1.62, 15.15)$ \citep{Bennett2008}, is aligned with those of OGLE-III.}
    \label{fig:cmd}
\end{figure*}

\begin{table*}[ht]
\centering
\caption{CMD Parameters, $\thetas$, $\thetae$, and $\murel$ for \eventa\ and \eventb}

\label{tab:cmd}

\renewcommand{\arraystretch}{1.15}

\begin{tabular}{l|cc|cc}
\hline
\hline
Parameter 
& \multicolumn{2}{c|}{\eventa} 
& \multicolumn{2}{c}{\eventb}
 \\
& A&B 

&$u_0>0$ & $u_0<0$

\\

\hline

$(V-I)_{\rm cl}$ 
& ... & ... &$2.00\pm0.01$ & $\leftarrow$  \\

$I_{\rm cl}$ 
& $15.43\pm0.03 $ 
& $\leftarrow$
& $15.43 \pm 0.04$ 
& $\leftarrow$   \\

$(V-I)_{\rm cl,0}$
& ...
& ...
&$1.06\pm0.03$
&$\leftarrow$\\

$I_{{\rm cl},0}$ 
& $14.41 \pm 0.04$ 
& $\leftarrow$
& $14.26 \pm 0.04$ 
& $\leftarrow$   \\

$(V-I)_{\rm S}$ 
& ... & ... &$1.70\pm0.04$ & $\leftarrow$  \\

$I_{\rm S}$ 
& $20.83 \pm 0.02$ 
& $20.95 \pm 0.02$ 
& $21.54 \pm 0.01$ 
& $21.45 \pm 0.01$ 

 \\

$(V-I)_{{\rm S},0}$ 
& $0.99 \pm 0.10$ 
& $1.04 \pm 0.11$ 
& $0.77 \pm 0.05$ 
& $\leftarrow$ 

 \\

$I_{{\rm S},0}$ 
& $19.81 \pm 0.05$ 
& $19.94 \pm 0.05$ 
& $20.36 \pm 0.06$ 
& $20.28 \pm 0.06$ 

 \\

$\thetas$ ($\mu$as) 
& $0.450\pm 0.044$ 
& $0.443 \pm 0.047$ 
& $0.288 \pm 0.016$ 
& $0.298 \pm 0.017$ 

 \\

$\thetae$ (mas) 
& $0.417 \pm 0.112$ 
& $0.382 \pm 0.098$  
& $0.456 \pm 0.104$ 
& $0.474 \pm 0.094$ 

 \\

$\murel$ (mas yr$^{-1}$) 
& $3.80 \pm 1.06$ 
& $3.23 \pm 0.87$
& $1.55 \pm 0.37$ 
& $1.67 \pm 0.34$ 
 \\

\hline
\hline
\end{tabular}
\end{table*}

\begin{table*}[ht]
\centering
\caption{CMD Parameters, $\thetas$, $\thetae$, and $\murel$ for \eventd}
\label{tab:cmd2}

\renewcommand{\arraystretch}{1.15}

\begin{tabular}{l|cccc}
\hline
\hline
Parameter 
& Close Small $\rho$ & Close Large $\rho$ & Wide Small $\rho$ & Wide Large $\rho$ \\
\hline

$(V-I)_{\rm cl}$ 
& $2.05\pm0.01$ & $\leftarrow$ & $\leftarrow$ & $\leftarrow$ \\

$I_{\rm cl}$ 
& $15.54 \pm 0.04$ 
& $\leftarrow$ & $\leftarrow$
& $\leftarrow$\\

$(V-I)_{\rm cl,0}$
& $1.06\pm0.03$
& $\leftarrow$ & $\leftarrow$
& $\leftarrow$\\

$I_{{\rm cl},0}$ 
& $14.60 \pm 0.04$ 
& $\leftarrow$ & $\leftarrow$
& $\leftarrow$\\

$(V-I)_{\rm S}$ 
& $1.75\pm0.03$& $\leftarrow$ & $\leftarrow$& $\leftarrow$ \\

$I_{\rm S}$ 
& $19.87 \pm 0.01$ 
& $19.93 \pm 0.01$ 
& $19.80 \pm 0.01$ 
& $19.90 \pm 0.01$ 
 \\

$(V-I)_{{\rm S},0}$ 
& $0.76 \pm 0.04$ 
& $\leftarrow$ & $\leftarrow$
& $\leftarrow$
 \\

$I_{{\rm S},0}$ 
& $18.92 \pm 0.06$ 
& $18.98 \pm 0.06$ 
& $18.85 \pm 0.06$ 
& $18.95 \pm 0.06$ 
 \\

$\thetas$ ($\mu$as) 
& $0.553 \pm 0.026$ 
& $0.538 \pm 0.026$ 
& $0.571 \pm 0.027$ 
& $0.545 \pm 0.026$ 
 \\

$\thetae$ (mas) 
& $>0.29$ 
& $0.103\pm0.024$ 
& $>0.23$ 
& $0.089 \pm 0.017$ 
 \\

$\murel$ (mas yr$^{-1}$) 
& $>6.4$ 
& $2.19 \pm 0.52$ 
& $>5.2$
& $1.92 \pm 0.38$ \\
\hline
\hline
\end{tabular}
\end{table*}

\begin{table*}[ht]
\centering
\caption{Lensing Physical Parameters from Bayesian Analyses}
\renewcommand{\arraystretch}{1.2}
\label{tab:bayes}
\begin{tabular}{l l| c c c c c}
\hline\hline
Event & Model & $M_{\rm host}$ ($M_{\odot}$) & $M_{\rm planet}$ ($M_{\rm Jupiter}$) & $D_{\rm L}$ (kpc) & $r_{\perp}$ (au) & $\mu_{\rm rel,hel}$ (mas yr$^{-1}$) \\
\hline

\multirow{2}{*}{KMT-2023-BLG-1810} 
& A
& $0.56^{+0.33}_{-0.27}$ 
& $7.24^{+4.34}_{-3.56}$ 
& $7.0^{+0.8}_{-1.6}$ 
& $3.4^{+0.8}_{-0.9}$ 
& $4.32^{+1.14}_{-1.02}$ \\

& B 
& $0.56^{+0.32}_{-0.25}$ 
& $6.17^{+3.58}_{-2.84}$ 
& $7.2^{+0.8}_{-1.2}$ 
& $2.7^{+0.7}_{-0.6}$ 
& $3.41^{+0.85}_{-0.78}$ \\

\hline

\multirow{2}{*}{KMT-2023-BLG-0084} 
& $u_0>0$ 
& $0.23^{+0.10}_{-0.07}$ 
& $2.50^{+1.50}_{-0.95}$ 
& $4.0^{+2.3}_{-0.8}$ 
& $2.5^{+0.7}_{-0.5}$ 
& $1.52^{+0.37}_{-0.39}$ \\

& $u_0<0$ 
& $0.20^{+0.06}_{-0.05}$ 
& $2.40^{+1.19}_{-0.86}$ 
& $3.9^{+0.8}_{-0.5}$ 
& $2.3^{+0.4}_{-0.4}$ 
& $1.61^{+0.31}_{-0.32}$ \\

\hline

\multirow{4}{*}{KMT-2023-BLG-1118}
& Close Small $\rho$
& $0.50^{+0.33}_{-0.26}$ 
& $2.86^{+2.11}_{-1.52}$ 
& $6.4^{+0.8}_{-1.5}$ 
& $2.2^{+0.5}_{-0.4}$ 
& $8.33^{+2.29}_{-1.36}$ \\

& Close Large $\rho$
& $0.11^{+0.15}_{-0.05}$ 
& $0.24^{+0.36}_{-0.18}$ 
& $7.2^{+0.7}_{-0.7}$ 
& $0.8^{+0.2}_{-0.2}$ 
& $2.53^{+0.51}_{-0.49}$ \\

& Wide Small $\rho$
& $0.45^{+0.33}_{-0.24}$ 
& $3.11^{+2.40}_{-1.66}$ 
& $6.5^{+0.7}_{-1.3}$ 
& $2.6^{+0.7}_{-0.6}$ 
& $7.60^{+2.50}_{-1.64}$ \\

& Wide Large $\rho$
& $0.09^{+0.12}_{-0.04}$ 
& $0.21^{+0.31}_{-0.12}$ 
& $7.2^{+0.7}_{-0.6}$ 
& $0.9^{+0.2}_{-0.1}$ 
& $2.15^{+0.38}_{-0.36}$ \\

\hline\hline
\end{tabular}
\end{table*}

\begin{figure*}
    \centering
    \includegraphics[width=0.99\textwidth]{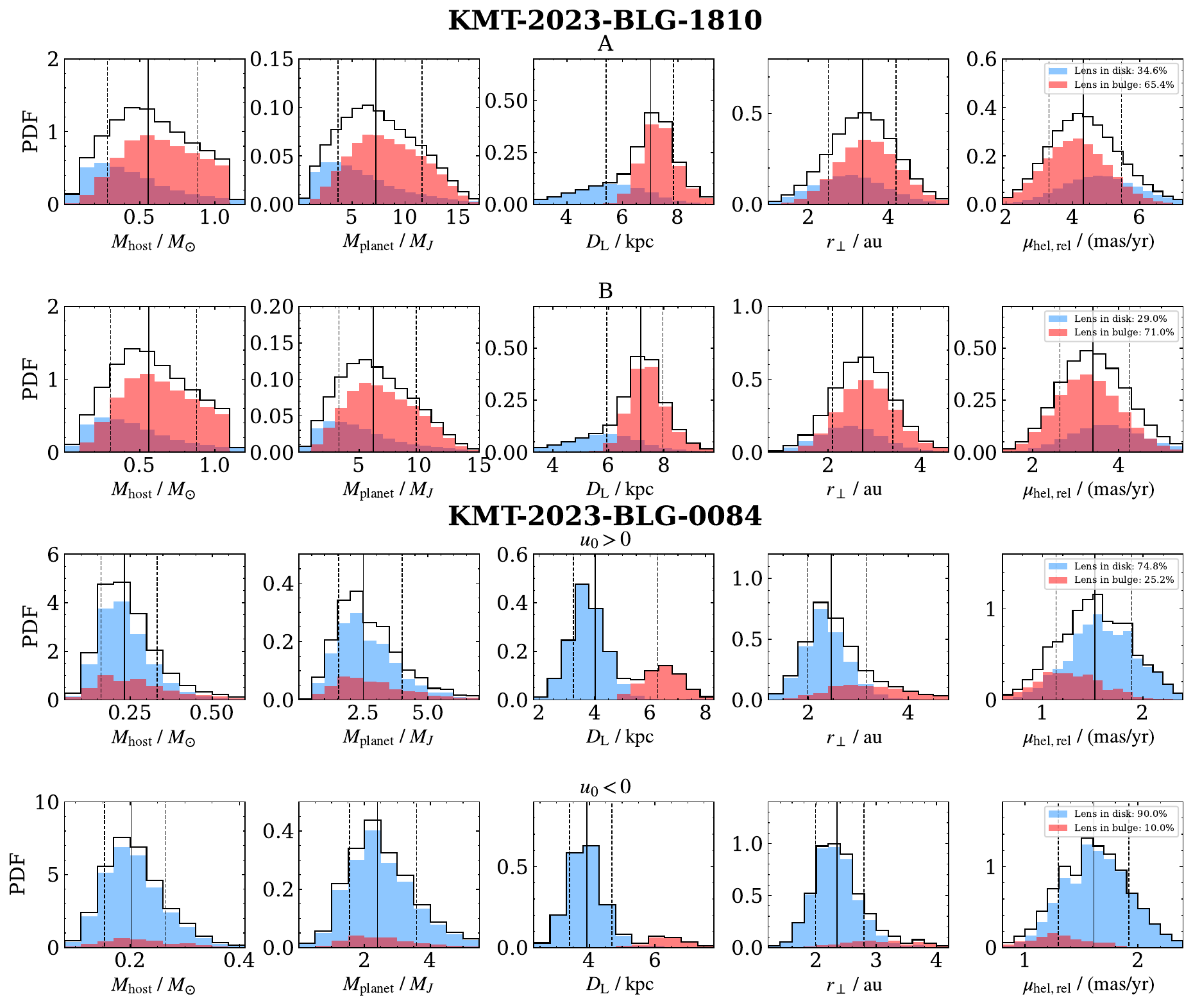}
    \caption{Posterior distributions from the Bayesian analysis of \eventa\ and \eventb\ for the host mass $M_{\rm host}$, planetary mass $M_{\rm planet}$, lens distance $D_{\rm L}$, projected planet-host separation $r_{\perp}$, and the heliocentric lens-source relative proper motion $µ_{\rm rel,hel}$. In each panel, the solid black curve denotes the median value, while the dashed black curves represent the 15.9\% and 84.1\% credible intervals. The contributions from disk and bulge lens populations are illustrated in blue and red, respectively.}
    \label{fig:bayes1}
\end{figure*}

\begin{figure*}
    \centering
    \includegraphics[width=0.99\textwidth]{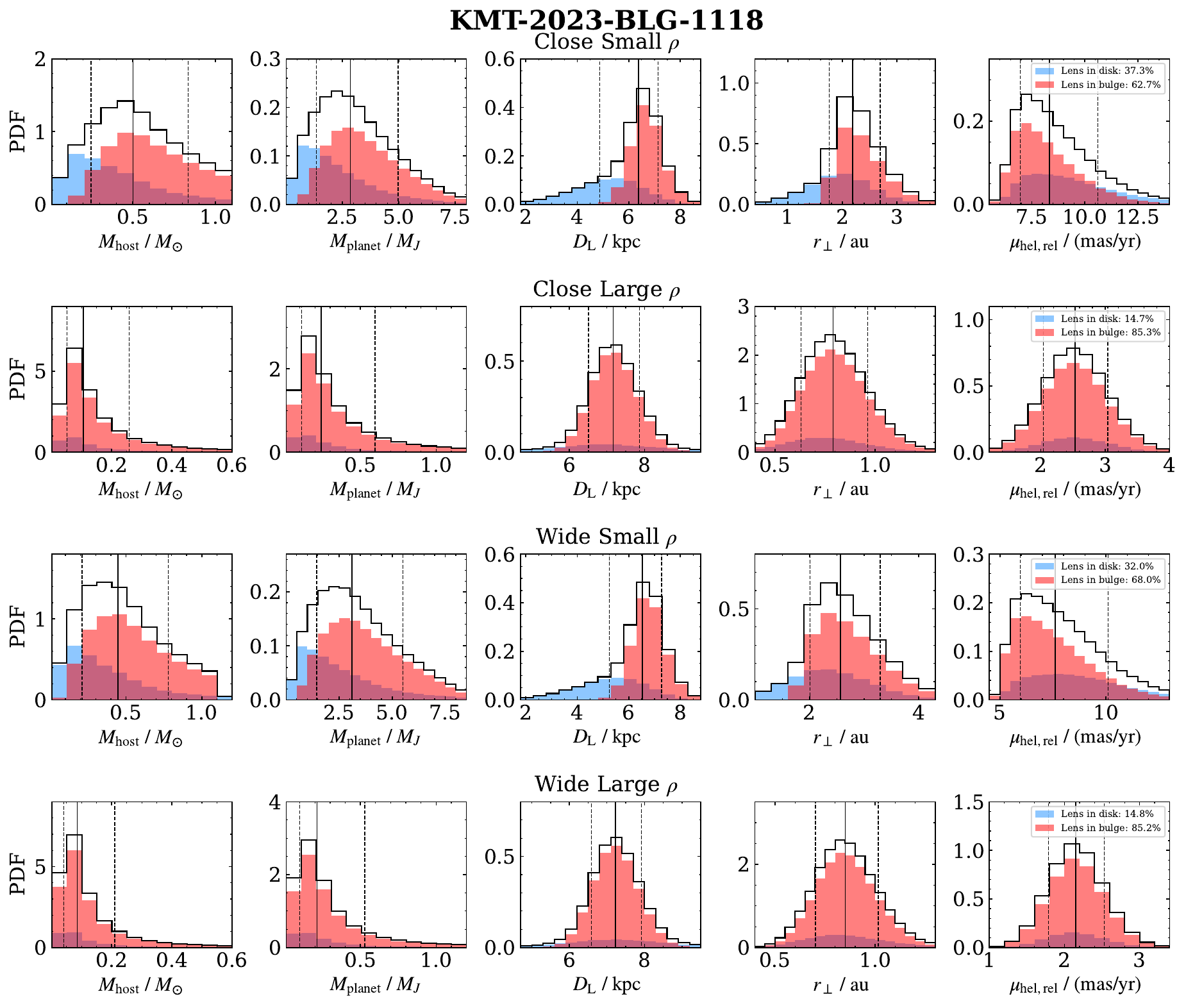}
    \caption{Posterior lens physical parameter distributions from the Bayesian analysis of \eventd. The
symbols are the same as those in Figure \ref{fig:bayes1}.}
    \label{fig:bayes2}
\end{figure*}

\begin{deluxetable}{lcccl}
\tablecaption{Known 2023 KMTNet Planetary Events from AnoamlyFinder}
\tablehead{
\colhead{Event Name$^{b}$} &
\colhead{$\log q$} &
\colhead{$s$} &
\colhead{Method} &
\colhead{Reference}
}
\startdata
KB231866$^{b}$  & $-4.81$ & $0.98$&by-eye & \cite{KB231866} \\
KB231746 & $-4.25$ & $0.96$&AF & \cite{KMT2023_mass1} \\
KB230416$^{b}$  & $-4.18$ & $1.00$&by-eye &\cite{KB230416_1454_1642}  \\
KB230382  & $-4.08$ & $1.09$&by-eye & Rybicki et al. in prep\\
KB230164  & $-3.97$ & $1.08$&AF & \cite{KMT2023_mass1} \\
KB231286  & $-3.72$ & $1.12$&AF & \cite{KMT2023_mass1} \\
KB230735  & $-3.71$ & $1.04$&by-eye & \cite{KB230469_0735} \\
KB230830$^{b}$  & $-3.63$ & $1.10$&by-eye & \cite{Han2324six} \\
KB230469  & $-3.60$ & $1.03$&by-eye & \cite{KB230469_0735} \\
KB230119  & $-3.43$ & $1.10$&by-eye & \cite{KB230119_1896} \\
KB231592  & $-3.01$ & $1.40$&by-eye & \cite{2023_prime}  \\
KB232209  & $-2.98$ & $1.00$&by-eye & \cite{KB172197_KB221790_KB222076_KB232209}  \\
KB231729  & $-2.71$ & $0.66$ & by-eye & Rybicki et al. in prep\\
KB231118$^{b}$ & $-2.66$ & $0.94$&AF & This work \\
OB230766  & $-2.59$ & $0.94$&AF &\cite{2023_prime}  \\
KB231454  & $-2.46$ & $0.90$&by-eye &\cite{KB230416_1454_1642}  \\
KB230548$^{b}$  & $-2.35$ & $2.43$ & by-eye & \cite{Han2324six} \\
OB230836$^{c}$  & $-2.23$ & $1.11$&by-eye &\cite{OB230836}  \\
KB231642  & $-2.23$ & $0.98$&by-eye &\cite{KB230416_1454_1642}  \\
KB230466  & $-2.16$ & $1.08$ &by-eye & \cite{KB200202_KB221551_KB230466_KB250121} \\
KB230084  & $-1.98$ & $0.86$&by-eye & This work \\
KB230598  & $-1.97$ & $0.89$&by-eye & Nunota et al. in prep\\
KB230949  & $-1.93$ & $0.73$ &by-eye & \cite{Han2324six}\\
KB231810  & $-1.90$ & $1.26$&AF & This work \\
KB231246  & $-1.81$ & $0.88$&by-eye & \cite{HanBD2325} \\ 
\hline
KB231896$^{b,d}$ & $-4.16$ & $1.27$&by-eye & \cite{KB230119_1896} \\
KB230486$^{b,e}$ & $-2.93$ & $0.88$&AF & \cite{2023_prime} \\
KB232218$^{d,e}$ & $-2.89$ & $1.37$&AF & This work \\
KB230792$^{e}$  & $-2.60$ & $0.71$&AF & \cite{2023_prime} \\
OB231043$^{e}$  & $-2.58$ & $0.76$&AF & \cite{2023_prime} \\
KB231593$^{e}$  & $-2.49$ & $0.97$&AF &  \cite{KMT2023_mass1} \\
KB230584$^{e}$  & $-2.20$ & $0.58$&AF & This work \\
KB230614$^{e}$  & $-1.93$ & $0.95$&AF & \cite{KMT2023_mass1} \\
KB231697$^{d,e}$  & $-1.58$ & $0.55$&by-eye & This work \\
\enddata

\tablecomments{
a: Event names are abbreviations, e.g., KMT-2023-BLG-1866 to KB231866.
b: large $q$ uncertainty according to the criteria of \cite{OB160007}.
c: planet in binary system.
d: planet/binary degeneracy.
e: 1L2S/2L1S degeneracy.
}
\label{tab:2023}
\end{deluxetable}

\section{Source and Lens Properties}\label{sec:lens}

\subsection{Preamble}
We estimate the lens properties for the three unambiguous planetary events in this section. From Equations~(\ref{equ:1}) and~(\ref{equ:3}), the lens mass $M_L$ and distance $D_L$ are given by \citep{Gould1992,Gould2000}
\begin{equation}
    M_{\rm L} = \frac{\thetae}{\kappa \pie}, \qquad
    D_{\rm L} = \frac{\mathrm{au}}{\pie \thetae + \pi_{\rm S}},
    \label{equ:5}
\end{equation}
where $\pi_{\rm S}$ is the source parallax. To determine the angular Einstein radius $\thetae$, we first estimate the angular source radius $\theta_*$ and then apply $\thetae = \theta_*/\rho$. 
We perform a color-magnitude diagram (CMD) analysis for the source star following \citet{Yoo2004}. 
The source is placed on the OGLE-III CMD \citep{OGLEIII} constructed from stars surrounding the event. 
The centroid of the red clump is measured as $(V-I, I)_{\rm cl}$ using the method of \citet{Nataf2013}. 
We adopt the intrinsic centroid $(V-I, I)_{\rm cl,0}$ from \citet{Bensby2013} and Table~1 of \citet{Nataf2013}.
The source apparent magnitude is determined from the light-curve modeling. For \eventb\ and \eventc, the source color is derived from a linear regression of the KMTC $I$- and $V$-band data. For \eventa, due to the low signal-to-noise ratio of the $V$-band data, we adopt the method of \cite{MB07192} to estimate the source color. Specifically, we match the \cite{HSTCMD} HST CMD to the KMTC CMD and then determine the source color from HST field stars whose magnitude lie within $5\sigma$ of the source brightness.

The angular source radius $\theta_*$ is derived using the color-surface brightness relations of \citet{Adams2018}. 
Finally, we obtain $\theta_{\rm E} = \theta_*/\rho$ and the relative proper motion $\mu_{\rm rel} = \theta_{\rm E}/t_{\rm E}$. Figure~\ref{fig:cmd} displays the CMD of the three secured planetary events. Table~\ref{tab:cmd} and~\ref{tab:cmd2} exhibits the CMD parameters and the resulting $\theta_*$, $\theta_{\rm E}$ and $\mu_{\rm{rel}}$.

For the microlensing parallax, only \eventb\ yields a well measurement of $\pi_{\rm E}$, whereas \eventa\ provides only a one-dimensional constraint on $\pi_{\rm E, \parallel}$. The remaining event, \eventd, does not yield a useful constraint on $\pi_{\rm E}$. We therefore infer the physical parameters of the planetary systems via a Bayesian analysis with a Galactic model prior, assuming that the planetary occurrence rate is independent of host-star properties (e.g., host mass). The Galactic model and analysis procedures follow those described in \citet{Yang2021_GalacticModel}, to which we refer the reader for details.

The resulting posterior distributions are shown in Figures~\ref{fig:bayes1} and~\ref{fig:bayes2} with the parameters in Table~\ref{tab:bayes}, 
including the host mass $M_{\rm host}$, planetary mass $M_{\rm planet}$, lens distance $D_{\rm L}$, projected separation $r_{\perp}$, and heliocentric lens-source relative proper motion $\mu_{\rm rel,hel}$.

\subsection{\eventa}
We construct the CMD using stars from the OGLE-III catalog within a $3'$ radius centered on the event position, together with the HST CMD \citep{HSTCMD}, aligned to the OGLE-III CMD using the position of the red clump. From the CMD analysis, we derive the angular Einstein radius of $\theta_{\rm E} = 0.417 \pm 0.112$~mas and $0.382 \pm 0.098$~mas for solutions ``A'' and ``B'', respectively.

The Bayesian analysis incorporates the one-dimensional parallax constraint ($\pi_{\rm E,\parallel}$). The host is likely a K- or M-dwarf orbited by a super-Jupiter. The projected planet-host separation is $3.4^{+0.8}_{-0.9}~\rm au$ for solution ``A'' and $2.7^{+0.7}_{-0.6}~\rm au$ for solution ``B''. Adopting a snow-line scaling of $a_{\rm SL} = 2.7 (M/M_{\odot})~\rm au$ \citep{snowline}, the planet is located well beyond the snow line in both solutions.

\subsection{\eventb}
The CMD was constructed using OGLE-III stars within a $3'$ radius of the source. As shown in Figure~\ref{fig:cmd}, the source is bluer than the bulge main-sequence branch, suggesting that it is likely a very metal-poor K dwarf. The CMD analysis yields $\theta_{\rm E} \sim 0.46$~mas, which, when combined with $\pi_{\rm E} \sim 0.31$, implies a low-mass M-dwarf host located in the Galactic disk.

This event has a relatively bright blend of $(V-I, I)_{\rm blend} = (1.701 \pm 0.033, 17.080 \pm 0.015)$. Based on its location on the CMD, the source color lies between the bulge giant branch and the foreground main-sequence locus. This suggests that it may be either a bulge subgiant or a disk star. However, the blend exhibits an astrometric offset of $0.41^{\prime\prime}$ from the source position, indicating that the majority of the blended light is probabaly not associated with either the lens or system.

Using a Galactic model prior, the Bayesian analysis indicates that the host is likely a $\sim 0.2\,M_{\odot}$ M dwarf. The inferred planetary masses are $2.50^{+1.50}_{-0.95}\,M_J$ for the $u_0>0$ solution and $2.40^{+1.19}_{-0.86}\,M_J$ for the $u_0<0$ solution, indicating a super-Jupiter companion. The projected planet-host separation is $\sim 2.4~\rm au$, placing the planet well beyond the snow line. The lens-source relative proper motion is low, about $1.5~\rm mas~yr^{-1}$. Therefore, although the lens and source may have similar brightness in the $K$ band, resolving them requires 30\,m-class telescopes.

\subsection{\eventd}

For this event, the two ``Small $\rho$'' solutions yield only upper limits on $\rho$. Combined with $\theta_*$ from the CMD analysis, these imply $3\sigma$ lower limits on $\theta_{\rm E}$ (0.2--0.3~mas) and $\mu_{\rm rel}$ (5--6~$\rm mas~yr^{-1}$). For the two ``Large $\rho$'' solutions, finite-source effects are well measured, yielding $\theta_{\rm E} \sim 0.1$~mas and $\mu_{\rm rel} \sim 2~\rm mas~yr^{-1}$. Because the two pairs of solutions have significantly different lens-source relative proper motions, they can be distinguished through future high-angular-resolution imaging that resolves the lens and source.

The Bayesian analysis reveals two distinct regimes of lens properties. For the two ``Small $\rho$'' solutions, the lens system is inferred to consist of an M- or K-dwarf host orbited by a super-Jupiter with a mass $M_{\rm planet} \sim 3\,M_J$ at a separation beyond the snow line. In contrast, the ``Large $\rho$'' solutions favor a late-type M dwarf with a mass near the hydrogen-burning limit, $M_{\rm host} \sim 0.10\,M_{\odot}$. In this scenario, the planet could be a Saturn-mass companion.

\section{Discussion}\label{sec:dis}

\begin{figure}[htb]
    \centering
    \includegraphics[width=0.47\textwidth]{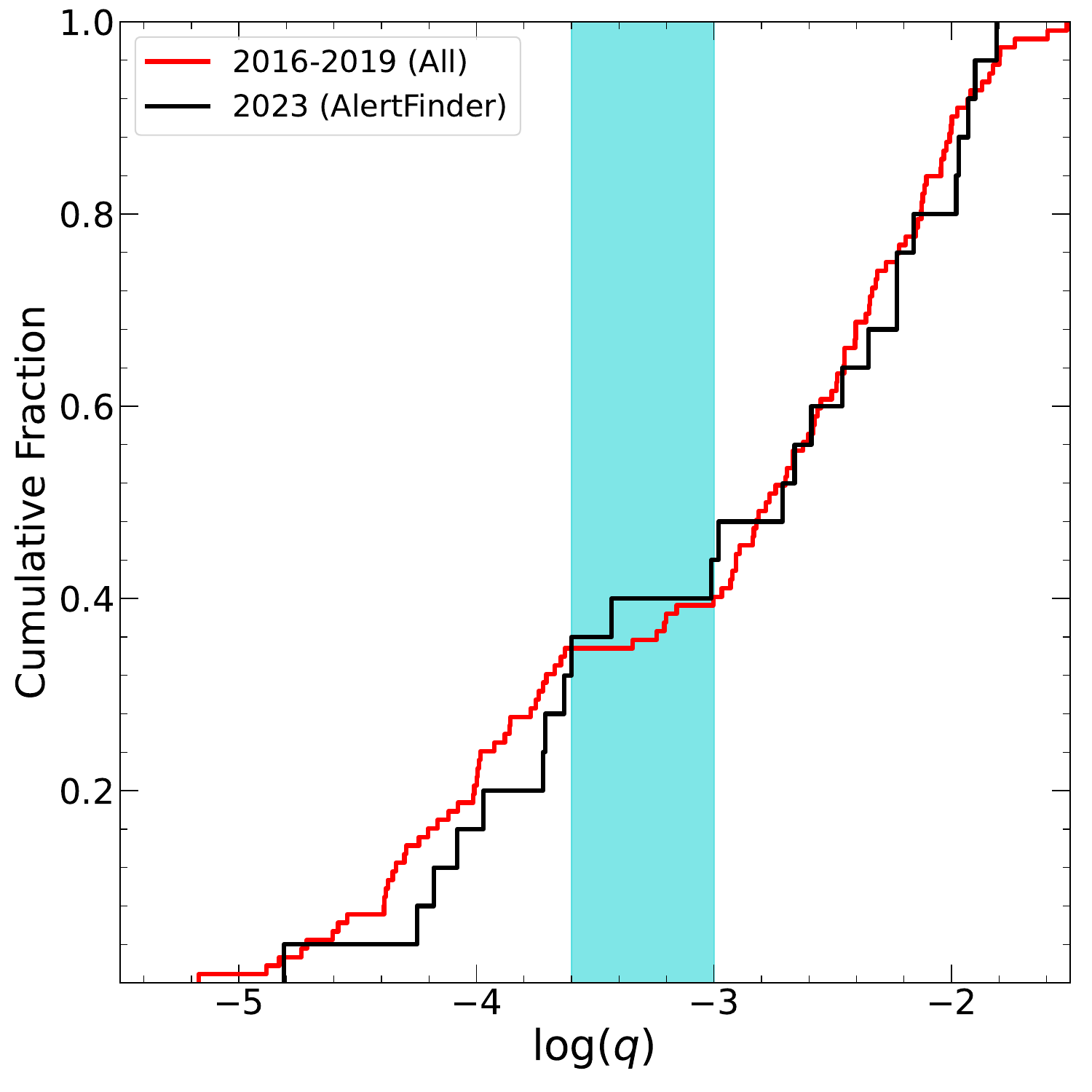}
    \caption{Cumulative distributions of $\log q$ for KMTNet AnomalyFinder planets from all 2016--2019 events (black curve) and from 2023 AlertFinder events (red curve), adapted from Figure 15 of \cite{2017_subprime}. The mass ratio of each planet is adopted from the best-fit model. The dark turquoise region indicates the ``sub-Saturn desert'' ($\log q = \left[-3.6, -3.0\right]$) indicated by \cite{OB160007}. }
\label{fig:cum}
\end{figure}

We have presented observations and analysis of six microlensing events located in the KMTNet subprime fields. Among them, three are securely confirmed as planetary events, for which a unique 2L1S interpretation is favored, with inferred mass ratios satisfying $\log q < -1.5$. The remaining three events exhibit the well-known 2L1S/1L2S degeneracy, and for \evente\ and \eventf, an additional planet/binary degeneracy is present. 

Table~\ref{tab:2023} lists all planetary events in the 2023 season that were alerted by the KMTNet AlertFinder system, including their event names, the parameters $q$ and $s$ for the preferred solutions, discovery methods, and references. We group them into two categories. The first category contains 25 unambiguous planetary events with $\log q < -1.5$. Another planet from the 2023 season, KMT-2023-BLG-1431 \citep{KB231431}, was discovered through follow-up observations of KMTNet high-magnification events \citep{KB200414} and is therefore not included in this sample. Of this sample, 19 were identified through by-eye searches, while 6 first were detected by the AnomalyFinder algorithm. The fraction of planets identified by the AnomalyFinder algorithm, $6/25 = 24\%$, is lower than that of the KMTNet AnomalyFinder planetary sample from the 2016--2019 seasons, $37/112 = 33\%$ \citep{2017_subprime}.

The second category consists of nine planets with ambiguous interpretations, including cases of the 2L1S/1L2S degeneracy (8 events) and the planet/binary degeneracy (3 events). In contrast to the unambiguous planetary sample, 7 out of 9 events in this category were first identified by the AnomalyFinder algorithm. This result is consistent with the findings of \cite{2023_prime} based on the \cite{OB160007} and 2023 prime-field samples, which show that AnomalyFinder is more sensitive to subtle anomalies and thus the 2L1S/1L2S and planet/binary degeneracies are more likely to arise.

The main scientific driver of this work is to expand the KMTNet planetary sample in order to reduce the uncertainty in the planetary mass-ratio function. Figure~\ref{fig:cum} shows the cumulative mass-ratio distributions for the KMTNet unambiguous AnomalyFinder planets from the 2016--2019 seasons and from the 2023 AlertFinder events. Overall, the two distributions are consistent with each other, with a Kolmogorov-Smirnov test yielding a p-value of 0.34. Regarding the ``sub-Saturn desert'' identified by \cite{OB160007}, only one of the 2023 AlertFinder planetary events, KMT-2023-BLG-0119 ($\log q = -3.43$, \citealt{KB230119_1896}), lies well within this desert. Two additional events, KMT-2023-BLG-0469 ($\log q = -3.601$, \citealt{KB230469_0735}) and KMT-2023-BLG-1592 ($\log q = -3.010$, \citealt{2023_prime}), are located near its boundary.

At present, among the 890 events newly identified by the end-of-year KMTNet EventFinder pipeline \citep{KMTeventfinder}, only two stellar binary events with prominent orbital effects \citep{Han2024orbital} and one free-floating planetary event have been published \citep{KB232669}. For comparison, in the 2019 KMTNet planetary sample, 9 out of 24 unambiguous AnomalyFinder planets were discovered among the new events detected by the end-of-year EventFinder pipeline \citep{2019_prime,2019_subprime}. Therefore, a systematic search using AnomalyFinder, along with a comprehensive analysis, is needed to complete the 2023 KMTNet planetary sample.

\bigskip

H.L., Z.L., W.Z., H.Y., Y.T., J.Z. and S.M. acknowledge support by the National Natural Science Foundation of China (Grant No. 12133005). This research has made use of the KMTNet system
operated by the Korea Astronomy and Space Science Institute
(KASI) at three host sites of CTIO in Chile, SAAO in South
Africa, and SSO in Australia. Data transfer from the host site to KASI was supported by the Korea Research Environment Open NETwork (KREONET). This research was supported by KASI under the R\&D program (project No. 2026-1-904-01) supervised by the Ministry of Science and ICT. The OGLE project has received funding from the Polish National Science Centre grant OPUS 2024/55/B/ST9/00447 awarded to A.U. The MOA project is supported by JSPS KAKENHI Grant Number JP24253004, JP26247023, JP16H06287 and JP22H00153. This work is part of the ET space mission which is funded by the China's Space Origins Exploration Program. H.Y. acknowledges support by the China Postdoctoral Science Foundation (No. 2024M762938). J.C.Y. acknowledges support from U.S. NASA Grant No. 80NSSC25K7146. J.C.Y. acknowledges support from a Scholarly Studies grant from the Smithsonian Institution. Work by C.H. was supported by the grants of National Research Foundation of Korea (2019R1A2C2085965 and 2020R1A4A2002885). R.P. acknowledges support by the Polish National Agency for Academic Exchange grant ``Polish Returns 2019.''

\software{pySIS \citep{pysis,Yang_TLC,Yang_TLC2}, numpy \citep{numpy}, emcee \citep{emcee2,emcee}, Matplotlib \citep{Matplotlib}, SciPy \citep{scipy}}

\bibliography{Zang.bib}

@ARTICLE{NASAExo,
       author = {{Christiansen}, Jessie L. and {McElroy}, Douglas L. and {Harbut}, Marcy and {Ciardi}, David R. and {Crane}, Megan and {Good}, John and {Hardegree-Ullman}, Kevin K. and {Kesseli}, Aurora Y. and {Lund}, Michael B. and {Lynn}, Meca and {Muthiar}, Ananda and {Nilsson}, Ricky and {Oluyide}, Toba and {Papin}, Michael and {Rivera}, Amalia and {Swain}, Melanie and {Susemiehl}, Nicholas D. and {Tam}, Raymond and {van Eyken}, Julian and {Beichman}, Charles},
        title = "{The NASA Exoplanet Archive and Exoplanet Follow-up Observing Program: Data, Tools, and Usage}",
      journal = {PSJ},
         year = 2025,
        month = aug,
       volume = {6},
       number = {8},
          eid = {186},
        pages = {186},
          doi = {10.3847/PSJ/ade3c2},
archivePrefix = {arXiv},
       eprint = {2506.03299},
 primaryClass = {astro-ph.EP},
       adsurl = {https://ui.adsabs.harvard.edu/abs/2025PSJ.....6..186C}
}

@ARTICLE{OB160007,
       author = {{Zang}, Weicheng and {Jung}, Youn Kil and {Yee}, Jennifer C. and {Hwang}, Kyu-Ha and {Yang}, Hongjing and {Udalski}, Andrzej and {Sumi}, Takahiro and {Gould}, Andrew and {Mao}, Shude and {Albrow}, Michael D. and {Chung}, Sun-Ju and {Han}, Cheongho and {Ryu}, Yoon-Hyun and {Shin}, In-Gu and {Shvartzvald}, Yossi and {Cha}, Sang-Mok and {Kim}, Dong-Jin and {Kim}, Hyoun-Woo and {Kim}, Seung-Lee and {Lee}, Chung-Uk and {Lee}, Dong-Joo and {Lee}, Yongseok and {Park}, Byeong-Gon and {Pogge}, Richard W. and {Zhang}, Xiangyu and {Kuang}, Renkun and {Wang}, Hanyue and {Zhang}, Jiyuan and {Hu}, Zhecheng and {Zhu}, Wei and {Mr{\'o}z}, Przemek and {Skowron}, Jan and {Poleski}, Rados{\l}aw and {Szyma{\'n}ski}, Micha{\l} K. and {Soszy{\'n}ski}, Igor and {Pietrukowicz}, Pawe{\l} and {Koz{\l}owski}, Szymon and {Ulaczyk}, Krzysztof and {Rybicki}, Krzysztof A. and {Iwanek}, Patryk and {Wrona}, Marcin and {Gromadzki}, Mariusz and {Abe}, Fumio and {Barry}, Richard and {Bennett}, David P. and {Bhattacharya}, Aparna and {Bond}, Ian A. and {Fujii}, Hirosane and {Fukui}, Akihiko and {Hamada}, Ryusei and {Hirao}, Yuki and {Silva}, Stela Ishitani and {Itow}, Yoshitaka and {Kirikawa}, Rintaro and {Koshimoto}, Naoki and {Matsubara}, Yutaka and {Miyazaki}, Shota and {Muraki}, Yasushi and {Olmschenk}, Greg and {Ranc}, Cl{\'e}ment and {Rattenbury}, Nicholas J. and {Satoh}, Yuki and {Suzuki}, Daisuke and {Tomoyoshi}, Mio and {Tristram}, Paul J. and {Vandorou}, Aikaterini and {Yama}, Hibiki and {Yamashita}, Kansuke},
        title = "{Microlensing events indicate that super-Earth exoplanets are common in Jupiter-like orbits}",
      journal = {Science},
         year = 2025,
        month = apr,
       volume = {388},
       number = {6745},
        pages = {400-404},
          doi = {10.1126/science.adn6088},
archivePrefix = {arXiv},
       eprint = {2504.20158},
 primaryClass = {astro-ph.EP},
       adsurl = {https://ui.adsabs.harvard.edu/abs/2025Sci...388..400Z}
}

@ARTICLE{MASADA,
       author = {{Gould}, Andrew},
        title = "{MASADA: From Microlensing Planet Mass-Ratio Function to Planet Mass Function}",
      journal = {arXiv e-prints},
         year = 2022,
        month = sep,
          eid = {arXiv:2209.12501},
        pages = {arXiv:2209.12501},
archivePrefix = {arXiv},
       eprint = {2209.12501},
 primaryClass = {astro-ph.EP},
       adsurl = {https://ui.adsabs.harvard.edu/abs/2022arXiv220912501G}
}

@ARTICLE{Yang2021_GalacticModel,
       author = {{Yang}, Hongjing and {Mao}, Shude and {Zang}, Weicheng and {Zhang}, Xiangyu},
        title = "{Microlensing predictions: impact of Galactic disc dynamical models}",
      journal = {\mnras},
         year = 2021,
        month = apr,
       volume = {502},
       number = {4},
        pages = {5631-5642},
          doi = {10.1093/mnras/stab441},
archivePrefix = {arXiv},
       eprint = {2010.16146},
 primaryClass = {astro-ph.GA},
       adsurl = {https://ui.adsabs.harvard.edu/abs/2021MNRAS.502.5631Y}
}

@Article{Matplotlib,
  Author    = {Hunter, J. D.},
  Title     = {Matplotlib: A 2D graphics environment},
  Journal   = {Computing in Science \& Engineering},
  Volume    = {9},
  Number    = {3},
  Pages     = {90--95},
  publisher = {IEEE COMPUTER SOC},
  doi       = {10.1109/MCSE.2007.55},
  year      = 2007
}

@Article{numpy,
 title         = {Array programming with {NumPy}},
 author        = {Charles R. Harris and K. Jarrod Millman and St{\'{e}}fan J.
                 van der Walt and Ralf Gommers and Pauli Virtanen and David
                 Cournapeau and Eric Wieser and Julian Taylor and Sebastian
                 Berg and Nathaniel J. Smith and Robert Kern and Matti Picus
                 and Stephan Hoyer and Marten H. van Kerkwijk and Matthew
                 Brett and Allan Haldane and Jaime Fern{\'{a}}ndez del
                 R{\'{i}}o and Mark Wiebe and Pearu Peterson and Pierre
                 G{\'{e}}rard-Marchant and Kevin Sheppard and Tyler Reddy and
                 Warren Weckesser and Hameer Abbasi and Christoph Gohlke and
                 Travis E. Oliphant},
 year          = {2020},
 month         = sep,
 journal       = {Nature},
 volume        = {585},
 number        = {7825},
 pages         = {357--362},
 doi           = {10.1038/s41586-020-2649-2},
 publisher     = {Springer Science and Business Media {LLC}},
 url           = {https://doi.org/10.1038/s41586-020-2649-2}
}

@ARTICLE{Nemiroff1994,
   author = {{Nemiroff}, R.~J. and {Wickramasinghe}, W.~A.~D.~T.},
    title = "{Finite source sizes and the information content of macho-type lens search light curves}",
  journal = {\apjl},
   eprint = {astro-ph/9401005},
     year = 1994,
    month = mar,
   volume = 424,
    pages = {L21-L23},
      doi = {10.1086/187265},
   adsurl = {http://adsabs.harvard.edu/abs/1994ApJ...424L..21N}
}

@ARTICLE{Gould2000,
   author = {{Gould}, A.},
    title = "{A Natural Formalism for Microlensing}",
  journal = {\apj},
   eprint = {astro-ph/0001421},
     year = 2000,
    month = oct,
   volume = 542,
    pages = {785-788},
      doi = {10.1086/317037},
   adsurl = {http://adsabs.harvard.edu/abs/2000ApJ...542..785G}
}

@ARTICLE{Nataf2013,
   author = {{Nataf}, D.~M. and {Gould}, A. and {Fouqu{\'e}}, P. and {Gonzalez}, O.~A. and 
	{Johnson}, J.~A. and {Skowron}, J. and {}, A. and {Szyma{\'n}ski}, M.~K. and 
	{Kubiak}, M. and {Pietrzy{\'n}ski}, G. and {Soszy{\'n}ski}, I. and 
	{Ulaczyk}, K. and {Wyrzykowski}, {\L}. and {Poleski}, R.},
    title = "{Reddening and Extinction toward the Galactic Bulge from OGLE-III: The Inner Milky Way's R$_{V}$ \~{} 2.5 Extinction Curve}",
  journal = {\apj},
archivePrefix = "arXiv",
   eprint = {1208.1263},
     year = 2013,
    month = jun,
   volume = 769,
      eid = {88},
    pages = {88},
      doi = {10.1088/0004-637X/769/2/88},
   adsurl = {http://adsabs.harvard.edu/abs/2013ApJ...769...88N}
}

@ARTICLE{1994ApJ...421L..75G,
   author = {{Gould}, A.},
    title = "{MACHO velocities from satellite-based parallaxes}",
  journal = {\apjl},
     year = 1994,
    month = feb,
   volume = 421,
    pages = {L75-L78},
      doi = {10.1086/187191},
   adsurl = {http://adsabs.harvard.edu/abs/1994ApJ...421L..75G}
}

@ARTICLE{Udalski2003,
   author = {{Udalski}, A.},
    title = "{The Optical Gravitational Lensing Experiment. Real Time Data Analysis Systems in the OGLE-III Survey}",
  journal = {\actaa},
   eprint = {astro-ph/0401123},
     year = 2003,
    month = dec,
   volume = 53,
    pages = {291-305},
   adsurl = {http://adsabs.harvard.edu/abs/2003AcA....53..291U}
}

@ARTICLE{Bond2001,
   author = {{Bond}, I.~A. and {Abe}, F. and {Dodd}, R.~J. and {Hearnshaw}, J.~B. and 
	{Honda}, M. and {Jugaku}, J. and {Kilmartin}, P.~M. and {Marles}, A. and 
	{Masuda}, K. and {Matsubara}, Y. and {Muraki}, Y. and {Nakamura}, T. and 
	{Nankivell}, G. and {Noda}, S. and {Noguchi}, C. and {Ohnishi}, K. and 
	{Rattenbury}, N.~J. and {Reid}, M. and {Saito}, T. and {Sato}, H. and 
	{Sekiguchi}, M. and {Skuljan}, J. and {Sullivan}, D.~J. and 
	{Sumi}, T. and {Takeuti}, M. and {Watase}, Y. and {Wilkinson}, S. and 
	{Yamada}, R. and {Yanagisawa}, T. and {Yock}, P.~C.~M.},
    title = "{Real-time difference imaging analysis of MOA Galactic bulge observations during 2000}",
  journal = {\mnras},
   eprint = {astro-ph/0102181},
     year = 2001,
    month = nov,
   volume = 327,
    pages = {868-880},
      doi = {10.1046/j.1365-8711.2001.04776.x},
   adsurl = {http://adsabs.harvard.edu/abs/2001MNRAS.327..868B}
}

@ARTICLE{Paczynski1986,
   author = {{Paczy{\'n}ski}, B.},
    title = "{Gravitational microlensing by the galactic halo}",
  journal = {\apj},
     year = 1986,
    month = may,
   volume = 304,
    pages = {1-5},
      doi = {10.1086/164140},
   adsurl = {http://adsabs.harvard.edu/abs/1986ApJ...304....1P}
}

@ARTICLE{Adams2018,
   author = {{Adams}, A.~D. and {Boyajian}, T.~S. and {von Braun}, K.},
    title = "{Predicting stellar angular diameters from V, I$_{C}$, H and K photometry}",
  journal = {\mnras},
archivePrefix = "arXiv",
   eprint = {1709.03902},
 primaryClass = "astro-ph.SR",
     year = 2018,
    month = jan,
   volume = 473,
    pages = {3608-3614},
      doi = {10.1093/mnras/stx2367},
   adsurl = {http://adsabs.harvard.edu/abs/2018MNRAS.473.3608A}
}

@ARTICLE{emcee,
   author = {{Foreman-Mackey}, D. and {Hogg}, D.~W. and {Lang}, D. and {Goodman}, J.
	},
    title = "{emcee: The MCMC Hammer}",
  journal = {\pasp},
archivePrefix = "arXiv",
   eprint = {1202.3665},
 primaryClass = "astro-ph.IM",
     year = 2013,
    month = mar,
   volume = 125,
    pages = {306},
      doi = {10.1086/670067},
   adsurl = {http://adsabs.harvard.edu/abs/2013PASP..125..306F}
}

@ARTICLE{OGLEIII,
   author = {{Szyma{\'n}ski}, M.~K. and {Udalski}, A. and {Soszy{\'n}ski}, I. and 
	{Kubiak}, M. and {Pietrzy{\'n}ski}, G. and {Poleski}, R. and 
	{Wyrzykowski}, {\L}. and {Ulaczyk}, K.},
    title = "{The Optical Gravitational Lensing Experiment. OGLE-III Photometric Maps of the Galactic Bulge Fields}",
  journal = {\actaa},
archivePrefix = "arXiv",
   eprint = {1107.4008},
 primaryClass = "astro-ph.SR",
     year = 2011,
    month = jun,
   volume = 61,
    pages = {83-102},
   adsurl = {http://adsabs.harvard.edu/abs/2011AcA....61...83S}
}

@ARTICLE{Udalski1994,
   author = {{Udalski}, A. and {Szymanski}, M. and {Kaluzny}, J. and {Kubiak}, M. and 
	{Mateo}, M. and {Krzeminski}, W. and {Paczynski}, B.},
    title = "{The Optical Gravitational Lensing Experiment. The Early Warning System: Real Time Microlensing}",
  journal = {\actaa},
   eprint = {astro-ph/9408026},
     year = 1994,
    month = jul,
   volume = 44,
    pages = {227-234},
   adsurl = {http://adsabs.harvard.edu/abs/1994AcA....44..227U}
}

@ARTICLE{KMT2016,
   author = {{Kim}, S.-L. and {Lee}, C.-U. and {Park}, B.-G. and {Kim}, D.-J. and 
	{Cha}, S.-M. and {Lee}, Y. and {Han}, C. and {Chun}, M.-Y. and 
	{Yuk}, I.},
    title = "{KMTNET: A Network of 1.6 m Wide-Field Optical Telescopes Installed at Three Southern Observatories}",
  journal = {Journal of Korean Astronomical Society},
     year = 2016,
    month = feb,
   volume = 49,
    pages = {37-44},
      doi = {10.5303/JKAS.2016.49.1.37},
   adsurl = {http://adsabs.harvard.edu/abs/2016JKAS...49...37K}
}

@ARTICLE{Shude1994,
   author = {{Witt}, H.~J. and {Mao}, S.},
    title = "{Can lensed stars be regarded as pointlike for microlensing by MACHOs?}",
  journal = {\apj},
     year = 1994,
    month = aug,
   volume = 430,
    pages = {505-510},
      doi = {10.1086/174426},
   adsurl = {http://adsabs.harvard.edu/abs/1994ApJ...430..505W}
}

@ARTICLE{Shude1991,
   author = {{Mao}, S. and {Paczynski}, B.},
    title = "{Gravitational microlensing by double stars and planetary systems}",
  journal = {\apjl},
     year = 1991,
    month = jun,
   volume = 374,
    pages = {L37-L40},
      doi = {10.1086/186066},
   adsurl = {http://adsabs.harvard.edu/abs/1991ApJ...374L..37M}
}

@ARTICLE{Andy1992,
   author = {{Gould}, A. and {Loeb}, A.},
    title = "{Discovering planetary systems through gravitational microlenses}",
  journal = {\apj},
     year = 1992,
    month = sep,
   volume = 396,
    pages = {104-114},
      doi = {10.1086/171700},
   adsurl = {http://adsabs.harvard.edu/abs/1992ApJ...396..104G}
}

@ARTICLE{OGLEIV,
   author = {{Udalski}, A. and {Szyma{\'n}ski}, M.~K. and {Szyma{\'n}ski}, G.
	},
    title = "{OGLE-IV: Fourth Phase of the Optical Gravitational Lensing Experiment}",
  journal = {\actaa},
archivePrefix = "arXiv",
   eprint = {1504.05966},
 primaryClass = "astro-ph.SR",
     year = 2015,
    month = mar,
   volume = 65,
    pages = {1-38},
   adsurl = {http://adsabs.harvard.edu/abs/2015AcA....65....1U}
}

@ARTICLE{Gaudi1998,
   author = {{Gaudi}, B.~S.},
    title = "{Distinguig Between Binary-Source and Planetary Microlensing Perturbations}",
  journal = {\apj},
     year = 1998,
    month = oct,
   volume = 506,
    pages = {533-539},
      doi = {10.1086/306256},
   adsurl = {http://adsabs.harvard.edu/abs/1998ApJ...506..533G}
}

@ARTICLE{pysis,
   author = {{Albrow}, M.~D. and {Horne}, K. and {Bramich}, D.~M. and {Fouqu{\'e}}, P. and 
	{Miller}, V.~R. and {Beaulieu}, J.-P. and {Coutures}, C. and 
	{Menzies}, J. and {Williams}, A. and {Batista}, V. and {Bennett}, D.~P. and 
	{Brillant}, S. and {Cassan}, A. and {Dieters}, S. and {Dominis Prester}, D. and 
	{Donatowicz}, J. and {Greenhill}, J. and {Kains}, N. and {Kane}, S.~R. and 
	{Kubas}, D. and {Marquette}, J.~B. and {Pollard}, K.~R. and 
	{Sahu}, K.~C. and {Tsapras}, Y. and {Wambsganss}, J. and {Zub}, M.
	},
    title = "{Difference imaging photometry of blended gravitational microlensing events with a numerical kernel}",
  journal = {\mnras},
archivePrefix = "arXiv",
   eprint = {0905.3003},
 primaryClass = "astro-ph.SR",
     year = 2009,
    month = aug,
   volume = 397,
    pages = {2099-2105},
      doi = {10.1111/j.1365-2966.2009.15098.x},
   adsurl = {http://adsabs.harvard.edu/abs/2009MNRAS.397.2099A}
}

@ARTICLE{Bozza2010,
   author = {{Bozza}, V.},
    title = "{Microlensing with an advanced contour integration algorithm: Green's theorem to third order, error control, optimal sampling and limb darkening}",
  journal = {\mnras},
archivePrefix = "arXiv",
   eprint = {1004.2796},
 primaryClass = "astro-ph.EP",
     year = 2010,
    month = nov,
   volume = 408,
    pages = {2188-2200},
      doi = {10.1111/j.1365-2966.2010.17265.x},
   adsurl = {http://adsabs.harvard.edu/abs/2010MNRAS.408.2188B}
}

@ARTICLE{MB12486,
   author = {{Hwang}, K.-H. and {Choi}, J.-Y. and {Bond}, I.~A. and {Sumi}, T. and 
	{Han}, C. and {Gaudi}, B.~S. and {Gould}, A. and {Bozza}, V. and 
	{Beaulieu}, J.-P. and {Tsapras}, Y. and {Abe}, F. and {Bennett}, D.~P. and 
	{Botzler}, C.~S. and {Chote}, P. and {Freeman}, M. and {Fukui}, A. and 
	{Fukunaga}, D. and {Harris}, P. and {Itow}, Y. and {Koshimoto}, N. and 
	{Ling}, C.~H. and {Masuda}, K. and {Matsubara}, Y. and {Muraki}, Y. and 
	{Namba}, S. and {Ohnishi}, K. and {Rattenbury}, N.~J. and {Saito}, T. and 
	{Sullivan}, D.~J. and {Sweatman}, W.~L. and {Suzuki}, D. and 
	{Tristram}, P.~J. and {Wada}, K. and {Yamai}, N. and {Yock}, P.~C.~M. and 
	{Yonehara}, A. and {MOA Collaboration} and {de Almeida}, L.~A. and 
	{DePoy}, D.~L. and {Dong}, S. and {Jablonski}, F. and {Jung}, Y.~K. and 
	{Kavka}, A. and {Lee}, C.-U. and {Park}, H. and {Pogge}, R.~W. and 
	{}, I.-G. and {Yee}, J.~C. and {{$\mu$}FUN Collaboration} and 
	{Albrow}, M.~D. and {Bachelet}, E. and {Batista}, V. and {Brillant}, S. and 
	{Caldwell}, J.~A.~R. and {Cassan}, A. and {Cole}, A. and {Corrales}, E. and 
	{Coutures}, C. and {Dieters}, S. and {Dominis Prester}, D. and 
	{Donatowicz}, J. and {Fouqu{\'e}}, P. and {Greenhill}, J. and 
	{J{\o}rgensen}, U.~G. and {Kane}, S.~R. and {Kubas}, D. and 
	{Marquette}, J.-B. and {Martin}, R. and {Meintjes}, P. and {Menzies}, J. and 
	{Pollard}, K.~R. and {Williams}, A. and {Wouters}, D. and {PLANET Collaboration} and 
	{Bramich}, D.~M. and {Dominik}, M. and {Horne}, K. and {Browne}, P. and 
	{Hundertmark}, M. and {Ipatov}, S. and {Kains}, N. and {Snodgrass}, C. and 
	{Steele}, I.~A. and {Street}, R.~A. and {RoboNet Collaboration}
	},
    title = "{Interpretation of a Short-term Anomaly in the Gravitational Microlensing Event MOA-2012-BLG-486}",
  journal = {\apj},
archivePrefix = "arXiv",
   eprint = {1308.5762},
 primaryClass = "astro-ph.SR",
     year = 2013,
    month = nov,
   volume = 778,
      eid = {55},
    pages = {55},
      doi = {10.1088/0004-637X/778/1/55},
   adsurl = {http://adsabs.harvard.edu/abs/2013ApJ...778...55H}
}

@ARTICLE{Yoo2004,
   author = {{Yoo}, J. and {DePoy}, D.~L. and {Gal-Yam}, A. and {Gaudi}, B.~S. and 
	{Gould}, A. and {Han}, C. and {Lipkin}, Y. and {Maoz}, D. and 
	{Ofek}, E.~O. and {Park}, B.-G. and {Pogge}, R.~W. and {Mu-Fun Collaboration} and 
	{Udalski}, A. and {Soszy{\'n}ski}, I. and {Wyrzykowski}, {\L}. and 
	{Kubiak}, M. and {Szyma{\'n}ski}, M. and {Pietrzy{\'n}ski}, G. and 
	{Szewczyk}, O. and {{\.Z}ebru{\'n}}, K. and {OGLE Collaboration}
	},
    title = "{OGLE-2003-BLG-262: Finite-Source Effects from a Point-Mass Lens}",
  journal = {\apj},
   eprint = {astro-ph/0309302},
     year = 2004,
    month = mar,
   volume = 603,
    pages = {139-151},
      doi = {10.1086/381241},
   adsurl = {http://adsabs.harvard.edu/abs/2004ApJ...603..139Y}
}

@ARTICLE{Wozniak2000,
   author = {{Wozniak}, P.~R.},
    title = "{Difference Image Analysis of the OGLE-II Bulge Data. I. The Method}",
  journal = {\actaa},
   eprint = {astro-ph/0012143},
     year = 2000,
    month = dec,
   volume = 50,
    pages = {421-450},
   adsurl = {http://adsabs.harvard.edu/abs/2000AcA....50..421W}
}

@ARTICLE{snowline,
   author = {{Kennedy}, G.~M. and {Kenyon}, S.~J.},
    title = "{Planet Formation around Stars of Various Masses: The Snow Line and the Frequency of Giant Planets}",
  journal = {\apj},
archivePrefix = "arXiv",
   eprint = {0710.1065},
     year = 2008,
    month = jan,
   volume = 673,
      eid = {502-512},
    pages = {502-512},
      doi = {10.1086/524130},
   adsurl = {http://adsabs.harvard.edu/abs/2008ApJ...673..502K}
}

@ARTICLE{Gouldpies2004,
   author = {{Gould}, A.},
    title = "{Resolution of the MACHO-LMC-5 Puzzle: The Jerk-Parallax Microlens Degeneracy}",
  journal = {\apj},
   eprint = {astro-ph/0311548},
     year = 2004,
    month = may,
   volume = 606,
    pages = {319-325},
      doi = {10.1086/382782},
   adsurl = {http://adsabs.harvard.edu/abs/2004ApJ...606..319G}
}

@ARTICLE{VBMicrolensing2025,
       author = {{Bozza}, V. and {Saggese}, V. and {Covone}, G. and {Rota}, P. and {Zhang}, J.},
        title = "{VBMicroLensing: Three algorithms for multiple lensing with contour integration}",
      journal = {\aap},
         year = 2025,
        month = feb,
       volume = {694},
          eid = {A219},
        pages = {A219},
          doi = {10.1051/0004-6361/202452648},
archivePrefix = {arXiv},
       eprint = {2410.13660},
 primaryClass = {astro-ph.IM},
       adsurl = {https://ui.adsabs.harvard.edu/abs/2025A&A...694A.219B}
}

@ARTICLE{KB231431,
       author = {{Bell}, Aislyn and {Zhang}, Jiyuan and {Zang}, Weicheng and {Jung}, Youn Kil and {Yee}, Jennifer C. and {Yang}, Hongjing and {Sumi}, Takahiro and {Udalski}, Andrzej and {Albrow}, Michael D. and {Chung}, Sun-Ju and {Gould}, Andrew and {Han}, Cheongho and {Hwang}, Kyu-Ha and {Ryu}, Yoon-Hyun and {Shin}, In-Gu and {Shvartzvald}, Yossi and {Cha}, Sang-Mok and {Kim}, Dong-Jin and {Kim}, Seung-Lee and {Lee}, Chung-Uk and {Lee}, Dong-Joo and {Lee}, Yongseok and {Park}, Byeong-Gon and {Pogge}, Richard W. and {KMTNet Collaboration} and {Tang}, Yunyi and {McCormick}, Jennie and {Dong}, Subo and {Liu}, Zhuokai and {de Almeida}, Leandro and {Mao}, Shude and {Maoz}, Dan and {Zhu}, Wei and {MAP and {\ensuremath{\mu}}FUN Follow-up Team} and {Abe}, Fumio and {Barry}, Richard and {Bennett}, David P. and {Bhattacharya}, Aparna and {Bond}, Ian A. and {Fujii}, Hirosane and {Fukui}, Akihiko and {Hamada}, Ryusei and {Hirao}, Yuki and {Silva}, Stela Ishitani and {Itow}, Yoshitaka and {Kirikawa}, Rintaro and {Kondo}, Iona and {Koshimoto}, Naoki and {Matsubara}, Yutaka and {Matsumoto}, Sho and {Miyazaki}, Shota and {Muraki}, Yasushi and {Okamura}, Arisa and {Lmschenk}, Greg and {Ranc}, Cl{\'e}ment and {Rattenbury}, Nicholas J. and {Satoh}, Yuki and {Suzuki}, Daisuke and {Toda}, Taiga and {Tomoyoshi}, Mio and {Tristram}, Paul J. and {Vandorou}, Aikaterini and {Yama}, Hibiki and {Yamashita}, Kansuke and {MOA Collaboration} and {Mr{\'o}z}, Przemek and {Skowron}, Jan and {Poleski}, Radoslaw and {Szyma{\'n}ski}, Micha{\l} K. and {Soszy{\'n}ski}, Igor and {Pietrukowicz}, Pawe{\l} and {Koz{\l}owski}, Szymon and {Ulaczyk}, Krzysztof and {Rybicki}, Krzysztof A. and {Iwanek}, Patryk and {Wrona}, Marcin and {Gromadzki}, Mariusz and {OGLE Collaboration}},
        title = "{KMT-2023-BLG-1431Lb: A New q < 10$^{‑4}$ Microlensing Planet from a Subtle Signature}",
      journal = {\pasp},
         year = 2024,
        month = may,
       volume = {136},
       number = {5},
          eid = {054402},
        pages = {054402},
          doi = {10.1088/1538-3873/ad48b8},
archivePrefix = {arXiv},
       eprint = {2311.13097},
 primaryClass = {astro-ph.EP},
       adsurl = {https://ui.adsabs.harvard.edu/abs/2024PASP..136e4402B}
}

@ARTICLE{MB09387,
       author = {{Batista}, V. and {Gould}, A. and {Dieters}, S. and {Dong}, S. and {Bond}, I. and {Beaulieu}, J.~P. and {Maoz}, D. and {Monard}, B. and {Christie}, G.~W. and {McCormick}, J. and {Albrow}, M.~D. and {Horne}, K. and {Tsapras}, Y. and {Burgdorf}, M.~J. and {Calchi Novati}, S. and {Skottfelt}, J. and {Caldwell}, J. and {Koz{\l}owski}, S. and {Kubas}, D. and {Gaudi}, B.~S. and {Han}, C. and {Bennett}, D.~P. and {An}, J. and {MOA Collaboration} and {Abe}, F. and {Botzler}, C.~S. and {Douchin}, D. and {Freeman}, M. and {Fukui}, A. and {Furusawa}, K. and {Hearnshaw}, J.~B. and {Hosaka}, S. and {Itow}, Y. and {Kamiya}, K. and {Kilmartin}, P.~M. and {Korpela}, A. and {Lin}, W. and {Ling}, C.~H. and {Makita}, B.~S. and {Masuda}, K. and {Matsubara}, Y. and {Miyake}, N. and {Muraki}, Y. and {Nagaya}, M. and {Nishimoto}, K. and {Ohnishi}, K. and {Okumura}, T. and {Perrott}, Y.~C. and {Rattenbury}, N. and {Saito}, To. and {Sullivan}, D.~J. and {Sumi}, T. and {Sweatman}, W.~L. and {Tristram}, P.~J. and {von Seggern}, E. and {Yock}, P.~C.~M. and {PLANET Collaboration} and {Brillant}, S. and {Calitz}, J.~J. and {Cassan}, A. and {Cole}, A. and {Cook}, K. and {Coutures}, C. and {Dominis Prester}, D. and {Donatowicz}, J. and {Greenhill}, J. and {Hoffman}, M. and {Jablonski}, F. and {Kane}, S.~R. and {Kains}, N. and {Marquette}, J. -B. and {Martin}, R. and {Martioli}, E. and {Meintjes}, P. and {Menzies}, J. and {Pedretti}, E. and {Pollard}, K. and {Sahu}, K.~C. and {Vinter}, C. and {Wambsganss}, J. and {Watson}, R. and {Williams}, A. and {Zub}, M. and {FUN Collaboration} and {Allen}, W. and {Bolt}, G. and {Bos}, M. and {DePoy}, D.~L. and {Drummond}, J. and {Eastman}, J.~D. and {Gal-Yam}, A. and {Gorbikov}, E. and {Higgins}, D. and {Janczak}, J. and {Kaspi}, S. and {Lee}, C. -U. and {Mallia}, F. and {Maury}, A. and {Monard}, L.~A.~G. and {Moorhouse}, D. and {Morgan}, N. and {Natusch}, T. and {Ofek}, E.~O. and {Park}, B. -G. and {Pogge}, R.~W. and {Polishook}, D. and {Santallo}, R. and {Shporer}, A. and {Spector}, O. and {Thornley}, G. and {Yee}, J.~C. and {MiNDSTEp Consortium} and {Bozza}, V. and {Browne}, P. and {Dominik}, M. and {Dreizler}, S. and {Finet}, F. and {Glitrup}, M. and {Grundahl}, F. and {Harps{\o}e}, K. and {Hessman}, F.~V. and {Hinse}, T.~C. and {Hundertmark}, M. and {J{\o}rgensen}, U.~G. and {Liebig}, C. and {Maier}, G. and {Mancini}, L. and {Mathiasen}, M. and {Rahvar}, S. and {Ricci}, D. and {Scarpetta}, G. and {Southworth}, J. and {Surdej}, J. and {Zimmer}, F. and {RoboNet Collaboration} and {Allan}, A. and {Bramich}, D.~M. and {Snodgrass}, C. and {Steele}, I.~A. and {Street}, R.~A.},
        title = "{MOA-2009-BLG-387Lb: a massive planet orbiting an M dwarf}",
      journal = {\aap},
         year = 2011,
        month = may,
       volume = {529},
          eid = {A102},
        pages = {A102},
          doi = {10.1051/0004-6361/201016111},
archivePrefix = {arXiv},
       eprint = {1102.0558},
 primaryClass = {astro-ph.EP},
       adsurl = {https://ui.adsabs.harvard.edu/abs/2011A&A...529A.102B}
}

@ARTICLE{Bensby2013,
   author = {{Bensby}, T. and {Yee}, J.~C. and {Feltzing}, S. and {Johnson}, J.~A. and 
	{Gould}, A. and {Cohen}, J.~G. and {Asplund}, M. and {Mel{\'e}ndez}, J. and 
	{Lucatello}, S. and {Han}, C. and {Thompson}, I. and {Gal-Yam}, A. and 
	{Udalski}, A. and {Bennett}, D.~P. and {Bond}, I.~A. and {Kohei}, W. and 
	{Sumi}, T. and {Suzuki}, D. and {Suzuki}, K. and {Takino}, S. and 
	{Tristram}, P. and {Yamai}, N. and {Yonehara}, A.},
    title = "{Chemical evolution of the Galactic bulge as traced by microlensed dwarf and subgiant stars. V. Evidence for a wide age distribution and a complex MDF}",
  journal = {\aap},
archivePrefix = "arXiv",
   eprint = {1211.6848},
     year = 2013,
    month = jan,
   volume = 549,
      eid = {A147},
    pages = {A147},
      doi = {10.1051/0004-6361/201220678},
   adsurl = {http://adsabs.harvard.edu/abs/2013A%26A...549A.147B}
}

@ARTICLE{KMTeventfinder,
   author = {{Kim}, D.-J. and {Kim}, H.-W. and {Hwang}, K.-H. and {Albrow}, M.~D. and 
	{Chung}, S.-J. and {Gould}, A. and {Han}, C. and {Jung}, Y.~K. and 
	{Ryu}, Y.-H. and {}, I.-G. and {Yee}, J.~C. and {Zhu}, W. and 
	{Cha}, S.-M. and {Kim}, S.-L. and {Lee}, C.-U. and {Lee}, D.-J. and 
	{Lee}, Y. and {Park}, B.-G. and {Pogge}, R.~W. and {The KMTNet Collaboration}
	},
    title = "{Korea Microlensing Telescope Network Microlensing Events from 2015: Event-finding Algorithm, Vetting, and Photometry}",
  journal = {\aj},
archivePrefix = "arXiv",
   eprint = {1703.06883},
 primaryClass = "astro-ph.EP",
     year = 2018,
    month = feb,
   volume = 155,
      eid = {76},
    pages = {76},
      doi = {10.3847/1538-3881/aaa47b},
   adsurl = {http://adsabs.harvard.edu/abs/2018AJ....155...76K}
}

@ARTICLE{Gould1992,
   author = {{Gould}, A.},
    title = "{Extending the MACHO search to about 10 exp 6 solar masses}",
  journal = {\apj},
     year = 1992,
    month = jun,
   volume = 392,
    pages = {442-451},
      doi = {10.1086/171443},
   adsurl = {http://adsabs.harvard.edu/abs/1992ApJ...392..442G}
}

@ARTICLE{Griest1998,
   author = {{Griest}, K. and {Safizadeh}, N.},
    title = "{The Use of High-Magnification Microlensing Events in Discovering Extrasolar Planets}",
  journal = {\apj},
   eprint = {astro-ph/9710342},
     year = 1998,
    month = jun,
   volume = 500,
    pages = {37-50},
      doi = {10.1086/305729},
   adsurl = {http://adsabs.harvard.edu/abs/1998ApJ...500...37G}
}

@ARTICLE{Yang_TLC,
       author = {{Yang}, Hongjing and {Yee}, Jennifer C. and {Hwang}, Kyu-Ha and {Qian}, Qiyue and {Bond}, Ian A. and {Gould}, Andrew and {Hu}, Zhecheng and {Zhang}, Jiyuan and {Mao}, Shude and {Zhu}, Wei and {Albrow}, Michael D. and {Chung}, Sun-Ju and {Kim}, Seung-Lee and {Park}, Byeong-Gon and {Han}, Cheongho and {Jung}, Youn Kil and {Ryu}, Yoon-Hyun and {Shin}, In-Gu and {Shvartzvald}, Yossi and {Cha}, Sang-Mok and {Kim}, Dong-Jin and {Kim}, Hyoun-Woo and {Lee}, Chung-Uk and {Lee}, Dong-Joo and {Lee}, Yongseok and {Pogge}, Richard W. and {Zang}, Weicheng and {Abe}, Fumio and {Barry}, Richard and {Bennett}, David P. and {Bhattacharya}, Aparna and {Donachie}, Martin and {Fujii}, Hirosane and {Fukui}, Akihiko and {Hirao}, Yuki and {Itow}, Yoshitaka and {Kirikawa}, Rintaro and {Kondo}, Iona and {Koshimoto}, Naoki and {Silva}, Stela Ishitani and {Li}, Man Cheung Alex and {Matsubara}, Yutaka and {Muraki}, Yasushi and {Suzuki}, Daisuke and {Tristram}, Paul J. and {Yonehara}, Atsunori and {Ranc}, Cl{\'e}ment and {Miyazaki}, Shota and {Olmschenk}, Greg and {Rattenbury}, Nicholas J. and {Satoh}, Yuki and {Shoji}, Hikaru and {Sumi}, Takahiro and {Tanaka}, Yuzuru and {Yamawaki}, Tsubasa},
        title = "{Systematic reanalysis of KMTNet microlensing events, paper I: Updates of the photometry pipeline and a new planet candidate}",
      journal = {\mnras},
         year = 2024,
        month = feb,
       volume = {528},
       number = {1},
        pages = {11-27},
          doi = {10.1093/mnras/stad3672},
archivePrefix = {arXiv},
       eprint = {2311.04876},
 primaryClass = {astro-ph.EP},
       adsurl = {https://ui.adsabs.harvard.edu/abs/2024MNRAS.528...11Y}
}

@ARTICLE{Yang_TLC2,
       author = {{Yang}, Hongjing and {Yee}, Jennifer C. and {Zhang}, Jiyuan and {Lee}, Chung-Uk and {Kim}, Dong-Jin and {Bond}, Ian A. and {Udalski}, Andrzej and {Hwang}, Kyu-Ha and {Zang}, Weicheng and {Qian}, Qiyue and {Gould}, Andrew and {Mao}, Shude and {Albrow}, Michael D. and {Chung}, Sun-Ju and {Han}, Cheongho and {Jung}, Youn Kil and {Ryu}, Yoon-Hyun and {Shin}, In-Gu and {Shvartzvald}, Yossi and {Cha}, Sang-Mok and {Kim}, Hyoun-Woo and {Kim}, Seung-Lee and {Lee}, Dong-Joo and {Lee}, Yongseok and {Park}, Byeong-Gon and {Pogge}, Richard W. and {KMTNet Collaboration} and {Abe}, Fumio and {Bando}, Ken and {Bennett}, David P. and {Bhattacharya}, Aparna and {Fukui}, Akihiko and {Hamada}, Ryusei and {Hamada}, Shunya and {Hamasaki}, Naoto and {Hirao}, Yuki and {Ishitani Silva}, Stela and {Itow}, Yoshitaka and {Koshimoto}, Naoki and {Matsubara}, Yutaka and {Miyazaki}, Shota and {Muraki}, Yasushi and {Nagai}, Tutumi and {Nunota}, Kansuke and {Olmschenk}, Greg and {Ranc}, Cl{\'e}ment and {Rattenbury}, Nicholas J. and {Satoh}, Yuki and {Sumi}, Takahiro and {Suzuki}, Daisuke and {Terry}, Sean K. and {Tristram}, Paul. J. and {Vandorou}, Aikaterini and {Yama}, Hibiki and {MOA Collaboration} and {Mr{\'o}z}, Przemek and {Skowron}, Jan and {Poleski}, Radoslaw and {Szyma{\'n}ski}, Micha{\l} K. and {Soszy{\'n}ski}, Igor and {Pietrukowicz}, Pawe{\l} and {Koz{\l}owski}, Szymon and {Ulaczyk}, Krzysztof and {Rybicki}, Krzysztof A. and {Iwanek}, Patryk and {Wrona}, Marcin and {OGLE Collaboration}},
        title = "{Systematic Reanalysis of KMTNet Microlensing Events. II. Two New Planets in Giant-source Events}",
      journal = {\aj},
         year = 2025,
        month = jun,
       volume = {169},
       number = {6},
          eid = {295},
        pages = {295},
          doi = {10.3847/1538-3881/adc73e},
archivePrefix = {arXiv},
       eprint = {2503.19471},
 primaryClass = {astro-ph.EP},
       adsurl = {https://ui.adsabs.harvard.edu/abs/2025AJ....169..295Y}
}

@ARTICLE{OB09020,
   author = {{Skowron}, J. and {Udalski}, A. and {Gould}, A. and {Dong}, S. and 
	{Monard}, L.~A.~G. and {Han}, C. and {Nelson}, C.~R. and {McCormick}, J. and 
	{Moorhouse}, D. and {Thornley}, G. and {Maury}, A. and {Bramich}, D.~M. and 
	{Greenhill}, J. and {Koz{\l}owski}, S. and {Bond}, I. and {Poleski}, R. and 
	{Wyrzykowski}, {\L}. and {Ulaczyk}, K. and {Kubiak}, M. and 
	{Szyma{\'n}ski}, M.~K. and {Pietrzy{\'n}ski}, G. and {Soszy{\'n}ski}, I. and 
	{OGLE Collaboration} and {Gaudi}, B.~S. and {Yee}, J.~C. and 
	{Hung}, L.-W. and {Pogge}, R.~W. and {DePoy}, D.~L. and {Lee}, C.-U. and 
	{Park}, B.-G. and {Allen}, W. and {Mallia}, F. and {Drummond}, J. and 
	{Bolt}, G. and {{$\mu$}FUN Collaboration} and {Allan}, A. and 
	{Browne}, P. and {Clay}, N. and {Dominik}, M. and {Fraser}, S. and 
	{Horne}, K. and {Kains}, N. and {Mottram}, C. and {Snodgrass}, C. and 
	{Steele}, I. and {Street}, R.~A. and {Tsapras}, Y. and {RoboNet Collaboration} and 
	{Abe}, F. and {Bennett}, D.~P. and {Botzler}, C.~S. and {Douchin}, D. and 
	{Freeman}, M. and {Fukui}, A. and {Furusawa}, K. and {Hayashi}, F. and 
	{Hearnshaw}, J.~B. and {Hosaka}, S. and {Itow}, Y. and {Kamiya}, K. and 
	{Kilmartin}, P.~M. and {Korpela}, A. and {Lin}, W. and {Ling}, C.~H. and 
	{Makita}, S. and {Masuda}, K. and {Matsubara}, Y. and {Muraki}, Y. and 
	{Nagayama}, T. and {Miyake}, N. and {Nishimoto}, K. and {Ohnishi}, K. and 
	{Perrott}, Y.~C. and {Rattenbury}, N. and {Saito}, T. and {Skuljan}, L. and 
	{Sullivan}, D.~J. and {Sumi}, T. and {Suzuki}, D. and {Sweatman}, W.~L. and 
	{Tristram}, P.~J. and {Wada}, K. and {Yock}, P.~C.~M. and {MOA Collaboration} and 
	{Beaulieu}, J.-P. and {Fouqu{\'e}}, P. and {Albrow}, M.~D. and 
	{Batista}, V. and {Brillant}, S. and {Caldwell}, J.~A.~R. and 
	{Cassan}, A. and {Cole}, A. and {Cook}, K.~H. and {Coutures}, C. and 
	{Dieters}, S. and {Dominis Prester}, D. and {Donatowicz}, J. and 
	{Kane}, S.~R. and {Kubas}, D. and {Marquette}, J.-B. and {Martin}, R. and 
	{Menzies}, J. and {Sahu}, K.~C. and {Wambsganss}, J. and {Williams}, A. and 
	{Zub}, M. and {PLANET Collaboration}},
    title = "{Binary Microlensing Event OGLE-2009-BLG-020 Gives Verifiable Mass, Distance, and Orbit Predictions}",
  journal = {\apj},
archivePrefix = "arXiv",
   eprint = {1101.3312},
 primaryClass = "astro-ph.SR",
     year = 2011,
    month = sep,
   volume = 738,
      eid = {87},
    pages = {87},
      doi = {10.1088/0004-637X/738/1/87},
   adsurl = {http://adsabs.harvard.edu/abs/2011ApJ...738...87S}
}

@ARTICLE{HSTCMD,
   author = {{Holtzman}, J.~A. and {Watson}, A.~M. and {Baum}, W.~A. and 
	{Grillmair}, C.~J. and {Groth}, E.~J. and {Light}, R.~M. and 
	{Lynds}, R. and {O'Neil}, Jr., E.~J.},
    title = "{The Luminosity Function and Initial Mass Function in the Galactic Bulge}",
  journal = {\aj},
   eprint = {astro-ph/9801321},
     year = 1998,
    month = may,
   volume = 115,
    pages = {1946-1957},
      doi = {10.1086/300336},
   adsurl = {http://adsabs.harvard.edu/abs/1998AJ....115.1946H}
}

@ARTICLE{MB07192,
   author = {{Bennett}, D.~P. and {Bond}, I.~A. and {Udalski}, A. and {Sumi}, T. and 
	{Abe}, F. and {Fukui}, A. and {Furusawa}, K. and {Hearnshaw}, J.~B. and 
	{Holderness}, S. and {Itow}, Y. and {Kamiya}, K. and {Korpela}, A.~V. and 
	{Kilmartin}, P.~M. and {Lin}, W. and {Ling}, C.~H. and {Masuda}, K. and 
	{Matsubara}, Y. and {Miyake}, N. and {Muraki}, Y. and {Nagaya}, M. and 
	{Okumura}, T. and {Ohnishi}, K. and {Perrott}, Y.~C. and {Rattenbury}, N.~J. and 
	{Sako}, T. and {Saito}, T. and {Sato}, S. and {Skuljan}, L. and 
	{Sullivan}, D.~J. and {Sweatman}, W.~L. and {Tristram}, P.~J. and 
	{Yock}, P.~C.~M. and {Kubiak}, M. and {Szyma{\'n}ski}, M.~K. and 
	{Pietrzy{\'n}ski}, G. and {Soszy{\'n}ski}, I. and {Szewczyk}, O. and 
	{Wyrzykowski}, {\L}. and {Ulaczyk}, K. and {Batista}, V. and 
	{Beaulieu}, J.~P. and {Brillant}, S. and {Cassan}, A. and {Fouqu{\'e}}, P. and 
	{Kervella}, P. and {Kubas}, D. and {Marquette}, J.~B.},
    title = "{A Low-Mass Planet with a Possible Sub-Stellar-Mass Host in Microlensing Event MOA-2007-BLG-192}",
  journal = {\apj},
archivePrefix = "arXiv",
   eprint = {0806.0025},
     year = 2008,
    month = sep,
   volume = 684,
    pages = {663-683},
      doi = {10.1086/589940},
   adsurl = {http://adsabs.harvard.edu/abs/2008ApJ...684..663B}
}

@ARTICLE{KB180029,
       author = {{Gould}, Andrew and {Ryu}, Yoon-Hyun and {Calchi Novati}, Sebastiano and
         {Zang}, Weicheng and {Albrow}, Michael D. and {Chung}, Sun-Ju and
         {Han}, Cheongho and {Hwang}, Kyu-Ha and {Jung}, Youn Kil and
         {Shin}, In-Gu and {Shvartzvald}, Yossi and {Yee}, Jennifer C. and
         {Cha}, Sang-Mok and {Kim}, Dong-Jin and {Kim}, Hyoun-Woo and
         {Kim}, Seung-Lee and {Lee}, Chung-Uk and {Lee}, Dong-Joo and
         {Lee}, Yongseok and {Park}, Byeong-Gon and {Pogge}, Richard W. and
         {Beichman}, Charles and {Bryden}, Geoff and {Carey}, Sean and
         {Gaudi}, B. Scott and {Henderson}, Calen B. and {Zhu}, Wei and
         {Fouque}, Pascal and {Penny}, Matthew T. and {Petric}, Andreea and
         {Burdullis}, Todd and {Mao}, Shude},
        title = "{KMT-2018-BLG-0029Lb: A Very Low Mass-Ratio Spitzer Microlens Planet}",
      journal = {Journal of Korean Astronomical Society},
         year = 2020,
        month = feb,
       volume = {53},
        pages = {9-26},
archivePrefix = {arXiv},
       eprint = {1906.11183},
 primaryClass = {astro-ph.EP},
       adsurl = {https://ui.adsabs.harvard.edu/abs/2020JKAS...53....9G}
}

@ARTICLE{OB181011,
       author = {{Han}, Cheongho and {Bennett}, David P. and {Udalski}, Andrzej and
         {Gould}, Andrew and {Bond}, Ian A. and {Shvartzvald}, Yossi and
         {Nikolaus}, Kay-Sebastian and {Hundertmark}, Markus and
         {Bozza}, Valerio and {Cassan}, Arnaud and {Hirao}, Yuki and
         {Bachelet}, Etienne and {Fouqu{\'e}}, Pascal and {Authors}, Leading and
         {Albrow}, Michael D. and {Chung}, Sun-Ju and {Hong}, Kyeongsoo and
         {Hwang}, Kyu-Ha and {Lee}, Chung-Uk and {Ryu}, Yoon-Hyun and
         {Shin}, In-Gu and {Yee}, Jennifer C. and {Jung}, Youn Kil and
         {Cha}, Sang-Mok and {Kim}, Doeon and {Kim}, Dong-Jin and
         {Kim}, Hyoun-Woo and {Kim}, Seung-Lee and {Lee}, Dong-Joo and
         {Lee}, Yongseok and {Park}, Byeong-Gon and {Pogge}, Richard W. and
         {The KMTNet Collaboration} and {Mr{\'o}z}, Przemek and
         {Szyma{\'n}ski}, Micha{\l} K. and {Skowron}, Jan and {Poleski}, Radek and
         {Soszy{\'n}ski}, Igor and {Pietrukowicz}, Pawe{\l} and
         {Koz{\l}owski}, Szymon and {Ulaczyk}, Krzysztof and
         {Rybicki}, Krzysztof A. and {Iwanek}, Patryk and {Wrona}, Marcin and
         {The OGLE Collaboration} and {Abe}, Fumio and {Barry}, Richard and
         {Bhattacharya}, Aparna and {Donachie}, Martin and {Fukui}, Akihiko and
         {Itow}, Yoshitaka and {Kawasaki}, Kohei and {Kondo}, Iona and
         {Koshimoto}, Naoki and {Li}, Man Cheung Alex and {Matsubara}, Yutaka and
         {Muraki}, Yasushi and {Miyazaki}, Shota and {Nagakane}, Masayuki and
         {Ranc}, Cl{\'e}ment and {Rattenbury}, Nicholas J. and
         {Suematsu}, Haruno and {Sullivan}, Denis J. and {Sumi}, Takahiro and
         {Suzuki}, Daisuke and {Tristram}, Paul J. and {Yonehara}, Atsunori and
         {The MOA Collaboration} and {Mao}, Shude and {Wang}, Tianshu and
         {Zang}, Weicheng and {Zhu}, Wei and {Penny}, Matthew T. and
         {The CFHT Collaboration} and {Beichman}, Charles A. and
         {Bryden}, Geoffery and {Calchi Novati}, Sebastiano and
         {Gaudi}, B. Scott and {Henderson}, Calen B. and {Jacklin}, Savannah and
         {Stassun}, Keivan G. and {The UKIRT Microlensing Team}},
        title = "{OGLE-2018-BLG-1011Lb,c: Microlensing Planetary System with Two Giant Planets Orbiting a Low-mass Star}",
      journal = {\aj},
         year = "2019",
        month = "Sep",
       volume = {158},
       number = {3},
          eid = {114},
        pages = {114},
          doi = {10.3847/1538-3881/ab2f74},
archivePrefix = {arXiv},
       eprint = {1907.01741},
 primaryClass = {astro-ph.EP},
       adsurl = {https://ui.adsabs.harvard.edu/abs/2019AJ....158..114H}
}

@ARTICLE{KMTAF,
       author = {{Kim}, Hyoun-Woo and {Hwang}, Kyu-Ha and {Shvartzvald}, Yossi and
         {Yee}, Jennifer C. and {Albrow}, Michael D. and {Cha}, Sang-Mok and
         {Chung}, Sun-Ju and {Gould}, Andrew and {Han}, Cheongho and
         {Jung}, Youn Kil and {Kim}, Dong-Jin and {Kim}, Seung-Lee and
         {Lee}, Chung-Uk and {Lee}, Dong-Joo and {Lee}, Yongseok and
         {Park}, Byeong-Gon and {Pogge}, Richard W. and {Ryu}, Yoon-Hyun and
         {Shin}, In-Gu and {Zang}, Weicheng},
        title = "{The Korea Microlensing Telescope Network (KMTNet) Alert Algorithm and Alert System}",
      journal = {arXiv e-prints},
         year = 2018,
        month = jun,
          eid = {arXiv:1806.07545},
        pages = {arXiv:1806.07545},
archivePrefix = {arXiv},
       eprint = {1806.07545},
 primaryClass = {astro-ph.IM},
       adsurl = {https://ui.adsabs.harvard.edu/abs/2018arXiv180607545K}
}

@ARTICLE{MB11293,
       author = {{Yee}, J.~C. and {Shvartzvald}, Y. and {Gal-Yam}, A. and {Bond}, I.~A. and
         {Udalski}, A. and {Koz{\l}owski}, S. and {Han}, C. and {Gould}, A. and
         {Skowron}, J. and {Suzuki}, D. and {Abe}, F. and {Bennett}, D.~P. and
         {Botzler}, C.~S. and {Chote}, P. and {Freeman}, M. and {Fukui}, A. and
         {Furusawa}, K. and {Itow}, Y. and {Kobara}, S. and {Ling}, C.~H. and
         {Masuda}, K. and {Matsubara}, Y. and {Miyake}, N. and {Muraki}, Y. and
         {Ohmori}, K. and {Ohnishi}, K. and {Rattenbury}, N.~J. and
         {Saito}, To. and {Sullivan}, D.~J. and {Sumi}, T. and {Suzuki}, K. and
         {Sweatman}, W.~L. and {Takino}, S. and {Tristram}, P.~J. and
         {Wada}, K. and {MOA Collaboration} and {Szyma{\'n}ski}, M.~K. and
         {Kubiak}, M. and {Pietrzy{\'n}ski}, G. and {Soszy{\'n}ski}, I. and
         {Poleski}, R. and {Ulaczyk}, K. and {Wyrzykowski}, {\L}. and
         {Pietrukowicz}, P. and {OGLE Collaboration} and {Allen}, W. and
         {Almeida}, L.~A. and {Batista}, V. and {Bos}, M. and {Christie}, G. and
         {DePoy}, D.~L. and {Dong}, Subo and {Drummond}, J. and {Finkelman}, I. and
         {Gaudi}, B.~S. and {Gorbikov}, E. and {Henderson}, C. and
         {Higgins}, D. and {Jablonski}, F. and {Kaspi}, S. and {Manulis}, I. and
         {Maoz}, D. and {McCormick}, J. and {McGregor}, D. and
         {Monard}, L.~A.~G. and {Moorhouse}, D. and {Mu{\~n}oz}, J.~A. and
         {Natusch}, T. and {Ngan}, H. and {Ofek}, E. and {Pogge}, R.~W. and
         {Santallo}, R. and {Tan}, T. -G. and {Thornley}, G. and {Shin}, I. -G. and
         {Choi}, J. -Y. and {Park}, S. -Y. and {Lee}, C. -U. and {Koo}, J. -R. and
         {{\ensuremath{\mu}}FUN Collaboration}},
        title = "{MOA-2011-BLG-293Lb: A Test of Pure Survey Microlensing Planet Detections}",
      journal = {\apj},
         year = 2012,
        month = aug,
       volume = {755},
       number = {2},
          eid = {102},
        pages = {102},
          doi = {10.1088/0004-637X/755/2/102},
archivePrefix = {arXiv},
       eprint = {1201.1002},
 primaryClass = {astro-ph.EP},
       adsurl = {https://ui.adsabs.harvard.edu/abs/2012ApJ...755..102Y}
}

@ARTICLE{2019_prime,
       author = {{Zang}, Weicheng and {Yang}, Hongjing and {Han}, Cheongho and {Lee}, Chung-Uk and {Udalski}, Andrzej and {Gould}, Andrew and {Mao}, Shude and {Zhang}, Xiangyu and {Zhu}, Wei and {Albrow}, Michael D. and {Chung}, Sun-Ju and {Hwang}, Kyu-Ha and {Jung}, Youn Kil and {Ryu}, Yoon-Hyun and {Shin}, In-Gu and {Shvartzvald}, Yossi and {Yee}, Jennifer C. and {Cha}, Sang-Mok and {Kim}, Dong-Jin and {Kim}, Hyoun-Woo and {Kim}, Seung-Lee and {Lee}, Dong-Joo and {Lee}, Yongseok and {Park}, Byeong-Gon and {Pogge}, Richard W. and {Mr{\'o}z}, Przemek and {Skowron}, Jan and {Poleski}, Radoslaw and {Szyma{\'n}ski}, Micha{\l} K. and {Soszy{\'n}ski}, Igor and {Pietrukowicz}, Pawe{\l} and {Koz{\l}owski}, Szymon and {Ulaczyk}, Krzysztof and {Rybicki}, Krzysztof A. and {Iwanek}, Patryk and {Wrona}, Marcin and {Gromadzki}, Mariusz},
        title = "{Systematic KMTNet planetary anomaly search. IV. Complete sample of 2019 prime-field}",
      journal = {\mnras},
         year = 2022,
        month = sep,
       volume = {515},
       number = {1},
        pages = {928-939},
          doi = {10.1093/mnras/stac1883},
archivePrefix = {arXiv},
       eprint = {2204.02017},
 primaryClass = {astro-ph.EP},
       adsurl = {https://ui.adsabs.harvard.edu/abs/2022MNRAS.515..928Z}
}

@ARTICLE{2017_subprime,
       author = {{Gui}, Yuqian and {Zang}, Weicheng and {Zhai}, Ruocheng and {Ryu}, Yoon-Hyun and {Udalski}, Andrzej and {Yang}, Hongjing and {Han}, Cheongho and {Mao}, Shude and {Albrow}, Michael D. and {Chung}, Sun-Ju and {Gould}, Andrew and {Hwang}, Kyu-Ha and {Jung}, Youn Kil and {Shin}, In-Gu and {Shvartzvald}, Yossi and {Yee}, Jennifer C. and {Cha}, Sang-Mok and {Kim}, Dong-Jin and {Kim}, Hyoun-Woo and {Kim}, Seung-Lee and {Lee}, Chung-Uk and {Lee}, Dong-Joo and {Lee}, Yongseok and {Park}, Byeong-Gon and {Pogge}, Richard W. and {KMTNet Collaboration} and {Mr{\'o}z}, Przemek and {Szyma{\'n}ski}, Micha{\l} K. and {Skowron}, Jan and {Poleski}, Rados{\l}aw and {Soszy{\'n}ski}, Igor and {Pietrukowicz}, Pawe{\l} and {Koz{\l}owski}, Szymon and {Ulaczyk}, Krzysztof and {Rybicki}, Krzysztof A. and {Iwanek}, Patryk and {Wrona}, Marcin and {Gromadzki}, Mariusz and {OGLE Collaboration} and {Wang}, Hanyue and {Zhang}, Jiyuan and {Kuang}, Renkun and {Qian}, Qiyue and {Zhu}, Wei and {MAP Collaboration}},
        title = "{Systematic KMTNet Planetary Anomaly Search. XII. Complete Sample of 2017 Subprime Field Planets}",
      journal = {\aj},
         year = 2024,
        month = aug,
       volume = {168},
       number = {2},
          eid = {49},
        pages = {49},
          doi = {10.3847/1538-3881/ad4ce5},
archivePrefix = {arXiv},
       eprint = {2504.20155},
 primaryClass = {astro-ph.EP},
       adsurl = {https://ui.adsabs.harvard.edu/abs/2024AJ....168...49G}
}

@ARTICLE{2023_prime,
       author = {{Li}, Zhixing and {Li}, Hongyu and {Zang}, Weicheng and {Ryu}, Yoon-Hyun and {Udalski}, Andrzej and {Sumi}, Takahiro and {Yang}, Hongjing and {Tang}, Yuchen and {Zhang}, Jiyuan and {Mao}, Shude and {Albrow}, Michael D. and {Chung}, Sun-Ju and {Gould}, Andrew and {Han}, Cheongho and {Hwang}, Kyu-Ha and {Jung}, Youn Kil and {Shin}, In-Gu and {Shvartzvald}, Yossi and {Yee}, Jennifer C. and {Cha}, Sang-Mok and {Kim}, Dong-Jin and {Kim}, Seung-Lee and {Lee}, Chung-Uk and {Lee}, Dong-Joo and {Lee}, Yongseok and {Park}, Byeong-Gon and {Pogge}, Richard W. and {Mr{\'o}z}, Przemek and {Szyma{\'n}ski}, Micha{\l} K. and {Skowron}, Jan and {Poleski}, Radoslaw and {Soszy{\'n}ski}, Igor and {Pietrukowicz}, Pawe{\l} and {Koz{\l}owski}, Szymon and {Rybicki}, Krzysztof A. and {Iwanek}, Patryk and {Ulaczyk}, Krzysztof and {Wrona}, Marcin and {Gromadzki}, Mariusz and {Mr{\'o}z}, Mateusz J. and {Abe}, Fumio and {Bando}, Ken and {Bennett}, David P. and {Bhattacharya}, Aparna and {Bond}, Ian A. and {Fukui}, Akihiko and {Hamada}, Ryusei and {Hamada}, Shunya and {Hamasak}, Naoto and {Hirao}, Yuki and {Ishitani Silva}, Stela and {Koshimoto}, Naoki and {Matsubara}, Yutaka and {Miyazaki}, Shota and {Muraki}, Yasushi and {Nagai}, Tutumi and {Nunota}, Kansuke and {Olmschenk}, Greg and {Ranc}, Cl{\'e}ment and {Rattenbury}, Nicholas J. and {Satoh}, Yuki and {Suzuki}, Daisuke and {Terry}, Sean and {Tristram}, Paul J. and {Vandorou}, Aikaterini and {Yama}, Hibiki},
        title = "{Mass Production of 2023 KMTNet Microlensing Planets. II: Two Planets and A Brown Dwarf}",
      journal = {arXiv e-prints},
         year = 2026,
        month = mar,
          eid = {arXiv:2603.13887},
        pages = {arXiv:2603.13887},
          doi = {10.48550/arXiv.2603.13887},
archivePrefix = {arXiv},
       eprint = {2603.13887},
 primaryClass = {astro-ph.EP},
       adsurl = {https://ui.adsabs.harvard.edu/abs/2026arXiv260313887L}
}

@ARTICLE{Bozza2018,
       author = {{Bozza}, Valerio and {Bachelet}, Etienne and {Bartoli{\'c}}, Fran and
         {Heintz}, Tyler M. and {Hoag}, Ava R. and {Hundertmark}, Markus},
        title = "{VBBINARYLENSING: a public package for microlensing light-curve computation}",
      journal = {\mnras},
         year = 2018,
        month = oct,
       volume = {479},
       number = {4},
        pages = {5157-5167},
          doi = {10.1093/mnras/sty1791},
archivePrefix = {arXiv},
       eprint = {1805.05653},
 primaryClass = {astro-ph.IM},
       adsurl = {https://ui.adsabs.harvard.edu/abs/2018MNRAS.479.5157B}
}

@ARTICLE{Poindexter2005,
       author = {{Poindexter}, Shawn and {Afonso}, Cristina and {Bennett}, David P. and
         {Glicenstein}, Jean-Francois and {Gould}, Andrew and
         {Szyma{\'n}ski}, Micha{\l} K. and {Udalski}, Andrzej},
        title = "{Systematic Analysis of 22 Microlensing Parallax Candidates}",
      journal = {\apj},
         year = "2005",
        month = "Nov",
       volume = {633},
       number = {2},
        pages = {914-930},
          doi = {10.1086/468182},
archivePrefix = {arXiv},
       eprint = {astro-ph/0506183},
 primaryClass = {astro-ph},
       adsurl = {https://ui.adsabs.harvard.edu/abs/2005ApJ...633..914P}
}

@ARTICLE{Jiang2004,
       author = {{Jiang}, Guangfei and {DePoy}, D.~L. and {Gal-Yam}, A. and
         {Gaudi}, B.~S. and {Gould}, A. and {Han}, C. and {Lipkin}, Y. and
         {Maoz}, D. and {Ofek}, E.~O. and {Park}, B. -G. and {Pogge}, R.~W. and
         {MuFun Collaboration} and {Udalski}, A. and {Kubiak}, M. and
         {Szyma{\'n}ski}, M.~K. and {Szewczyk}, O. and {{\.Z}ebru{\'n}}, K. and
         {Wyrzykowski}, {\L}. and {Soszy{\'n}ski}, I. and {Pietrzy{\'n}ski}, G. and
         {OGLE Collaboration} and {Albrow}, M.~D. and {Beaulieu}, J. -P. and
         {Caldwell}, J.~A.~R. and {Cassan}, A. and {Coutures}, C. and
         {Dominik}, M. and {Donatowicz}, J. and {Fouqu{\'e}}, P. and
         {Greenhill}, J. and {Hill}, K. and {Horne}, K. and
         {J{\o}rgensen}, S.~F. and {J{\o}rgensen}, U.~G. and {Kane}, S. and
         {Kubas}, D. and {Martin}, R. and {Menzies}, J. and {Pollard}, K.~R. and
         {Sahu}, K.~C. and {Wambsganss}, J. and {Watson}, R. and {Williams}, A. and
         {PLANET Collaboration}},
        title = "{OGLE-2003-BLG-238: Microlensing Mass Estimate of an Isolated Star}",
      journal = {\apj},
         year = "2004",
        month = "Dec",
       volume = {617},
       number = {2},
        pages = {1307-1315},
          doi = {10.1086/425678},
archivePrefix = {arXiv},
       eprint = {astro-ph/0404394},
 primaryClass = {astro-ph},
       adsurl = {https://ui.adsabs.harvard.edu/abs/2004ApJ...617.1307J}
}

@ARTICLE{KB200414,
       author = {{Zang}, Weicheng and {Han}, Cheongho and {Kondo}, Iona and {Yee}, Jennifer C. and {Lee}, Chung-Uk and {Gould}, Andrew and {Mao}, Shude and {de Almeida}, Leandro and {Shvartzvald}, Yossi and {Zhang}, Xiangyu and {Albrow}, Michael D. and {Chung}, Sun-Ju and {Hwang}, Kyu-Ha and {Jung}, Youn Kil and {Ryu}, Yoon-Hyun and {Shin}, In-Gu and {Cha}, Sang-Mok and {Kim}, Dong-Jin and {Kim}, Hyoun-Woo and {Kim}, Seung-Lee and {Lee}, Dong-Joo and {Lee}, Yongseok and {Park}, Byeong-Gon and {Pogge}, Richard W. and {Drummond}, John and {Tan}, Thiam-Guan and {Nascimento J{\'u}nior}, Jos{\'e} Dias do and {Maoz}, Dan and {Penny}, Matthew T. and {Zhu}, Wei and {Bond}, Ian A. and {Abe}, Fumio and {Barry}, Richard and {Bennett}, David P. and {Bhattacharya}, Aparna and {Donachie}, Martin and {Fujii}, Hirosane and {Fukui}, Akihiko and {Hirao}, Yuki and {Itow}, Yoshitaka and {Kirikawa}, Rintaro and {Koshimoto}, Naoki and {Alex Li}, Man Cheung and {Matsubara}, Yutaka and {Muraki}, Yasushi and {Miyazaki}, Shota and {Olmschenk}, Greg and {Ranc}, Cl{\'e}ment and {Rattenbury}, Nicholas J. and {Satoh}, Yuki and {Shoji}, Hikaru and {Silva}, Stela Ishitani and {Sumi}, Takahiro and {Suzuki}, Daisuke and {Tanaka}, Yuzuru and {Tristram}, Paul J. and {Yamawaki}, Tsubasa and {Yonehara}, Atsunori and {Petric}, Andreea and {Burdullis}, Todd and {Fouqu{\'e}}, Pascal},
        title = "{An Earth-mass planet in a time of COVID-19: KMT-2020-BLG-0414Lb}",
      journal = {Research in Astronomy and Astrophysics},
         year = 2021,
        month = nov,
       volume = {21},
       number = {9},
          eid = {239},
        pages = {239},
          doi = {10.1088/1674-4527/21/9/239},
archivePrefix = {arXiv},
       eprint = {2103.01896},
 primaryClass = {astro-ph.EP},
       adsurl = {https://ui.adsabs.harvard.edu/abs/2021RAA....21..239Z}
}

@ARTICLE{GG1997,
       author = {{Gaudi}, B. Scott and {Gould}, Andrew},
        title = "{Planet Parameters in Microlensing Events}",
      journal = {\apj},
         year = 1997,
        month = sep,
       volume = {486},
       number = {1},
        pages = {85-99},
          doi = {10.1086/304491},
archivePrefix = {arXiv},
       eprint = {astro-ph/9610123},
 primaryClass = {astro-ph},
       adsurl = {https://ui.adsabs.harvard.edu/abs/1997ApJ...486...85G}
}

@ARTICLE{emcee2,
       author = {{Goodman}, Jonathan and {Weare}, Jonathan},
        title = "{Ensemble samplers with affine invariance}",
      journal = {Communications in Applied Mathematics and Computational Science},
         year = 2010,
        month = jan,
       volume = {5},
       number = {1},
        pages = {65-80},
          doi = {10.2140/camcos.2010.5.65},
       adsurl = {https://ui.adsabs.harvard.edu/abs/2010CAMCS...5...65G}
}

@ARTICLE{OB191053,
       author = {{Zang}, Weicheng and {Hwang}, Kyu-Ha and {Udalski}, Andrzej and {Wang}, Tianshu and {Zhu}, Wei and {Sumi}, Takahiro and {Yee}, Jennifer C. and {Gould}, Andrew and {Mao}, Shude and {Zhang}, Xiangyu and {Albrow}, Michael D. and {Chung}, Sun-Ju and {Han}, Cheongho and {Jung}, Youn Kil and {Ryu}, Yoon-Hyun and {Shin}, In-Gu and {Shvartzvald}, Yossi and {Cha}, Sang-Mok and {Kim}, Dong-Jin and {Kim}, Hyoun-Woo and {Kim}, Seung-Lee and {Lee}, Chung-Uk and {Lee}, Dong-Joo and {Lee}, Yongseok and {Park}, Byeong-Gon and {Pogge}, Richard W. and {Mr{\'o}z}, Przemek and {Skowron}, Jan and {Poleski}, Radoslaw and {Szyma{\'n}ski}, Micha{\l} K. and {Soszy{\'n}ski}, Igor and {Pietrukowicz}, Pawe{\l} and {Koz{\l}owski}, Szymon and {Ulaczyk}, Krzysztof and {Rybicki}, Krzysztof A. and {Iwanek}, Patryk and {Wrona}, Marcin and {Gromadzki}, Mariusz and {Bond}, Ian A. and {Abe}, Fumio and {Barry}, Richard and {Bennett}, David P. and {Bhattacharya}, Aparna and {Donachie}, Martin and {Fujii}, Hirosane and {Fukui}, Akihiko and {Hirao}, Yuki and {Itow}, Yoshitaka and {Kirikawa}, Rintaro and {Kondo}, Iona and {Koshimoto}, Naoki and {Li}, Man Cheung Alex and {Matsubara}, Yutaka and {Muraki}, Yasushi and {Miyazaki}, Shota and {Olmschenk}, Greg and {Ranc}, Cl{\'e}ment and {Rattenbury}, Nicholas J. and {Satoh}, Yuki and {Shoji}, Hikaru and {Ishitani Silva}, Stela and {Suzuki}, Daisuke and {Tanaka}, Yuzuru and {Tristram}, Paul J. and {Yamawaki}, Tsubasa and {Yonehara}, Atsunori and {Beichman}, Charles A. and {Bryden}, Geoffery and {Calchi Novati}, Sebastiano and {Carey}, Sean and {Gaudi}, B. Scott and {Henderson}, Calen B. and {Johnson}, Samson and {Spitzer Team}},
        title = "{Systematic KMTNet Planetary Anomaly Search. I. OGLE-2019-BLG-1053Lb, a Buried Terrestrial Planet}",
      journal = {\aj},
         year = 2021,
        month = oct,
       volume = {162},
       number = {4},
          eid = {163},
        pages = {163},
          doi = {10.3847/1538-3881/ac12d4},
archivePrefix = {arXiv},
       eprint = {2103.11880},
 primaryClass = {astro-ph.EP},
       adsurl = {https://ui.adsabs.harvard.edu/abs/2021AJ....162..163Z}
}

@ARTICLE{Guo2026,
       author = {{Guo}, Kangrou and {Ida}, Shigeru and {Ogihara}, Masahiro},
        title = "{Formation of Free-floating Planets via Ejection: Population Synthesis with a Realistic IMF and Comparison to Microlensing Observations}",
      journal = {\apj},
         year = 2026,
        month = jan,
       volume = {997},
       number = {1},
          eid = {34},
        pages = {34},
          doi = {10.3847/1538-4357/ae0cc8},
archivePrefix = {arXiv},
       eprint = {2511.03246},
 primaryClass = {astro-ph.EP},
       adsurl = {https://ui.adsabs.harvard.edu/abs/2026ApJ...997...34G}
}

@ARTICLE{KB232669,
       author = {{Jung}, Youn Kil and {Hwang}, Kyu-Ha and {Yang}, Hongjing and {Gould}, Andrew and {Yee}, Jennifer C. and {Han}, Cheongho and {Albrow}, Michael D. and {Chung}, Sun-Ju and {Ryu}, Yoon-Hyun and {Shin}, In-Gu and {Shvartzvald}, Yossi and {Zang}, Weicheng and {Cha}, Sang-Mok and {Kim}, Dong-Jin and {Kim}, Seung-Lee and {Lee}, Chung-Uk and {Lee}, Dong-Joo and {Lee}, Yongseok and {Park}, Byeong-Gon and {Pogge}, Richard W.},
        title = "{KMT-2023-BLG-2669: Ninth Free-floating Planet Candidate with {\ensuremath{\theta}} $_{E}$ Measurements}",
      journal = {\aj},
         year = 2024,
        month = oct,
       volume = {168},
       number = {4},
          eid = {152},
        pages = {152},
          doi = {10.3847/1538-3881/ad6b12},
archivePrefix = {arXiv},
       eprint = {2405.16857},
 primaryClass = {astro-ph.EP},
       adsurl = {https://ui.adsabs.harvard.edu/abs/2024AJ....168..152J}
}

@ARTICLE{Han2024orbital,
       author = {{Han}, Cheongho and {Udalski}, Andrzej and {Bond}, Ian A. and {Lee}, Chung-Uk and {Gould}, Andrew and {Albrow}, Michael D. and {Chung}, Sun-Ju and {Hwang}, Kyu-Ha and {Jung}, Youn Kil and {Kim}, Hyoun-Woo and {Ryu}, Yoon-Hyun and {Shvartzvald}, Yossi and {Shin}, In-Gu and {Yee}, Jennifer C. and {Yang}, Hongjing and {Zang}, Weicheng and {Cha}, Sang-Mok and {Kim}, Doeon and {Kim}, Dong-Jin and {Kim}, Seung-Lee and {Lee}, Dong-Joo and {Lee}, Yongseok and {Park}, Byeong-Gon and {Pogge}, Richard W. and {Mr{\'o}z}, Przemek and {Szyma{\'n}ski}, Micha{\l} K. and {Skowron}, Jan and {Poleski}, Rados{\l}aw and {Soszy{\'n}ski}, Igor and {Pietrukowicz}, Pawe{\l} and {Koz{\l}owski}, Szymon and {Rybicki}, Krzysztof A. and {Iwanek}, Patryk and {Ulaczyk}, Krzysztof and {Wrona}, Marcin and {Gromadzki}, Mariusz and {Mr{\'o}z}, Mateusz J. and {Abe}, Fumio and {Barry}, Richard and {Bennett}, David P. and {Bhattacharya}, Aparna and {Fujii}, Hirosame and {Fukui}, Akihiko and {Hamada}, Ryusei and {Hirao}, Yuki and {Silva}, Stela Ishitani and {Itow}, Yoshitaka and {Kirikawa}, Rintaro and {Koshimoto}, Naoki and {Matsubara}, Yutaka and {Miyazaki}, Shota and {Muraki}, Yasushi and {Olmschenk}, Greg and {Ranc}, Cl{\'e}ment and {Rattenbury}, Nicholas J. and {Satoh}, Yuki and {Sumi}, Takahiro and {Suzuki}, Daisuke and {Tomoyoshi}, Mio and {Tristram}, Paul J. and {Vandorou}, Aikaterini and {Yama}, Hibiki and {Yamashita}, Kansuke},
        title = "{OGLE-2018-BLG-0971, MOA-2023-BLG-065, and OGLE-2023-BLG-0136: Microlensing events with prominent orbital effects}",
      journal = {\aap},
         year = 2024,
        month = jun,
       volume = {686},
          eid = {A234},
        pages = {A234},
          doi = {10.1051/0004-6361/202349063},
archivePrefix = {arXiv},
       eprint = {2404.05912},
 primaryClass = {astro-ph.SR},
       adsurl = {https://ui.adsabs.harvard.edu/abs/2024A&A...686A.234H}
}

@ARTICLE{KB172197_KB221790_KB222076_KB232209,
       author = {{Han}, Cheongho and {Albrow}, Michael D. and {Lee}, Chung-Uk and {Chung}, Sun-Ju and {Gould}, Andrew and {Hwang}, Kyu-Ha and {Jung}, Youn Kil and {Ryu}, Yoon-Hyun and {Shvartzvald}, Yossi and {Shin}, In-Gu and {Yee}, Jennifer C. and {Yang}, Hongjing and {Zang}, Weicheng and {Cha}, Sang-Mok and {Kim}, Doeon and {Kim}, Dong-Jin and {Kim}, Seung-Lee and {Lee}, Dong-Joo and {Lee}, Yongseok and {Park}, Byeong-Gon and {Pogge}, Richard W.},
        title = "{Four Planets Found through Microlensing Events Involving Faint Source Stars}",
      journal = {\aj},
         year = 2025,
        month = jun,
       volume = {169},
       number = {6},
          eid = {288},
        pages = {288},
          doi = {10.3847/1538-3881/adc5e7},
       adsurl = {https://ui.adsabs.harvard.edu/abs/2025AJ....169..288H}
}

@ARTICLE{Han2324six,
       author = {{Han}, Cheongho and {Lee}, Chung-Uk and {Udalski}, Andrzej and {Bond}, Ian A. and {Albrow}, Michael D. and {Chung}, Sun-Ju and {Gould}, Andrew and {Jung}, Youn Kil and {Hwang}, Kyu-Ha and {Ryu}, Yoon-Hyun and {Shvartzvald}, Yossi and {Shin}, In-Gu and {Yee}, Jennifer C. and {Zang}, Weicheng and {Yang}, Hongjing and {Cha}, Sang-Mok and {Kim}, Doeon and {Kim}, Dong-Jin and {Kim}, Seung-Lee and {Lee}, Dong-Joo and {Lee}, Yongseok and {Park}, Byeong-Gon and {Pogge}, Richard W. and {Mr{\'o}z}, Przemek and {Szyma{\'n}ski}, Micha{\l} K. and {Skowron}, Jan and {Poleski}, Rados{\l}aw and {Soszy{\'n}ski}, Igor and {Pietrukowicz}, Pawe{\l} and {Koz{\l}owski}, Szymon and {Rybicki}, Krzysztof A. and {Iwanek}, Patryk and {Ulaczyk}, Krzysztof and {Wrona}, Marcin and {Gromadzki}, Mariusz and {Mr{\'o}z}, Mateusz J. and {Jaroszy{\'n}ski}, Micha{\l} and {Kiraga}, Marcin and {Abe}, Fumio and {Bennett}, David P. and {Bhattacharya}, Aparna and {Fukui}, Akihiko and {Hamada}, Ryusei and {Ishitani Silva}, Stela and {Hirao}, Yuki and {Koshimoto}, Naoki and {Matsubara}, Yutaka and {Miyazaki}, Shota and {Muraki}, Yasushi and {Nagai}, Tutumi and {Nunota}, Kansuke and {Olmschenk}, Greg and {Ranc}, Cl{\'e}ment and {Rattenbury}, Nicholas J. and {Satoh}, Yuki and {Sumi}, Takahiro and {Suzuki}, Daisuke and {Terry}, Sean K. and {Tristram}, Paul J. and {Vandorou}, Aikaterini and {Yama}, Hibiki},
        title = "{Six microlensing planets detected via sub-day signals during the 2023─2024 season}",
      journal = {\aap},
         year = 2025,
        month = oct,
       volume = {702},
          eid = {A152},
        pages = {A152},
          doi = {10.1051/0004-6361/202554557},
archivePrefix = {arXiv},
       eprint = {2509.05522},
 primaryClass = {astro-ph.EP},
       adsurl = {https://ui.adsabs.harvard.edu/abs/2025A&A...702A.152H}
}

@ARTICLE{KB200202_KB221551_KB230466_KB250121,
       author = {{Han}, Cheongho and {Lee}, Chung-Uk and {Albrow}, Michael D. and {Chung}, Sun-Ju and {Gould}, Andrew and {Jung}, Youn Kil and {Ha Hwang}, Kyu- and {Ryu}, Yoon-Hyun and {Shvartzvald}, Yossi and {Shin}, In-Gu and {Yee}, Jennifer C. and {Zang}, Weicheng and {Yang}, Hongjing and {Kim}, Doeon and {Kim}, Dong-Jin and {Cha}, Sang-Mok and {Kim}, Seung-Lee and {Lee}, Dong-Joo and {Lee}, Yongseok and {Park}, Byeong-Gon and {Pogge}, Richard W. and {KMTNet Collaboration}},
        title = "{Four Cold Super-Jupiters Revealed by Extended and Complex Microlensing Signals}",
      journal = {\pasp},
         year = 2026,
        month = feb,
       volume = {138},
       number = {2},
          eid = {024401},
        pages = {024401},
          doi = {10.1088/1538-3873/ae3ae0},
archivePrefix = {arXiv},
       eprint = {2601.13450},
 primaryClass = {astro-ph.EP},
       adsurl = {https://ui.adsabs.harvard.edu/abs/2026PASP..138b4401H}
}

@ARTICLE{-4planet,
       author = {{Zang}, Weicheng and {Jung}, Youn Kil and {Yang}, Hongjing and {Zhang}, Xiangyu and {Udalski}, Andrzej and {Yee}, Jennifer C. and {Gould}, Andrew and {Mao}, Shude and {Albrow}, Michael D. and {Chung}, Sun-Ju and {Han}, Cheongho and {Hwang}, Kyu-Ha and {Ryu}, Yoon-Hyun and {Shin}, In-Gu and {Shvartzvald}, Yossi and {Cha}, Sang-Mok and {Kim}, Dong-Jin and {Kim}, Hyoun-Woo and {Kim}, Seung-Lee and {Lee}, Chung-Uk and {Lee}, Dong-Joo and {Lee}, Yongseok and {Park}, Byeong-Gon and {Pogge}, Richard W. and {KMTNet Collaboration} and {Mr{\'o}z}, Przemek and {Skowron}, Jan and {Poleski}, Radoslaw and {Szyma{\'n}ski}, Micha{\l} K. and {Soszy{\'n}ski}, Igor and {Pietrukowicz}, Pawe{\l} and {Koz{\l}owski}, Szymon and {Ulaczyk}, Krzysztof and {Rybicki}, Krzysztof A. and {Iwanek}, Patryk and {Wrona}, Marcin and {Gromadzki}, Mariusz and {OGLE Collaboration} and {Wang}, Hanyue and {Zhang}, Jiyuan and {Zhu}, Wei and {MAP Collaboration}},
        title = "{Systematic KMTNet Planetary Anomaly Search. VII. Complete Sample of q < 10$^{-4}$ Planets from the First 4 yr Survey}",
      journal = {\aj},
         year = 2023,
        month = mar,
       volume = {165},
       number = {3},
          eid = {103},
        pages = {103},
          doi = {10.3847/1538-3881/acb34b},
archivePrefix = {arXiv},
       eprint = {2210.12344},
 primaryClass = {astro-ph.EP},
       adsurl = {https://ui.adsabs.harvard.edu/abs/2023AJ....165..103Z}
}

@ARTICLE{2019_subprime,
       author = {{Jung}, Youn Kil and {Zang}, Weicheng and {Wang}, Hanyue and {Han}, Cheongho and {Gould}, Andrew and {Udalski}, Andrzej and {Albrow}, Michael D. and {Chung}, Sun-Ju and {Hwang}, Kyu-Ha and {Ryu}, Yoon-Hyun and {Shin}, In-Gu and {Shvartzvald}, Yossi and {Yang}, Hongjing and {Yee}, Jennifer C. and {Cha}, Sang-Mok and {Kim}, Dong-Jin and {Kim}, Seung-Lee and {Lee}, Chung-Uk and {Lee}, Dong-Joo and {Lee}, Yongseok and {Park}, Byeong-Gon and {Pogge}, Richard W. and {KMTNet Collaboration} and {Szyma{\'n}ski}, Micha{\l} K. and {Skowron}, Jan and {Poleski}, Radek and {Soszy{\'n}ski}, Igor and {Pietrukowicz}, Pawe{\l} and {Koz{\l}owski}, Szymon and {Ulaczyk}, Krzysztof and {Rybicki}, Krzysztof A. and {Iwanek}, Patryk and {Wrona}, Marcin and {OGLE Collaboration} and {Green}, Jonathan and {Hennerley}, Steve and {Marmont}, Andrew and {Mao}, Shude and {Maoz}, Dan and {McCormick}, Jennie and {Natusch}, Tim and {Penny}, Matthew T. and {Porritt}, Ian and {Zhu}, Wei and {Tsinghua Team} and {FUN Follow-Up Team}},
        title = "{Systematic KMTNet Planetary Anomaly Search. VIII. Complete Sample of 2019 Subprime Field Planets}",
      journal = {\aj},
         year = 2023,
        month = jun,
       volume = {165},
       number = {6},
          eid = {226},
        pages = {226},
          doi = {10.3847/1538-3881/accb8f},
archivePrefix = {arXiv},
       eprint = {2302.13544},
 primaryClass = {astro-ph.EP},
       adsurl = {https://ui.adsabs.harvard.edu/abs/2023AJ....165..226J}
}

@ARTICLE{2016_prime,
       author = {{Shin}, In-Gu and {Yee}, Jennifer C. and {Zang}, Weicheng and {Yang}, Hongjing and {Hwang}, Kyu-Ha and {Han}, Cheongho and {Gould}, Andrew and {Udalski}, Andrzej and {Bond}, Ian A. and {Albrow}, Michael D. and {Chung}, Sun-Ju and {Jung}, Youn Kil and {Ryu}, Yoon-Hyun and {Shvartzvald}, Yossi and {Cha}, Sang-Mok and {Kim}, Dong-Jin and {Kim}, Seung-Lee and {Lee}, Chung-Uk and {Lee}, Dong-Joo and {Lee}, Yongseok and {Park}, Byeong-Gon and {Pogge}, Richard W. and {Mr{\'o}z}, Przemek and {Szyma{\'n}ski}, Micha{\l} K. and {Skowron}, Jan and {Poleski}, Rados{\l}aw and {Soszy{\'n}ski}, Igor and {Pietrukowicz}, Pawe{\l} and {Koz{\l}owski}, Szymon and {Rybicki}, Krzysztof A. and {Iwanek}, Patryk and {Ulaczyk}, Krzysztof and {Wrona}, Marcin and {Gromadzki}, Mariusz and {Abe}, Fumio and {Barry}, Richard and {Bennett}, David P. and {Bhattacharya}, Aparna and {Fujii}, Hirosane and {Fukui}, Akihiko and {Hamada}, Ryusei and {Hirao}, Yuki and {Silva}, Stela Ishitani and {Itow}, Yoshitaka and {Kirikawa}, Rintaro and {Kondo}, Iona and {Koshimoto}, Naoki and {Matsubara}, Yutaka and {Miyazaki}, Shota and {Muraki}, Yasushi and {Olmschenk}, Greg and {Ranc}, Cl{\'e}ment and {Rattenbury}, Nicholas J. and {Satoh}, Yuki and {Sumi}, Takahiro and {Suzuki}, Daisuke and {Tomoyoshi}, Mio and {Tristram}, Paul J. and {Vandorou}, Aikaterini and {Yama}, Hibiki and {Yamashita}, Kansuke},
        title = "{Systematic KMTNet Planetary Anomaly Search. IX. Complete Sample of 2016 Prime-field Planets}",
      journal = {\aj},
         year = 2023,
        month = sep,
       volume = {166},
       number = {3},
          eid = {104},
        pages = {104},
          doi = {10.3847/1538-3881/ace96d},
archivePrefix = {arXiv},
       eprint = {2303.16881},
 primaryClass = {astro-ph.EP},
       adsurl = {https://ui.adsabs.harvard.edu/abs/2023AJ....166..104S}
}

@ARTICLE{Bennett2008,
       author = {{Bennett}, D.~P. and {Bond}, I.~A. and {Udalski}, A. and {Sumi}, T. and {Abe}, F. and {Fukui}, A. and {Furusawa}, K. and {Hearnshaw}, J.~B. and {Holderness}, S. and {Itow}, Y. and {Kamiya}, K. and {Korpela}, A.~V. and {Kilmartin}, P.~M. and {Lin}, W. and {Ling}, C.~H. and {Masuda}, K. and {Matsubara}, Y. and {Miyake}, N. and {Muraki}, Y. and {Nagaya}, M. and {Okumura}, T. and {Ohnishi}, K. and {Perrott}, Y.~C. and {Rattenbury}, N.~J. and {Sako}, T. and {Saito}, To. and {Sato}, S. and {Skuljan}, L. and {Sullivan}, D.~J. and {Sweatman}, W.~L. and {Tristram}, P.~J. and {Yock}, P.~C.~M. and {Kubiak}, M. and {Szyma{\'n}ski}, M.~K. and {Pietrzy{\'n}ski}, G. and {Soszy{\'n}ski}, I. and {Szewczyk}, O. and {Wyrzykowski}, {\L}. and {Ulaczyk}, K. and {Batista}, V. and {Beaulieu}, J.~P. and {Brillant}, S. and {Cassan}, A. and {Fouqu{\'e}}, P. and {Kervella}, P. and {Kubas}, D. and {Marquette}, J.~B.},
        title = "{A Low-Mass Planet with a Possible Sub-Stellar-Mass Host in Microlensing Event MOA-2007-BLG-192}",
      journal = {\apj},
         year = 2008,
        month = sep,
       volume = {684},
       number = {1},
        pages = {663-683},
          doi = {10.1086/589940},
archivePrefix = {arXiv},
       eprint = {0806.0025},
 primaryClass = {astro-ph},
       adsurl = {https://ui.adsabs.harvard.edu/abs/2008ApJ...684..663B}
}

@ARTICLE{scipy,
       author = {{Virtanen}, Pauli and {Gommers}, Ralf and {Oliphant}, Travis E. and {Haberland}, Matt and {Reddy}, Tyler and {Cournapeau}, David and {Burovski}, Evgeni and {Peterson}, Pearu and {Weckesser}, Warren and {Bright}, Jonathan and {van der Walt}, St{\'e}fan J. and {Brett}, Matthew and {Wilson}, Joshua and {Millman}, K. Jarrod and {Mayorov}, Nikolay and {Nelson}, Andrew R.~J. and {Jones}, Eric and {Kern}, Robert and {Larson}, Eric and {Carey}, C.~J. and {Polat}, {\.I}lhan and {Feng}, Yu and {Moore}, Eric W. and {VanderPlas}, Jake and {Laxalde}, Denis and {Perktold}, Josef and {Cimrman}, Robert and {Henriksen}, Ian and {Quintero}, E.~A. and {Harris}, Charles R. and {Archibald}, Anne M. and {Ribeiro}, Ant{\^o}nio H. and {Pedregosa}, Fabian and {van Mulbregt}, Paul and {SciPy 1. 0 Contributors}},
        title = "{SciPy 1.0: fundamental algorithms for scientific computing in Python}",
      journal = {Nature Methods},
         year = 2020,
        month = feb,
       volume = {17},
        pages = {261-272},
          doi = {10.1038/s41592-019-0686-2},
archivePrefix = {arXiv},
       eprint = {1907.10121},
 primaryClass = {cs.MS},
       adsurl = {https://ui.adsabs.harvard.edu/abs/2020NatMe..17..261V}
}

@ARTICLE{HanBD2325,
       author = {{Han}, Cheongho and {Udalski}, Andrzej and {Bond}, Ian A. and {Lee}, Chung-Uk and {Albrow}, Michael D. and {Chung}, Sun-Ju and {Gould}, Andrew and {Jung}, Youn Kil and {Hwang}, Kyu-Ha and {Ryu}, Yoon-Hyun and {Shvartzvald}, Yossi and {Shin}, In-Gu and {Yee}, Jennifer C. and {Zang}, Weicheng and {Yang}, Hongjing and {Kim}, Doeon and {Kim}, Dong-Jin and {Kim}, Seung-Lee and {Lee}, Dong-Joo and {Cha}, Sang-Mok and {Lee}, Yongseok and {Park}, Byeong-Gon and {Pogge}, Richard W. and {Mr{\'o}z}, Przemek and {Szyma{\'n}ski}, Micha{\l} K. and {Skowron}, Jan and {Poleski}, Rados{\l}aw and {Soszy{\'n}ski}, Igor and {Pietrukowicz}, Pawe{\l} and {Koz{\l}owski}, Szymon and {Rybicki}, Krzysztof A. and {Iwanek}, Patryk and {Ulaczyk}, Krzysztof and {Wrona}, Marcin and {Gromadzki}, Mariusz and {Mr{\'o}z}, Mateusz J. and {Abe}, Fumio and {Bennett}, David P. and {Bhattacharya}, Aparna and {Hamada}, Ryusei and {Hirao}, Yuki and {Idei}, Asahi and {Ishitani Silva}, Stela and {Makida}, Shuma and {Miyazaki}, Shota and {Muraki}, Yasushi and {Nagai}, Tutumi and {Nagano}, Togo and {Nakayama}, Seiya and {Nishio}, Mayu and {Nunota}, Kansuke and {Ogawa}, Ryo and {Oishi}, Ryunosuke and {Okumoto}, Yui and {Olmschenk}, Greg and {Ranc}, Cl{\'e}ment and {Rattenbury}, Nicholas J. and {Satoh}, Yuki and {Sumi}, Takahiro and {Suzuki}, Daisuke and {Tamaoki}, Takuto and {Terry}, Sean K. and {Tristram}, Paul J. and {Vandorou}, Aikaterini and {Yama}, Hibiki},
        title = "{Candidate Microlensing Brown Dwarfs in Binary Lens Systems from the 2023--2025 Observing Seasons}",
      journal = {arXiv e-prints},
         year = 2026,
        month = apr,
          eid = {arXiv:2604.07932},
        pages = {arXiv:2604.07932},
          doi = {10.48550/arXiv.2604.07932},
archivePrefix = {arXiv},
       eprint = {2604.07932},
 primaryClass = {astro-ph.SR},
       adsurl = {https://ui.adsabs.harvard.edu/abs/2026arXiv260407932H}
}

\end{CJK*}
\end{document}